\documentclass[twocolumn,trackchanges]{aastex63}

\newcommand{\TESS}{\emph{TESS}}
\newcommand{\fermi}{\emph{Fermi}}
\newcommand{\swift}{\emph{Swift}}

\accepted{2020 July 14}
\submitjournal{ApJ}

\turnoffeditone

\shorttitle{Short-Timescale Variability of BL Lac}
\shortauthors{Weaver et al.}

\begin{document}

\title{Multi-Wavelength Variability of BL Lacertae Measured with High Time Resolution}

\correspondingauthor{Zachary R. Weaver}
\email{zweaver@bu.edu}


\author[0000-0001-6314-0690]{Weaver, Z.R.}
\affiliation{Institute for Astrophysical Research, Boston University, 725 Commonwealth Avenue, Boston, MA 02215, USA}

\author[0000-0003-1318-8535]{Williamson, K.E.}
\affiliation{Institute for Astrophysical Research, Boston University, 725 Commonwealth Avenue, Boston, MA 02215, USA}

\author[0000-0001-6158-1708]{Jorstad, S.G.}
\affiliation{Institute for Astrophysical Research, Boston University, 725 Commonwealth Avenue, Boston, MA 02215, USA}
\affiliation{Astronomical Institute, St. Petersburg State University, Universitetskij Pr. 28, Petrodvorets, St. Petersburg 198504, Russia}

\author[0000-0001-7396-3332]{Marscher, A.P.}
\affiliation{Institute for Astrophysical Research, Boston University, 725 Commonwealth Avenue, Boston, MA 02215, USA}

\author[0000-0002-4640-4356]{Larionov, V.M.}
\affiliation{Astronomical Institute, St. Petersburg State University, Universitetskij Pr. 28, Petrodvorets, St. Petersburg 198504, Russia}
\affiliation{Main (Pulkovo) Astronomical Observatory of RAS, Pulkovskoye shosse 60, St. Petersburg 196149, Russia}

\author[0000-0003-1784-2784]{Raiteri, C.M.}
\affiliation{INAF, Osservatorio Astrofisico di Torino, Via Osservatorio 20, I-10025 Pino Torinese, Italy}

\author[0000-0003-1743-6946]{Villata, M.}
\affiliation{INAF, Osservatorio Astrofisico di Torino, Via Osservatorio 20, I-10025 Pino Torinese, Italy}

\author[0000-0002-0433-9656]{Acosta-Pulido, J.A.}
\affiliation{Instituto de Astrof\'isica de Canarias, La Laguna (Canary Islands), Spain}
\affiliation{Departamento de Astrof\'isica, Universidad de La Laguna (ULL), E-38206 La Laguna, Tenerife, Spain}

\author{Bachev, R.}
\affiliation{Institute of Astronomy and National Astronomical Observatory, Bulgarian Academy of Sciences, 72 Tsarigradsko shosse Blvd., 1784 Sofia, Bulgaria}

\author{Baida, G.V}
\affiliation{Crimean Astrophysical Observatory RAS, P/O Nauchny, 298409, Russia}

\author[0000-0002-9844-1730]{Balonek, T.J.}
\affiliation{Department of Physics and Astronomy, Colgate University, 13 Oak Drive, Hamilton, New York 13346, USA}

\author[0000-0003-1018-2613]{Ben\'itez, E.}
\affiliation{Instituto de Astronom\'ia, Universidad Nacional Aut\'onoma de M\'exico, Apdo. Postal 70-264, CDMX 04510, M\'exico}

\author{Borman, G.A.}
\affiliation{Crimean Astrophysical Observatory RAS, P/O Nauchny, 298409, Russia}

\author{Bozhilov, V.}
\affiliation{Department of Astronomy, Faculty of Physics, University of Sofia, BG-1164 Sofia, Bulgaria}

\author[0000-0001-5843-5515]{Carnerero, M.I.}
\affiliation{INAF, Osservatorio Astrofisico di Torino, Via Osservatorio 20, I-10025 Pino Torinese, Italy}

\author{Carosati, D.}
\affiliation{EPT Observatories, Tijarafe, E-38780 La Palma, Spain}
\affiliation{INAF, TNG Fundacion Galileo Galilei, E-38712 La Palma, Spain}

\author{Chen, W.P.}
\affiliation{Graduate Institute of Astronomy, National Central University, 300 Jhongda Road, Zhongli, Taoyuan, 32001, Taiwan}

\author{Damljanovic, G.}
\affiliation{Astronomical Observatory, Volgina 7, 11060, Belgrade, Serbia}

\author{Dhiman, V.}
\affiliation{Aryabhatta Research Institute of Observational Sciences (ARIES), Manora Peak, Nainital - 263 001, India}

\author{Dougherty, D.J.}
\affiliation{Department of Physics and Astronomy, Colgate University, 13 Oak Drive, Hamilton, New York 13346, USA}

\author{Ehgamberdiev, S.A.}
\affiliation{Ulugh Beg Astronomical Institute, Maidanak Observatory, Uzbekistan}

\author[0000-0002-3953-6676]{Grishina, T.S.}
\affiliation{Astronomical Institute, St. Petersburg State University, Universitetskij Pr. 28, Petrodvorets, St. Petersburg 198504, Russia}

\author{Gupta, A.C.}
\affiliation{Aryabhatta Research Institute of Observational Sciences (ARIES), Manora Peak, Nainital - 263 001, India}

\author[0000-0002-0330-1648]{Hart, M.}
\affiliation{Institute for Astrophysical Research, Boston University, 725 Commonwealth Avenue, Boston, MA 02215, USA}

\author[0000-0002-4711-7658]{Hiriart, D.}
\affiliation{Instituto de Astronom\'ia, Universidad Nacional Aut\'onoma de M\'exico, Ensenada, Baja California, M\'exico}

\author{Hsiao, H.Y.}
\affiliation{Graduate Institute of Astronomy, National Central University, 300 Jhongda Road, Zhongli, Taoyuan, 32001, Taiwan}

\author{Ibryamov, S.}
\affiliation{Department of Physics and Astronomy, Faculty of Natural Sciences, University of Shumen, 115 Universitetska Str., 9712 Shumen, Bulgaria}

\author{Joner, M.}
\affiliation{Department of Physics and Astronomy, Brigham Young University, Provo, UT 84602, USA}

\author{Kimeridze, G.N.}
\affiliation{Abastumani Observatory, Mt. Kanobili, 0301 Abastumani, Georgia}

\author[0000-0001-9518-337X]{Kopatskaya, E.N.}
\affiliation{Astronomical Institute, St. Petersburg State University, Universitetskij Pr. 28, Petrodvorets, St. Petersburg 198504, Russia}

\author{Kurtanidze, O.M.}
\affiliation{Abastumani Observatory, Mt. Kanobili, 0301 Abastumani, Georgia}
\affiliation{Engelhardt Astronomical Observatory, Kazan Federal University, Tatarstan, Russia}
\affiliation{Landessternwarte, Zentrum f\"ur Astronomie der Universit\"at Heidelberg, K\"onigstuhl 12, 69117 Heidelberg, Germany}

\author{Kurtanidze, S.O.}
\affiliation{Abastumani Observatory, Mt. Kanobili, 0301 Abastumani, Georgia}
\affiliation{Samtskhe-Javakheti State University, 92 Shota Rustaveli St. Akhaltsikhe, Georgia}
\affiliation{Landessternwarte, Zentrum f\"ur Astronomie der Universit\"at Heidelberg, K\"onigstuhl 12, 69117 Heidelberg, Germany}

\author[0000-0002-2471-6500]{Larionova, E.G.}
\affiliation{Astronomical Institute, St. Petersburg State University, Universitetskij Pr. 28, Petrodvorets, St. Petersburg 198504, Russia}

\author{Matsumoto, K.}
\affiliation{Astronomical Institute, Osaka Kyoiku University, Osaka, 582-8582, Japan}

\author{Matsumura, R.}
\affiliation{Astronomical Institute, Osaka Kyoiku University, Osaka, 582-8582, Japan}

\author{Minev, M.}
\affiliation{Department of Astronomy, Faculty of Physics, University of Sofia, BG-1164 Sofia, Bulgaria}

\author{Mirzaqulov, D.O.}
\affiliation{Ulugh Beg Astronomical Institute, Maidanak Observatory, Uzbekistan}

\author[0000-0002-9407-7804]{Morozova, D.A.}
\affiliation{Astronomical Institute, St. Petersburg State University, Universitetskij Pr. 28, Petrodvorets, St. Petersburg 198504, Russia}

\author[0000-0001-9858-4355]{Nikiforova, A.A.}
\affiliation{Astronomical Institute, St. Petersburg State University, Universitetskij Pr. 28, Petrodvorets, St. Petersburg 198504, Russia}
\affiliation{Main (Pulkovo) Astronomical Observatory of RAS, Pulkovskoye shosse 60, St. Petersburg 196149, Russia}

\author{Nikolashvili, M.G.}
\affiliation{Abastumani Observatory, Mt. Kanobili, 0301 Abastumani, Georgia}
\affiliation{Landessternwarte, Zentrum f\"ur Astronomie der Universit\"at Heidelberg, K\"onigstuhl 12, 69117 Heidelberg, Germany}

\author{Ovcharov, E.}
\affiliation{Department of Astronomy, Faculty of Physics, University of Sofia, BG-1164 Sofia, Bulgaria}

\author{Rizzi, N.} 
\affiliation{Osservatorio Astronomico Sirio, Grotte di Castellana, Italy}

\author{Sadun, A.}
\affiliation{Department of Physics, University of Colorado, Denver, CO 80217, USA}

\author[0000-0003-4147-3851]{Savchenko, S.S.}
\affiliation{Astronomical Institute, St. Petersburg State University, Universitetskij Pr. 28, Petrodvorets, St. Petersburg 198504, Russia}

\author{Semkov, E.}
\affiliation{Institute of Astronomy and National Astronomical Observatory, Bulgarian Academy of Sciences, 72 Tsarigradsko shosse Blvd., 1784 Sofia, Bulgaria}

\author{Slater, J.J.}
\affiliation{Department of Physics and Astronomy, Colgate University, 13 Oak Drive, Hamilton, New York 13346, USA}

\author{Smith, K.L.}
\affiliation{KIPAC at SLAC, Stanford University, Menlo Park, CA 94025, USA}

\author{Stojanovic, M.}
\affiliation{Astronomical Observatory, Volgina 7, 11060, Belgrade, Serbia}

\author{Strigachev, A.}
\affiliation{Institute of Astronomy and National Astronomical Observatory, Bulgarian Academy of Sciences, 72 Tsarigradsko shosse Blvd., 1784 Sofia, Bulgaria}

\author[0000-0002-9907-9876]{Troitskaya, Yu.V.}
\affiliation{Astronomical Institute, St. Petersburg State University, Universitetskij Pr. 28, Petrodvorets, St. Petersburg 198504, Russia}

\author[0000-0002-4218-0148]{Troitsky, I.S.}
\affiliation{Astronomical Institute, St. Petersburg State University, Universitetskij Pr. 28, Petrodvorets, St. Petersburg 198504, Russia}

\author{Tsai, A.L.}
\affiliation{Graduate Institute of Astronomy, National Central University, 300 Jhongda Road, Zhongli, Taoyuan, 32001, Taiwan}

\author{Vince, O.}
\affiliation{Astronomical Observatory, Volgina 7, 11060, Belgrade, Serbia}

\author{Valcheva, A.}
\affiliation{Department of Astronomy, Faculty of Physics, University of Sofia, BG-1164 Sofia, Bulgaria}

\author[0000-0002-8293-0214]{Vasilyev, A.A.}
\affiliation{Astronomical Institute, St. Petersburg State University, Universitetskij Pr. 28, Petrodvorets, St. Petersburg 198504, Russia}

\author{Zaharieva, E.}
\affiliation{Department of Astronomy, Faculty of Physics, University of Sofia, BG-1164 Sofia, Bulgaria}

\author{Zhovtan, A.V.}
\affiliation{Crimean Astrophysical Observatory RAS, P/O Nauchny, 298409, Russia}


\begin{abstract}
        In an effort to locate the sites of emission at different frequencies and physical processes causing variability in blazar jets, we have obtained high time-resolution observations of BL Lacertae over a wide wavelength range: with the \emph{Transiting Exoplanet Survey Satellite} (TESS) at 6,000-10,000 \AA\ with 2-minute cadence; with the Neil Gehrels \swift\ satellite at optical, UV, and X-ray bands; with the Nuclear Spectroscopic Telescope Array at hard X-ray bands; with the \fermi\ Large Area Telescope at $\gamma$-ray energies; and with the Whole Earth Blazar Telescope for measurement of the optical flux density and polarization. All light curves are correlated, with similar structure on timescales from hours to days. The shortest timescale of variability at optical frequencies observed with \TESS\ is $\sim 0.5$ hr. The most common timescale is $13\pm1$~hr, comparable with the minimum timescale of X-ray variability, 14.5 hr. The multi-\edit1{wavelength} 
        variability properties cannot be explained by a change solely in the Doppler factor of the emitting plasma. The polarization behavior implies that there are both ordered and turbulent components to the magnetic field in the jet. Correlation analysis indicates that the X-ray variations lag behind the $\gamma$-ray and optical light curves by up to $\sim 0.4$ days. The timescales of variability, cross-frequency lags, and polarization properties can be explained by turbulent plasma that is energized by a shock in the jet and subsequently loses energy to synchrotron and inverse Compton radiation in a magnetic field of strength $\sim3$ G. 
\end{abstract}



\section{Introduction}
\label{sec:Introduction}

Blazars are a class of active galactic nuclei (AGN) that possess extreme characteristics across the electromagnetic spectrum \citep{Angel1980}. They are the most common extragalactic sources of $\gamma$-ray photons with energies
$\geq 0.1$ GeV \citep[VHE: $E>100$ GeV, e.g.,][]{Abdollahi2020, Magic2019}. They are thought to be powered by relativistic jets of high-energy plasma flowing away from the central engine at nearly the speed of light \citep[e.g.,][]{Lister2016, Jorstad2017}, with trajectories closely aligned to the line of sight. The observed phenomena include ultraluminous emission \citep[apparent luminosity as high as $\sim10^{50}$ erg s$^{-1}$, e.g.,][]{Abdo2010intro, Abdo2011, Giommi2012, Senturk2013},
high amplitudes of variability on timescales as short as several minutes at various wavebands \citep[e.g.,][]{HESS2010, Jorstad2013, Weaver2019}, and high degrees of optical linear polarization \cite[which can exceed 40\%; e.g.,][]{Smith2016}.
Both theoretical work \citep[e.g.,][]{Konigl1981, Marscher1987} and observations \citep[e.g.,][]{Hartman1999,Jorstad2001,Lister2011} have found a tight connection between the high-energy and radio emission from the jets.

Blazars are split into two classes: flat-spectrum radio quasars (FSRQs) and BL Lacertae objects (BLs), based on their optical emission-line properties and compact radio morphologies \citep{Weymann1991, Stickel1991,Urry1995,Wardle1984}. The spectral energy distributions (SEDs) of both blazar types generally consist of two humps. The first, located between $10^{13}$ and $10^{17}$ Hz, is attributed to synchrotron radiation by relativistic electrons, and the second, peaking between 1 MeV and 100 GeV, is commonly interpreted as inverse Compton scattering of infrared/optical/UV photons by the same population of electrons \citep{Sikora2009}.
BLs are further divided by their synchrotron peaks into high (HBL), intermediate (IBL), and low (LBL) frequency peaking varieties, with $\nu_{\text{peak}}< 10^{14}$ Hz for LBLs, $10^{14}$- $10^{15}$ Hz for IBLs, and $> 10^{15}$ Hz for HBLs \citep{Padovani1995, Abdo2010intro2}.

BL Lacertae \citep[hereafter BL Lac, redshift $z = 0.069$;][]{Miller1978} is the prototype of BL Lac objects. It is usually classified as an LBL \citep{Nilsson2018}, but is sometimes listed as an IBL \citep{Ackermann2011, Hervet2016}. The blazar has been a target of numerous multi-wavelength observing campaigns \citep[e.g.,][]{Hagenthorn2002, Gaur2015, Agarwal2015, Wierzcholska2015,  Abeysekara2018, Bhatta2018, Magic2019},
including several carried out under the Whole Earth Blazar Telescope (WEBT) GLAST-AGILE Support Program \citep[GASP; e.g.,][]{Villata2002, Villata2004, Villata2004b, Villata2009, Bottcher2003, Bach2006, Raiteri2009, Raiteri2010}.

\citet{Abdo2011BLLac} described a campaign in which BL Lac was in a low, relatively quiescent $\gamma$-ray state. The low-level $\gamma$-ray emission was explained as inverse Compton (IC) scattering of photons originating outside the jet (external IC radiation, EIC) in addition to IC scattering of in-jet photons (synchrotron self-Compton emission, SSC) \citep[e.g.,][]{Madejski1999, Bottcher2000}. Based on \fermi, \swift, Submillimeter Array, and WEBT observations from 2008 to 2012, as well as data from other studies, \citet{Raiteri2013} interpreted the variability of emission from BL Lac in terms of changes in the orientation of the emitting regions, possibly caused by a shock oriented perpendicular to the jet axis.

More recently, \citet{Wehrle2016} extended the period analyzed by \cite{Raiteri2013} by one year to include an extended interval of erratic changes in $\gamma$-ray flux. Their study filled gaps in the SED with \emph{Herschel} far-infrared and Nuclear Spectroscopic Telescope Array (NuSTAR) hard X-ray observations. They described the flaring nature of the source in terms of turbulent plasma flowing across quasi-stationary shocks within 5 pc of the supermassive black hole, with high-energy electrons accelerated at the shock fronts.

\edit1{A number of studies have been performed to analyze the optical polarization behavior of BL Lac \citep[e.g.,][]{Hagenthorn2002, Sakimoto2013}. A two-component system was proposed to describe changes of polarization parameters with flux and time: a long-lived underlying source of polarized radiation (perhaps variable on timescales of years) plus several short-lived components (associated with flares) with randomly oriented polarization directions and high degrees of polarization. \cite{BH2009} have employed a Monte Carlo method to simulate simultaneous photometric and polarimetric data of BL Lac over a period of 22 years within such a system. These authors have found that the observed
photometric and polarimetric variability of BL Lac can be explained within a model
containing a steady component with a high degree of polarization, $\sim$40\%, and
a position angle of polarization along the parsec-scale jet direction, plus 10$\pm$5
components with variable polarization.}

Exploration of the complex emission mechanisms and physical processes that operate in blazar jets requires observations of variability on both long and short timescales. For example, time-series studies of long-term light curves show that for many blazars there is a correlation between $\gamma$-ray and optical variations, with time delays ranging from zero to several days \citep[e.g.,][]{Chatterjee2012, Jorstad2013, Raiteri2013, DAmmando2019}. However, the uncertainties are often comparable with the delays themselves. This limitation can be overcome with short-cadence observations in order to increase the precision of correlation analyses and search for patterns and characteristic timescales of variations at different wavelengths \citep[e.g.,][]{Uttley2002}.

Until recently, intensive monitoring campaigns to identify short-timescale variability have been limited to relatively brief time spans, usually with gaps in temporal coverage. The \emph{Kepler} mission has provided optical short-cadence light curves over spans of weeks and without gaps for several AGN \citep[e.g.,][]{Mushotzky2011, Edelson2014, SmithKL2015, SmithKL2018, Aranzana2018}. The resulting data allow for precise time-series analyses whose value can be amplified by the addition of simultaneous, well-sampled light curves at other wavelengths. The \emph{Transiting Exoplanet Survey Satellite} \citep[\TESS,][]{Ricker2015} is capable of producing short-cadence, unbiased light curves of many more AGNs as it performs a nearly all-sky survey. It samples fluxes over a wide optical to near-IR band with a default cadence of 30 minutes, shortened to 2 minutes for selected objects.

In this paper, we report the results of an observing campaign that combines continuous monitoring of BL Lac with \TESS\, along with broad multi-wavelength coverage from other space- and ground-based facilities. These include
\TESS, the \fermi-Large Area Telescope (LAT), NuSTAR, the Neil Gehrels \swift\ satellite, and ground based telescopes within the WEBT collaboration. The paper is structured as follows. In \S\ref{sec:ObsandData} we describe the reduction of observations at all wavelengths. The light curves are presented in \S\ref{sec:LCs}, first for the entire WEBT 3-month campaign and then for overlapping observations with the other telescopes. We investigate short-term variability observed with \TESS\ at optical and NuSTAR at X-ray frequencies, as well as \emph{R}-band optical polarization, in \S\ref{sec:ShortVariability}. We analyze the optical behavior \edit1{in different bands} 
in \S\ref{sec:MulticolorBehavior}. The polarization behavior over the entire time span, observed with a subset of WEBT telescopes throughout the campaign, is presented in \S\ref{sec:polarization}.
In \S\ref{sec:Correlations} we perform a correlation analysis between the \TESS\ and other light curves to identify potential time lags.
We review and discuss our results and offer theoretical interpretations in $\S$\ref{sec:discuss}, and summarize our conclusions in \S\ref{sec:Conclusions}.


\section{Observations and Data Reduction}
\label{sec:ObsandData}

In order to investigate the short timescale variability of BL Lac, we organized a multi-wavelength observing campaign around the \TESS\ observations, which took place as part of observing sector 16 at a 2-minute cadence, from 2019 September 12 to October 6 (MJD: 58738-58762). We have retrieved $\sim 3$ months of $\gamma$-ray data measured by the \fermi-LAT, and obtained four-\edit1{band}
, \emph{BVRI} optical flux and \emph{R}-band polarization data with numerous WEBT-affiliated ground-based telescopes, over the time period from 2019 August 5 to November 2 (MJD: 58700-58789). We also obtained observations at X-ray frequencies with the NuSTAR satellite and at X-ray, UV, and optical frequencies with the \swift\ satellite for 5 days during the \TESS\ monitoring, 2019 September 14-19 (MJD: 58740-58745). In this section we discuss the processing of the various data. 


\subsection{Gamma-ray Data}
\label{subsec:GammaRayDataReduction}

The \fermi\ LAT \citep{Atwood2009} surveys the entire sky every $\sim 3$ hours in the energy range 0.1-300 GeV, with data archived for public access. We retrieved P8R3 photon and spacecraft data centered on BL Lac (4FGL J2202.7+4216). The data were reduced using version v1.0.10 of the \fermi\ Science Tools, background models from the iso\_P8R3\_SOURCE\_V2\_v1.txt isotropic template, and the gll\_iem\_v07 Galactic diffuse emission model.\footnote{Provided at \url{https://fermi.gsfc.nasa.gov/ssc/data/access/lat/BackgroundModels.html}.}
We utilized analysis cuts of \texttt{evclass = 128}, \texttt{evtype = 3}, and \texttt{zmax=90} for an unbinned likelihood analysis of the photon data, which we restricted to an energy range of 0.1-200 GeV.

The $\gamma$-ray emission from BL Lac and other point sources within a $25\degr$ radius of BL Lac was represented by spectral models listed in the 4FGL catalogue of sources detected by the LAT \citep{Fermi2019}. Specifically, the number of photons $N$ per unit energy $E$ of BL Lac was modeled as a log-parabola of the form

\begin{equation}
    \frac{\text{d}N}{\text{d}E} = N_0 \left( \frac{E}{E_{\text{b}}} \right)^{-(\alpha + \beta\log{E/E_{\text{b}}})}\ .
\end{equation}

\noindent During the analysis, the spectral parameters of BL Lac were kept fixed at their 4FGL catalogue values: $\alpha = 2.1755$, $\beta = 6.0062 \times 10^{-2}$, and break energy $E_{\text{b}} = 7.47961 \times 10^2$ MeV. This was necessary because of the relatively low flux, which precludes an accurate spectral analysis. The prefactor $N_0$ was allowed to vary for BL Lac, as well as for all cataloged sources within $5\degr$ and bright ($F_\gamma > 10^{-11}$ erg cm$^{-2}$ s$^{-1}$) sources within $10\degr$.

In the initial analysis, we integrated the observations over 6 hours. Contiguous periods of upper limits for the 6-hr binning were recalculated with 12-hr binning. This procedure yielded a $\gamma$-ray light curve with 322 measurements of BL Lac over the time period from 2019 August 5 to November 2. The source was considered detected if the test statistic (TS) provided by the maximum-likelihood analysis exceeded 10, which corresponds to approximately a $3\sigma$ detection level \citep{Nolan2012}. If TS $<10$, we calculated\ 2$\sigma$ upper limits with the \fermi\ Python script. Of the 322 measurements, 126 were detections and 196 were upper limits.


\subsection{X-ray Data}
\label{subsec:XRayDataReduction}

\subsubsection{\swift\ X-ray Data}
\label{subsubsec:SwiftXRayDataReduction}

The Neil Gehrels Swift Observatory (\swift) X-ray Telescope \citep[XRT,][]{Burrows2005} observes over the 0.3-10 keV band. We obtained 40 observations over 5 days from 2019 September 14 to 19, averaging one observation every three hours, for a total exposure time of $\sim46$ ks. All observations were made in photon counting mode.

\begin{deluxetable*}{ccllrcc}[t]
    \tablecaption{Summary of \swift\ 0.3-10 keV modelling to calculate $N_H$.\label{tab:SwiftNHModelling}}
    \tablewidth{0pt}
    \tablehead{
    \colhead{MJD Start} & \colhead{Exposure Time} & \colhead{$N_H$} & \colhead{$\Gamma$} & \colhead{Flux} & \colhead{D.o.F.} & \colhead{$\chi_{\nu}^2$}  \\
    \colhead{} & \colhead{[sec]} & \colhead{$\times 10^{21}$ cm$^{-2}$} & \colhead{} & \colhead{$\times 10^{-12}$ erg cm$^{-2}$ s$^{-1}$} &\colhead{} & \colhead{}
    }
    \startdata
    58740.36 & 8910 & $2.60^{+0.28}_{-0.27}$ & $2.474^{+0.108}_{-0.105}$ & $11.90^{+0.58}_{-0.62}$ & 481 & 1.372 \\
    58741.28 & 9752 & $2.31^{+0.34}_{-0.32}$ & $2.334^{+0.133}_{-0.127}$ & $ 7.54^{+0.37}_{-0.43}$ & 454 & 1.063 \\
    58742.35 & 9382 & $2.19^{+0.51}_{-0.48}$ & $1.943^{+0.171}_{-0.162}$ & $ 4.88^{+0.50}_{-0.40}$ & 401 & 1.253 \\
    58743.27 & 9652 & $1.79^{+0.44}_{-0.41}$ & $1.930^{+0.151}_{-0.145}$ & $ 5.64^{+0.45}_{-0.55}$ & 406 & 1.002 \\
    58744.27 & 8296 & $2.16^{+0.35}_{-0.33}$ & $2.194^{+0.132}_{-0.126}$ & $ 8.73^{+0.59}_{-0.69}$ & 450 & 1.045 \\
    \tableline
    58740.36 & 8910 & 2.7 & $2.504^{+0.056}_{-0.057}$ & $11.80^{+0.53}_{-0.50}$ & 482 & 1.355 \\
    58741.28 & 9752 & 2.7 & $2.462^{+0.072}_{-0.071}$ & $ 7.27^{+0.40}_{-0.37}$ & 455 & 1.022 \\
    58742.35 & 9382 & 2.7 & $2.081^{+0.101}_{-0.100}$ & $ 4.67^{+0.45}_{-0.36}$ & 402 & 1.253 \\
    58743.27 & 9652 & 2.7 & $2.181^{+0.093}_{-0.092}$ & $ 5.22^{+0.41}_{-0.35}$ & 407 & 0.994 \\
    58744.27 & 8296 & 2.7 & $2.362^{+0.075}_{-0.074}$ & $ 8.31^{+0.63}_{-0.38}$ & 451 & 1.010 \\
    \tableline
    58740.36 & 8910 & 3.4 & $2.730^{+0.062}_{-0.061}$ & $11.20^{+0.35}_{-0.43}$ & 482 & 1.302 \\
    58741.28 & 9752 & 3.4 & $2.687^{+0.078}_{-0.077}$ & $ 6.85^{+0.30}_{-0.32}$ & 455 & 1.009 \\
    58742.35 & 9382 & 3.4 & $2.254^{+0.111}_{-0.110}$ & $ 4.43^{+0.26}_{-0.31}$ & 402 & 1.274 \\
    58743.27 & 9652 & 3.4 & $2.360^{+0.101}_{-0.099}$ & $ 4.96^{+0.33}_{-0.35}$ & 407 & 1.026 \\
    58744.27 & 8296 & 3.4 & $2.567^{+0.083}_{-0.082}$ & $ 7.85^{+0.35}_{-0.39}$ & 451 & 1.017 \\
    \tableline
    58740.36 & 8910 & $2.49^{+0.15}_{-0.14}$ & 2.419 & $12.10^{+0.40}_{-0.36}$ & 482 & 1.411 \\
    58741.28 & 9752 & $2.49^{+0.18}_{-0.18}$ & 2.419 & $ 7.32^{+0.28}_{-0.24}$ & 455 & 1.033 \\
    58742.35 & 9382 & $3.37^{+0.36}_{-0.32}$ & 2.419 & $ 4.04^{+0.27}_{-0.21}$ & 402 & 1.265 \\
    58743.27 & 9652 & $2.97^{+0.30}_{-0.28}$ & 2.419 & $ 4.66^{+0.24}_{-0.22}$ & 407 & 0.977 \\
    58744.27 & 8296 & $2.67^{+0.20}_{-0.19}$ & 2.419 & $ 8.05^{+0.26}_{-0.41}$ & 451 & 0.985 \\
    \enddata
    \tablecomments{In section 1 of this table, all parameters were allowed to vary. In sections 2 and 3, $N_{\text{H}}$ was fixed to the listed values. In section 4, the photon index $\Gamma$ was fixed at the average value of Sections 2 and 3, while $N_{\text{H}}$ was allowed to vary.}
\end{deluxetable*}

We used v6.26.1 of the \texttt{HEAsoft} package and CALDB v20190412 to process the data. We defined a circular source region with a 70\arcsec\ radius and an annular background region with inner radius 88\arcsec\ and outer radius 118\arcsec\ (selected to avoid contaminating sources), both centered on BL Lac. Using the standard reduction protocol, we first cleaned the data and created an exposure map with the \texttt{xrtpipeline} tool. The image and spectra were extracted using \texttt{XSELECT}, and the ancillary response file was generated with \texttt{xrtmkarf}. Because we used Cash statistics \citep{Cash1979, Humphrey2009} in the form of the modified C-statistic \texttt{cstat} to fit the data in \texttt{XSPEC}, we grouped our data by single photons in \texttt{grppha}. The spectrum was then fit in \texttt{XSPEC} and further evaluated using a $\chi^2$ test. 

\begin{deluxetable}{lr}[t!]
    \tablecaption{Summary of \swift\ XRT 0.3-10 keV Observations.\label{tab:SwiftXRTSummary}}
    \tablewidth{0pt}
    \tablehead{
    \colhead{Statistic} & \colhead{Value}
    }
    \startdata
    Number of observations & 40 \\
    Total exposure time &  \\
    (seconds) & 45,993\\
    \hline
    Averages per Observation: &  \\
    Count Rate [cts/s] & 0.20 \\
    Counts per Obs. & 233 \\
    Photon Index & 2.33 \\
    Minimum Photon Index & 1.79 \\
    Maximum Photon Index & 2.72 \\
    Flux [erg cm$^{-2}$ s$^{-1}$]& $7.12 \times 10^{-12}$ \\
    Min. Flux [erg cm$^{-2}$ s$^{-1}$] & $3.06 \times 10^{-12}$ \\
    Max. Flux [erg cm$^{-2}$ s$^{-1}$] & $1.83 \times 10^{-11}$ \\
    \enddata
\end{deluxetable}

The total hydrogen column density toward BL Lac consists of the atomic hydrogen column density, $N_{\text{HI}} = 1.74\times 10^{21}$ cm$^{-2}$\citep{Kalberla2005} and molecular column density, $N_{\text {H,mol}}$. 
The value of the latter ranges among various studies of the X-ray spectrum from $0.5\times 10^{21}$ cm$^{-2}$ \citep[considered as a possibility by][]{Madejski1999} to $1.7\times 10^{21}$ cm$^{-2}$\citep{Raiteri2009}. 
An even higher value of $N_{\text {H,mol}} =2.8 \times 10^{21}$ cm$^{-2}$, based on a CO emission line from a molecular cloud in the direction of BL Lac \citep{Bania1991}, has been proposed as well \citep[see][for more examples]{Raiteri2009}. 
In addition, \citet{Moore1995} have found changes by 14\% in the equivalent width of H$_2$CO absorption lines along the line of sight to BL Lac on a timescale of $\sim$2~yr, which suggests that $N_{\text {H,mol}}$ could be variable.

In order to estimate the value of $N_{\text H}$ that best fits our data, we have combined our XRT observations over five sets of 24~hr each. These sets were obtained by summing the individual exposure maps using \texttt{XIMAGE}, followed by application of \texttt{XSELECT} to sum the individual event files. The ancillary response file for each set of observations was generated using \texttt{xrtmkarf} with the corresponding summed exposure map. 

\begin{figure}
  \begin{center}
    \includegraphics[width=0.45\textwidth]{{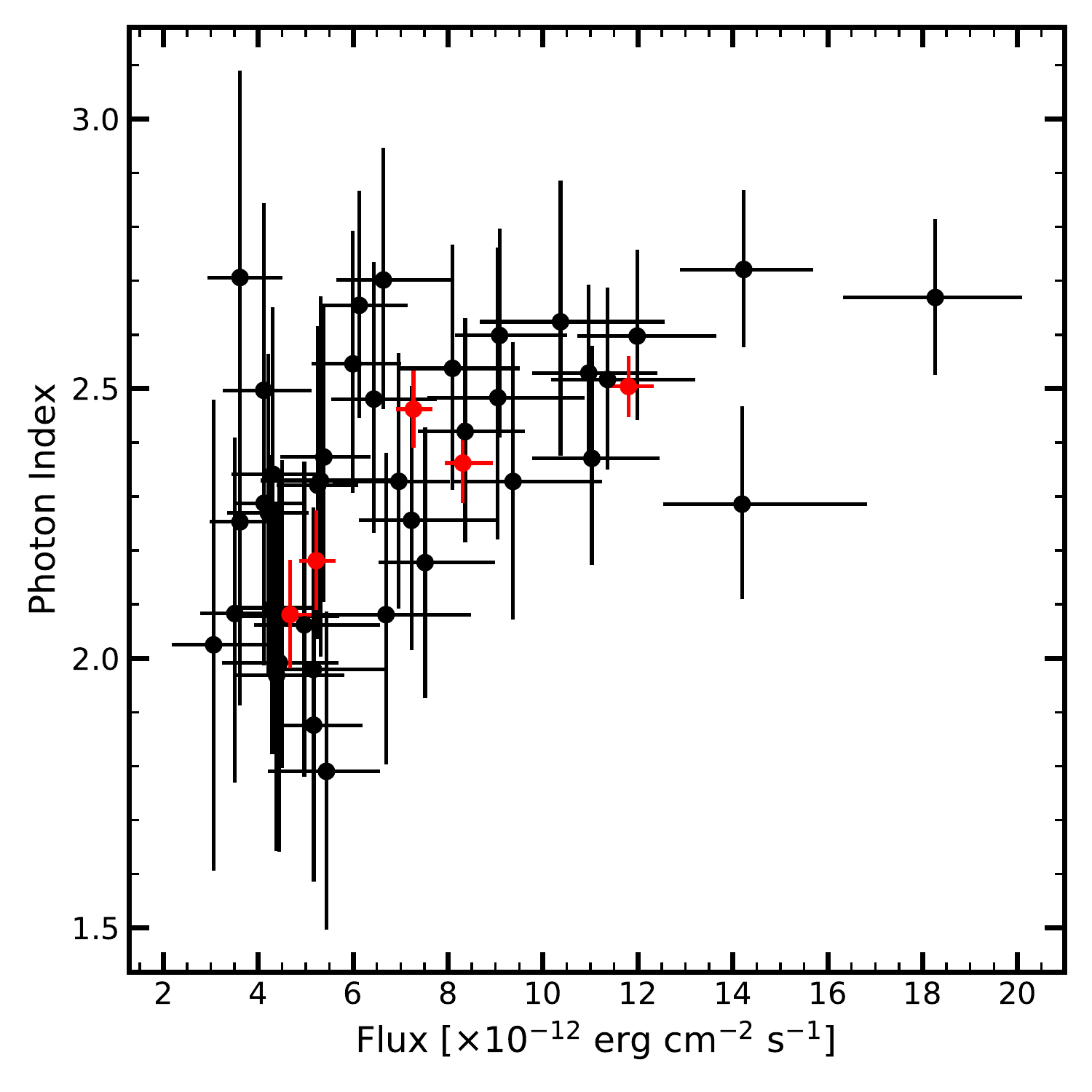}}
    \caption{\swift\ 0.3-10 keV photon index vs flux for all observations (black circles) and observations binned over 24 hours with a fixed column density $N_{\text{H}} = 2.7\times10^{21}$ cm$^{-2}$ (red circles). \label{fig:XRTPhIndvsFlux}}
  \end{center}
\end{figure}

We modeled the daily combined data at 0.3 - 10 keV in \texttt{XSPEC}
using an absorbed simple power law with all parameters free. The results are presented in Table \ref{tab:SwiftNHModelling}, which yields $\langle N_{\text H}\rangle = (2.21\pm 0.29)\times10^{21}$ cm$^{-2}$ over five days. We subsequently modeled each set with a power law for two different values of $N_{\text{H}}$ (fixed during each model fit), $2.7\times10^{21}$ cm$^{-2}$ and $3.4\times10^{21}$ cm$^{-2}$. These are the closest values to those estimated from atomic and molecular line observations (see above), and correspond to those used for BL Lac by \citet{Madejski1999}, \citet{Raiteri2009}, and \citet{Wehrle2016}. The results of the modeling are given in sections 2 and 3 of Table \ref{tab:SwiftNHModelling}, from which we find that there is no statistically significant difference between the models as judged by the reduced $\chi^2$. We then averaged the photon indices obtained over the five days and two fixed values of $N_{\text{H}}$, which resulted in $\Gamma$=2.419. Fixing $\Gamma$ at this value, we performed a search for the best-fit value of $N_{\text{H}}$. The results of this search are listed in section 4 of Table \ref{tab:SwiftNHModelling}. The reduced $\chi^2$ values are similar to those of the previous three model fits. The last procedure results in an average value of $N_{\text{H}}=(2.80\pm 0.32)\times10^{21}$ cm$^{-2}$ over five days, which is in good agreement with the value adopted by \cite{Madejski1999}. Based on these considerations, we have modeled the X-ray data presented below using the same fixed hydrogen column density as adopted by \cite{Madejski1999}, $N_{\text{H}} = 2.7\times10^{21}$ cm$^{-2}$. 

We have modeled the 40 \swift\ XRT observations at 0.3-10~keV with an absorbed simple power law.  
A summary of the results is provided in Table~\ref{tab:SwiftXRTSummary}. Figure~\ref{fig:XRTPhIndvsFlux} reveals that the photon index becomes steeper at higher flux levels. The dependence is more apparent for the daily averaged X-ray data.

\subsubsection{NuSTAR Data}
\label{subsubsec:NuSTARDataReduction}

The Nuclear Spectroscopic Telescope Array (NuSTAR) observes in the 3-79 keV energy band \citep{Harrison2013}. Two independent, co-aligned telescopes (FPMA and FPMB) observe as photon counting modules, with each module consisting of a $2\times2$ array of four detectors. Observations of a source can be obtained throughout the satellite's $\sim95$-min orbit, excluding dead-time while the observatory is slewing, performing calibration or house-keeping activities, passing through the South Atlantic Anomaly (SAA), or occulted by the Earth. We obtained five days of continuous observations (ID 60501024002) from 2019 September 14 05:36:09 to September 19 06:01:09 UT, for a total dead-time corrected exposure time of $\sim197$ ks spanning 75 orbits.

\begin{deluxetable}{lr}[t!]
    \tablecaption{Summary of NuSTAR 3-79 keV Observations.\label{tab:NuSTARObs}}
    \tablewidth{0pt}
    \tablehead{
    \colhead{Statistic} & \colhead{Value}
    }
    \startdata
    Number of orbits & 75 \\
    Total dead-time corrected &  \\
    exposure [seconds] & 196,938 \\
    \hline
    Averages per Orbit: & \\
    FPMA count rate [cts/s] & 0.14 \\
    FPMA counts per orbit & 366 \\
    FPMB count rate [cts/s] & 0.13 \\
    FPMB counts per orbit & 332 \\
    Photon Index & 1.872 \\
    Minimum Photon Index & 1.563 \\
    Maximum Photon Index & 2.147 \\
    Flux [erg cm$^{-2}$ s$^{-1}$]& $1.371\times10^{-11}$ \\
    Min. Flux [erg cm$^{-2}$ s$^{-1}$] & $0.862\times10^{-11}$ \\
    Max. Flux [erg cm$^{-2}$ s$^{-1}$] & $1.972\times10^{-11}$ \\
    \enddata
\end{deluxetable}

We processed the data using the NuSTAR Data Analysis Software (NuSTARDAS), downloaded as part of v6.25 of the \texttt{HEAsoft} package, with CALDB version 20190627. Upon examination of the provided SAA filtering reports, we chose to use the ``strict" SAA calculation mode and SAA passage algorithm 1, with no tentacle correction. The \texttt{nupipeline} was run with the SCIENCE observing mode and required the creation of an exposure map. The results were processed through \texttt{nuproducts}. Using SAOImageDS9\footnote{\url{https://ds9.si.edu/site/Home.html}}, we defined a 70\arcsec\ region for each telescope centered on the source and a 70\arcsec\ background region on the same detector as the source (but sufficiently distant to avoid contamination). As for the \swift\ data, we used Cash statistics to fit the data in \texttt{XSPEC}, and so grouped our data by single photons.

We divided the observations first by orbit, defined to begin with the satellite's emergence from the SAA as indicated on the Good Time Intervals (GTI) file. We used \texttt{XSELECT} to generate the GTI that subsequently was fed into the \texttt{nuproducts} process. The data were then examined in \texttt{XSPEC} and modeled as an absorbed simple power-law.
We simultaneously fit the FPMA and FPMB files with a cross-normalization factor frozen to unity for FPMA and allowed to vary for FPMB. The fit was evaluated with the $\chi^2$ test statistic.

\begin{figure}[t]
    \begin{center}
        \includegraphics[width=0.40\textwidth]{{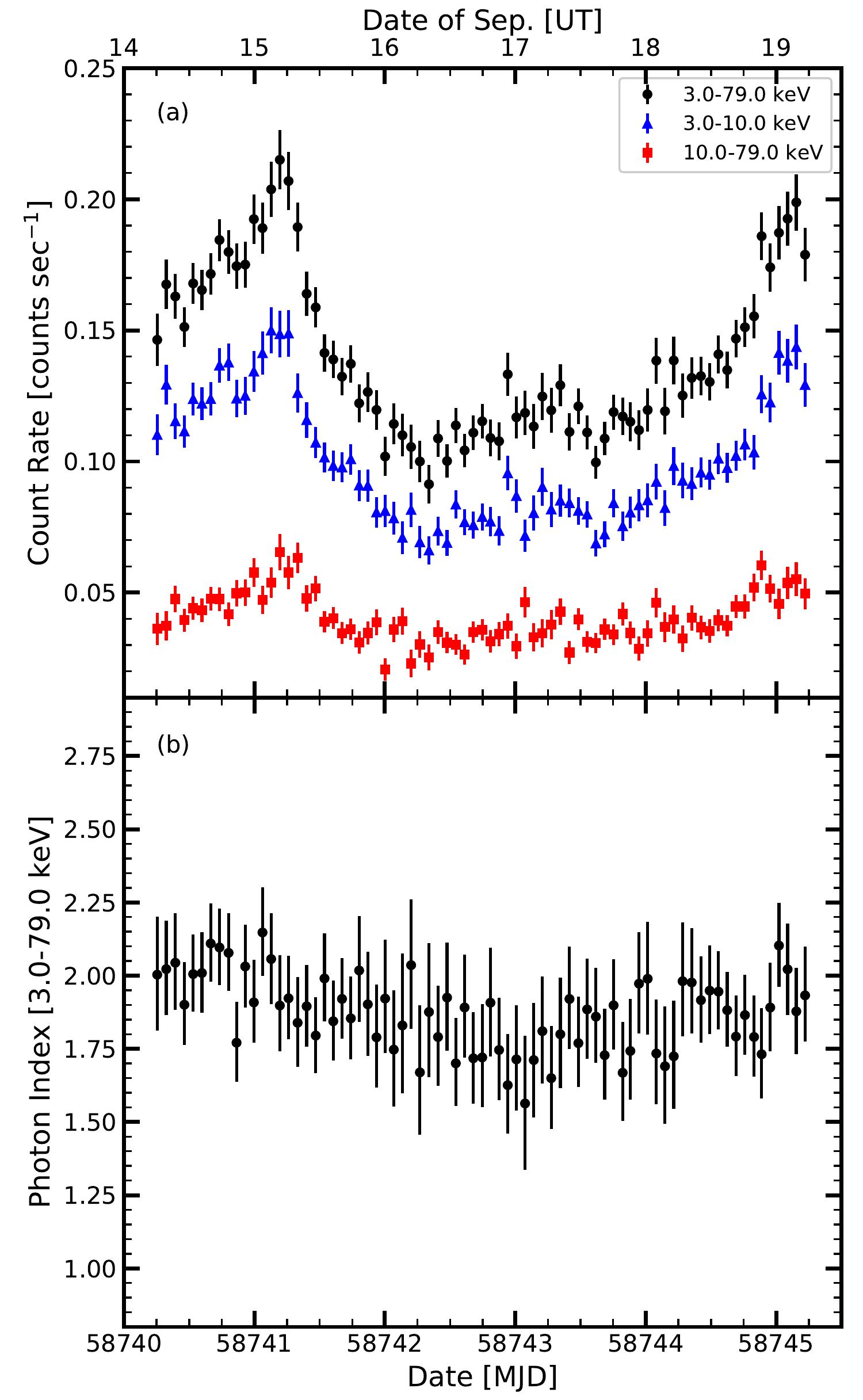}}
        \caption{\emph{(a)} Count rate per orbit for NuSTAR FPMA. The count rate for FPMB is similar for every orbit. \emph{(b)} Photon index per orbit for NuSTAR FPMA (the photon index for FPMB is similar). \label{fig:NuSTARcounts}}
    \end{center}
\end{figure}

The NuSTAR observations are summarized in Table~\ref{tab:NuSTARObs}. The average dead-time corrected exposure time per orbit was 2,626 seconds.
Figure~\ref{fig:NuSTARcounts}\emph{(a)} shows the light curve for the full energy band, as well as two narrower, soft and hard, bands (3-10 and 10-79 keV, respectively). The full spectrum is dominated by flux from the lower energies, while the counts for the higher energies are too low to be analyzed by single orbits. Fig.~\ref{fig:NuSTARcounts}\emph{(b)} shows the photon index for the full spectrum. There are enough counts in the soft band to allow the photon index to vary; this is not the case for the hard band.

\begin{figure}[t]
    \begin{center}
        \includegraphics[height=0.35\textwidth]{{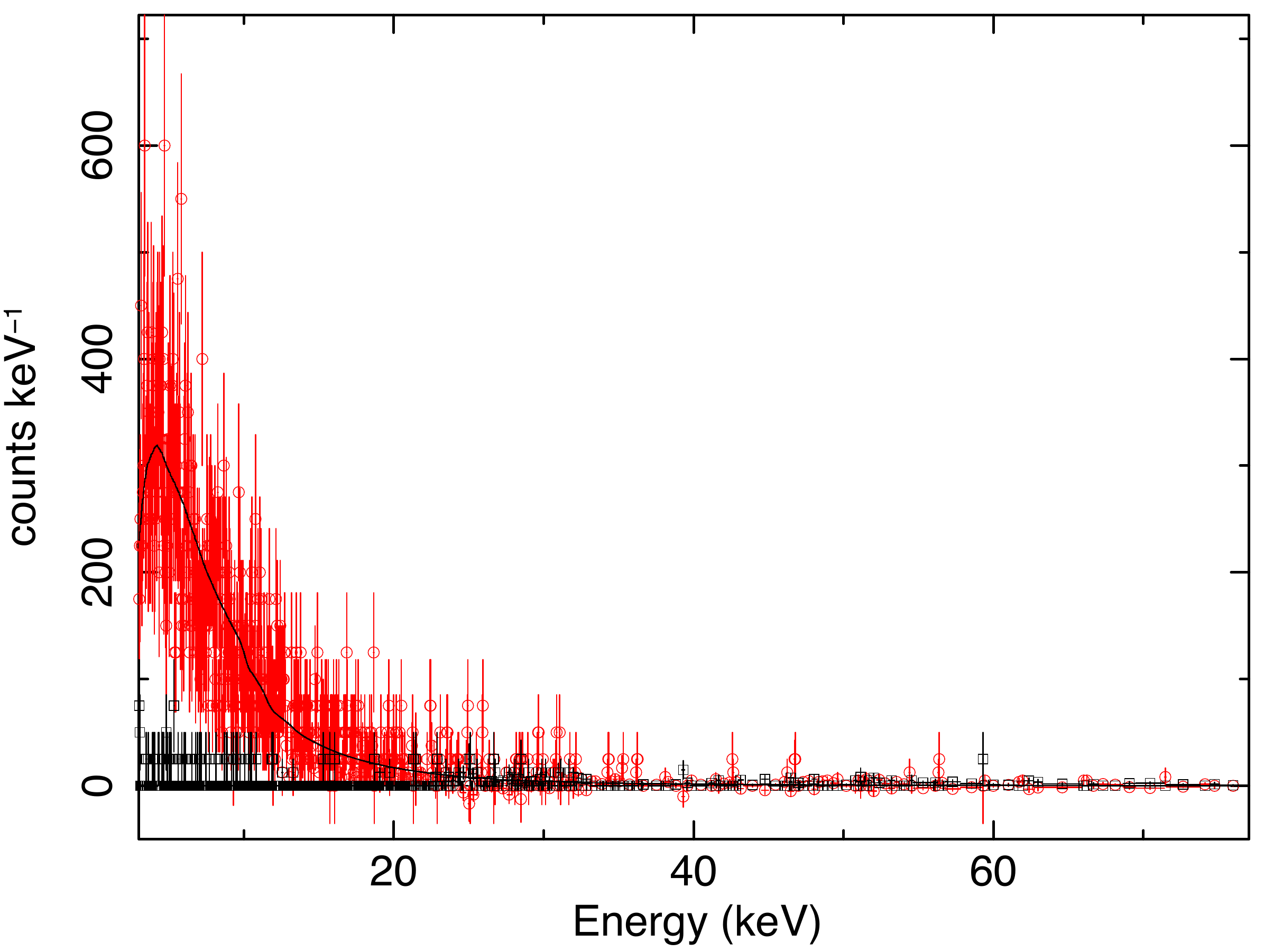}}
        \caption{Counts per unit energy observed by FPMA for a group of five orbits during the latter half of 2019 September 15 (MJD: 58741; orbits 16-20). Source counts are shown in red circles, and the background counts are shown with black squares. The solid line is the model fit obtained with \texttt{XSPEC}.\label{fig:NuSTARFPMAEnergy}}
    \end{center}
\end{figure}

\begin{figure*}[t]
    \begin{center}
        \includegraphics[width=0.75\textwidth]{{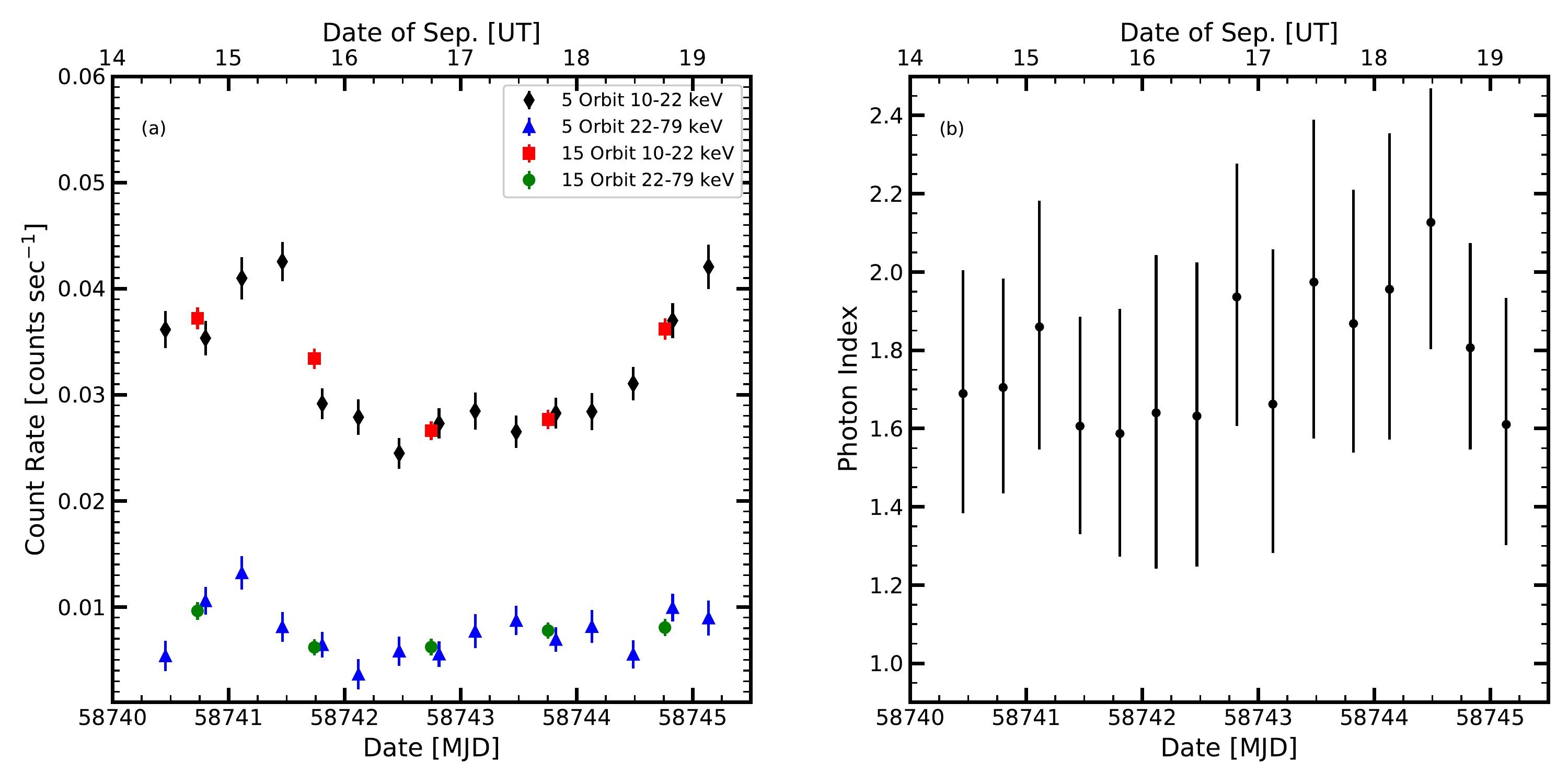}}
        \caption{\emph{(a)} NuSTAR light curves at 10-22 and 22-79 keV binned over 5 and 15 orbits, respectively. \emph{(b)} Photon index at 10-22 keV band with 5-orbit binning. \label{fig:NuSTARCountsBinned}}
    \end{center}
\end{figure*}

To analyze the hard energy band (10-79 keV), we divided the observations into groups of both 5 and 15 orbits. Figure~\ref{fig:NuSTARFPMAEnergy} shows the counts per unit energy of an average group of five orbits (orbits 16-20). For clarity, only data from FPMA is shown; data from FPMB are similar. The source counts are significantly above the background level for the lower-energy portion of the spectrum, but steadily deteriorate toward harder energies. Based on the results of all groups of orbits, we have separated the hard spectrum into two bands, 10-22 keV and 22-79 keV. Figure~\ref{fig:NuSTARCountsBinned}\emph{(a)} presents the 10-22 keV light curves, with data grouped over 5 and 15 orbits, while Figure~\ref{fig:NuSTARCountsBinned}\emph{(b)} plots the photon index at 10-22 keV with data grouped over 5 orbits.
For the 22-79 keV band grouped over 5 orbits, the counts are so few that it is necessary to fix the photon index at the value determined by the full 10-79 keV range, rather than allow the photon index to vary. Using this method permits us to use the 5-orbit binned light curve to enhance the time-resolution of the hard X-ray data.

\subsubsection{Simultaneous \swift\ and NuSTAR Data Reduction} 
\label{subsubsec:SimultaneousNuSTARSwift}

The photon indices for the \edit1{\swift}\ and \edit1{NuSTAR} observations provided in \edit1{Table~\ref{tab:SwiftXRTSummary} and Table~\ref{tab:NuSTARObs}}, respectively, suggest that, 
in general, the photon index of the 0.3-10~keV emission is steeper than that of the 3-79~keV emission. This implies a break in the X-ray spectrum. We simultaneously fit the NuSTAR and \swift\ XRT data that are contemporaneous within a given day. This resulted in 15 FPMA$/$FPMB and 8 XRT observations per 24-hour period, with five such daily sets over our observations. 

\begin{figure}
  \begin{center}
    \includegraphics[width=0.45\textwidth]{{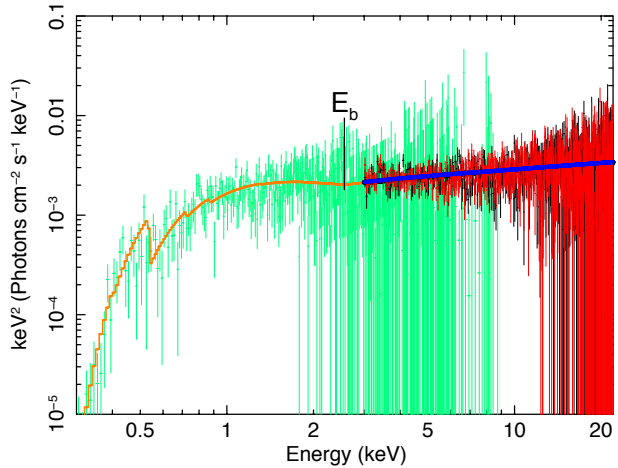}}
    \caption{Simultaneous broken power-law fit, with photoelectric absorption, to the \swift\ XRT and NuSTAR X-ray data for Day 5 (fit parameters are given in Table~\ref{tab:SwiftNuSTARBPLCombo}). 
    \swift\ data are shown in light green, while the NuSTAR FPMA and FPMB are shown in black and red, respectively. The simultaneous fit is shown as the solid line in orange and blue for the \swift\ and NuSTAR data, respectively. $E_{\text{b}}$ marks a location of the break energy of the model. Fits to the spectra on the other four days are similar,  although with variations in $E_{\text{b}}$.
    \label{fig:SwiftNuSTARComboFig}}
  \end{center}
\end{figure}

We used \texttt{XSPEC} to simultaneously fit the FPMA, FPMB, and XRT data sets for each day. We employed 2 models: a single power law and a broken power law, each with photoelectric absorption, and fit the energy range from 0.3 to 79 keV. The results are given in Tables~\ref{tab:SwiftNuSTARSPLCombo} and \ref{tab:SwiftNuSTARBPLCombo}, respectively. There are no statistically significant differences in the reduced $\chi^2$ values between the single and broken power-law models. Figure~\ref{fig:SwiftNuSTARComboFig} gives an example of modeling the data  by a broken power-law model. The results presented in Table~\ref{tab:SwiftNuSTARBPLCombo} show that the spectral indices both before and after the break do not exhibit a dependence on flux and that, overall,
$\Gamma_1$= 2.40$\pm$0.14 and $\Gamma_2$= 1.72$\pm$0.05 over the 5 days of observation. However, the break energy tends to increase with flux, with the best-fit models suggesting a break at the highest flux level, $\sim$6~keV, while $E_{\text b}\sim$ 2~keV at lower flux levels. 

\begin{deluxetable*}{cccccccc}[t]
    \tablecaption{Simultaneous single power-law fits to 24-hour NuSTAR and \swift\ XRT spectra.\label{tab:SwiftNuSTARSPLCombo}}
    \tablewidth{0pt}
    \tablehead{
    \colhead{MJD Start\tablenotemark{a}} & \colhead{Exposure\tablenotemark{a}} & \colhead{Exposure\tablenotemark{b}} & \colhead{$\Gamma$} & \colhead{Flux\tablenotemark{c}} & \colhead{Count Rate} & \colhead{d.o.f.} & \colhead{$\chi_\nu^2$} \\
    \colhead{} & \colhead{[ksec]} & \colhead{[ksec]} & \colhead{} & \colhead{} & \colhead{[counts s$^{-1}$]} & \colhead{} & \colhead{}
    }
    \startdata
    58740.24 & 39.6 & 8.9 & $2.151^{+0.021}_{-0.021}$ & $18.02^{+0.52}_{-0.42}$ & $0.678 \pm 0.007$ & 2387 & 1.110 \\
    58741.25 & 39.4 & 9.8 & $1.973^{+0.024}_{-0.025}$ & $15.42^{+0.50}_{-0.40}$ & $0.477 \pm 0.006$ & 2246 & 1.044 \\
    58742.26 & 39.3 & 9.4 & $1.728^{+0.029}_{-0.029}$ & $14.24^{+0.63}_{-0.41}$ & $0.327 \pm 0.004$ & 2191 & 1.056 \\
    58743.26 & 39.2 & 9.7 & $1.804^{+0.028}_{-0.028}$ & $14.06^{+0.54}_{-0.45}$ & $0.349 \pm 0.005$ & 2188 & 0.977 \\
    58744.27 & 39.4 & 8.3 & $1.949^{+0.024}_{-0.024}$ & $17.33^{+0.55}_{-0.49}$ & $0.526 \pm 0.006$ & 2305 & 0.978 \\
    \enddata
    \tablecomments{Single power law: $S(E)=KE^{-\Gamma}$, where $S$ is photon flux density in photons cm$^{-2}$ s$^{-1}$ keV$^{-1}$ and $E$ is in keV. }
    \tablenotetext{a}{Of the NuSTAR data.}
    \tablenotetext{b}{Of the \swift\ data.}
    \tablenotetext{c}{In units of $10^{-12}$ erg cm$^{-2}$ s$^{-1}$}
\end{deluxetable*}

\begin{deluxetable*}{cccccccc}[t]
    \tablecaption{Simultaneous broken power-law fits to 24-hour NuSTAR and \swift\ XRT spectra.\label{tab:SwiftNuSTARBPLCombo}}
    \tablewidth{0pt}
    \tablehead{
    \colhead{MJD Start\tablenotemark{a}} & \colhead{$\Gamma_1$} & \colhead{$E_{\text{b}}$} & \colhead{$\Gamma_2$} & \colhead{Flux\tablenotemark{b}} & \colhead{Count Rate} & \colhead{d.o.f.} & \colhead{$\chi_\nu^2$} \\
    \colhead{} &  \colhead{} & \colhead{[keV]} & \colhead{} & \colhead{} & \colhead{[counts s$^{-1}$]} & \colhead{} & \colhead{}
    }
    \startdata
    58740.24 & $2.363^{+0.038}_{-0.033}$ & $5.864^{+0.551}_{-0.678}$ & $1.716^{+0.067}_{-0.062}$ & $23.04^{+1.17}_{-1.10}$ & $0.175 \pm 0.002$ & 2385 & 1.089 \\
    58741.25 & $2.558^{+0.106}_{-0.126}$ & $2.430^{+0.475}_{-0.264}$ & $1.763^{+0.034}_{-0.036}$ & $17.43^{+0.76}_{-0.51}$ & $0.138 \pm 0.002$ & 2244 & 1.011 \\
    58742.26 & $2.146^{+0.771}_{-0.202}$ & $2.090^{+2.200}_{-1.053}$ & $1.651^{+0.061}_{-0.039}$ & $15.01^{+1.49}_{-0.45}$ & $0.110 \pm 0.002$ & 2189 & 1.070 \\
    58743.26 & $2.462^{+0.202}_{-0.188}$ & $1.770^{+0.392}_{-0.297}$ & $1.706^{+0.035}_{-0.036}$ & $14.96^{+0.96}_{-0.47}$ & $0.117 \pm 0.002$ & 2186 & 0.976 \\
    58744.27 & $2.449^{+0.167}_{-0.131}$ & $2.516^{+0.641}_{-0.483}$ & $1.787^{+0.038}_{-0.034}$ & $18.98^{+0.71}_{-0.65}$ & $0.154 \pm 0.002$ & 2303 & 0.965 \\
    \enddata
    \tablecomments{Broken power law: $S(E)=KE^{-\Gamma_1}$ if $E\leq E_{\text{b}}$ and
    $S(E)=KE_{\text{b}}^{(\Gamma_2-\Gamma_1)}E^{-\Gamma_2}$ if $E>E_{\text{b}}$, with $E$ in keV. \\
    Exposure times are the same as in Table~\ref{tab:SwiftNuSTARSPLCombo}.}
    \tablenotetext{a}{Of the NuSTAR data}
    \tablenotetext{b}{In units of $10^{-12}$ erg cm$^{-2}$ s$^{-1}$}
\end{deluxetable*}

\subsection{UV and Optical Data}
\label{subsec:UVOPTDataReduction}


\subsubsection{\swift\ Ultraviolet and Optical Data}
\label{subsubsec:SwiftUVOTDataReduction}

The \swift\ satellite also provides UV and optical data via the UV/Optical Telescope \citep[UVOT,][]{Roming2005}. We retrieved the data from the HEASARC Archive and reduced them with v6.26.1 of the \texttt{HEAsoft} software and  CALDB v20170922. We defined a 5\arcsec\ circular region centered on the source and a 20\arcsec\ circular aperture on a source-free region of the image to represent the background. We ran the tool \texttt{uvotunicorr} if an aspect correction was not applied. Multiple extensions within a particular file were summed using \texttt{uvotimsum}, then processed with \texttt{uvotsource}, setting the detection significance to $\sigma=5$. Images were  retained if they had an exposure time $\geq 40$ sec and a magnitude error $\sigma_{\text{mag}} < 0.2$. None of the observations suffered a high coincidence loss. We used the count-rate-to-flux conversion factors reported by \citet{Breeveld2011} for $\gamma$-ray burst models, which correspond to continuum spectra similar to those of blazars. We have corrected for Galactic extinction using the values found by \citet{Raiteri2009}. Our aperture size introduced a flux contamination from the host galaxy of $\sim 0.5$ times the host galaxy flux density \citep{Raiteri2010}, which we have subtracted from the source flux density. All Galactic extinction, magnitude-to-flux-density conversion factors, and host galaxy flux densities are given in Table~\ref{tab:BLLacOptConversion}.  

\begin{deluxetable}{lccl}[t]
    \tablecaption{UV and optical correction factors used in this work.\label{tab:BLLacOptConversion}}
    \tablewidth{0pt}
    \tablehead{
    \colhead{Filter} & \colhead{Extinction} & \colhead{Absolute Flux} & \colhead{Host Galaxy} \\
    \colhead{} & \colhead{[mag]} & \colhead{Density\tablenotemark{a}} & \colhead{Flux Density} \\
    \colhead{} & \colhead{} & \colhead{$\left [10^{-20}\right .$ erg } & \colhead{[mJy]}\\
    \colhead{} & \colhead{} & \colhead{cm$^{-2}$ s$^{-1}$ Hz$\left . ^{-1}\right ]$} & \colhead{}
    }
    \startdata
    \emph{UVW2} & 2.92 & 0.738 & 0.017 \\
    \emph{UVM2} & 3.04 & 0.689 & 0.020 \\
    \emph{UVW1} & 2.40 & 0.942 & 0.026 \\
    \emph{u} & 1.79 & 1.307 & 0.036 \\
    \emph{b} & 1.44 & 3.476 & 1.30\\
    \emph{v} & 1.10 & 3.420 & 2.89 \\
    \tableline
    \emph{B} & 1.42 & 4.063 & 1.30 \\
    \emph{V} & 1.08 & 3.636 & 2.89 \\
    \emph{R} & 0.90 & 3.064 & 4.23 \\
    \emph{I} & 0.64 & 2.416 & 5.90 \\
    \enddata
    \tablenotetext{a}{For a zero-magnitude star}
    \tablerefs{\citet{Bessel1998}, \citet{Raiteri2010}, \citet{Wehrle2016}}
\end{deluxetable}


\subsubsection{WEBT Optical Data}
\label{subsubsec:WEBTOpticalDataReduction}

\begin{deluxetable*}{lcrr}
  \tablecaption{WEBT-affiliated ground-based telescopes used in this work.\label{tab:OpticalTelescopes}}
  \tablewidth{0pt}
  \tablehead{
  \colhead{Observatory}  & \colhead{Bands} & \colhead{Number of} & \colhead{Marker}\\
  \colhead{}  & \colhead{} & \colhead{Observations\tablenotemark{a}} & \colhead{Style\tablenotemark{b}}
  }
  \startdata
  ARIES                       & \emph{BVRI} & 2, 2, 2, 2               & blue $\bullet$ \\
  Abastumani                  & \emph{R}    & 144                      & green $\bullet$ \\
  Belogradchik                & \emph{VRI}  & 14, 16, 15               & red $\bullet$ \\
  Burke-Gaffney               & \emph{R}    & 1                        & cyan $\bullet$ \\
  Crimean (AZT-8; AP7p)       & \emph{BVRI} & 61, 60, 30, 63           & magenta $\bullet$ \\
  Crimean\tablenotemark{c} (AZT-8; ST-7)     & \emph{BVRI}    & 32, 31, 115 (55), 34  & orange $\bullet$ \\
  Foggy Bottom                & \emph{R}    & 249                      & blue $\blacklozenge$ \\
   Las Cumbres                & \emph{R}    & 34                       & red $\blacktriangle$\\
  Lulin                       & \emph{R}    & 45                       & black $\bullet$ \\
  Mt. Maidanak                & \emph{BVRI} & 133, 135, 136, 136       & blue $\blacksquare$ \\
  Osaka Kyoiku                & \emph{R}    & 19                       & green $\blacksquare$\\
  Perkins\tablenotemark{c}    & \emph{BVRI} & 112, 116, 193 (193), 110 & red $\blacksquare$\\
  Rozhen (200 cm; 50/70 cm)   & \emph{BVRI} & 7, 8, 23, 8              & cyan $\blacksquare$\\
  San Pedro Martir\tablenotemark{c}   & \emph{R}    & 14 (14)                  & magenta $\blacksquare$ \\
  Sirio                       & \emph{R}    & 2                        & orange $\blacksquare$ \\
  Skinakas                    & \emph{BVRI} & 124, 123, 124, 124       & black $\blacksquare$ \\
  Skinakas\tablenotemark{c} (Robopol)      & \emph{R}    & 5 (5)                    & blue $\blacktriangle$ \\
  St. Petersburg\tablenotemark{c} (LX-200) & \emph{BVRI}    & 15, 17, 48 (37), 47   & green $\blacktriangle$\\
  Tijarafe                    & \emph{R}    & 219                      & cyan $\blacktriangle$\\
  Vidojevica (140 cm; 60 cm)  & \emph{BVRI} & 3, 3, 3, 3               & magenta $\blacktriangle$\\
  West Mountain               & \emph{V}        & 13                   & orange $\blacktriangle$\\
  \enddata
    \tablenotetext{a}{Listed for each filter. Number in parentheses refers to polarimetry measurements for that filter.}
    \tablenotetext{b}{For use in Figures~\ref{fig:EntireCampaignLC}-\ref{fig:WeekLC2}.}
    \tablenotetext{c}{Photometry and polarimetry}
\end{deluxetable*}

The Whole Earth Blazar Telescope (WEBT) was formed in 1997 as a network of optical, near-infrared, and radio observatories working together to obtain continuous well-sampled monitoring of the flux and polarization of blazars. In 2007 the WEBT started the GLAST-AGILE Support Program \citep[GASP; e.g.,][]{Villata2008, Villata2009a}. The GASP-WEBT data reported here correspond to four-\edit1{band} 
optical photometry (\emph{BVRI}) and \emph{R}-band polarimetry measured from 2019 August 05 to 2019 November 02. The data were checked for consistency between different observers and telescopes \citep[following the standard WEBT prescription, e.g.,][]{Villata2002}.
Table~\ref{tab:OpticalTelescopes} lists the observatories that participated in the campaign, while Table~\ref{tab:BLLacCompStar} gives magnitudes of comparison stars used in the photometric analysis. The data were corrected for Galactic extinction and contamination from the host galaxy, assuming contamination of $\sim60\%$ of the total host flux density, as suggested by \cite{Raiteri2010} for a circular aperture with a radius of 8\arcsec\ employed for BL Lac photometry. Galactic extinction along the line of sight to BL Lac was calculated according to \citet{Cardelli1989}, with $R_V = 3.1$ and $A_B = 1.42$ from \citet{Schlegel1998}.\footnote{\edit1{We adopt this Galactic extinction value in order to conform with previous studies \citep[e.g.,][]{Raiteri2010, Wehrle2016}. We note that a revised Galactic extinction value $A_B = 1.192$ has been proposed by \citet{Schlafly2011}.}} 
Table~\ref{tab:BLLacOptConversion} gives the extinction, absolute flux density conversion coefficient, and host galaxy total flux density for each filter.

\edit1{As in \citet{Raiteri2010}, a comparison between the \swift\ UVOT \emph{b} and \emph{v} data and WEBT \emph{B} and \emph{V} data revealed an offset between the space-based and ground-based magnitudes. We used the offset determined by \citet{Raiteri2010}, with
 $B-b = 0.10$, and $V - v = -0.05$.}

\begin{deluxetable*}{lcccccl}[t]
    \tablecaption{Magnitudes and distances of primary comparison stars in the BL Lac field.\label{tab:BLLacCompStar}}
   \tablewidth{0pt}
    \tablehead{
    \colhead{Star} & \colhead{\emph{B}\tablenotemark{a}} &\colhead{\emph{V}\tablenotemark{b}} & \colhead{\emph{R}\tablenotemark{b}} &\colhead{\emph{I}\tablenotemark{b}} & \colhead{$\varpi$} & \colhead{Distance} \\
    \colhead{} & \colhead{} & \colhead{} & \colhead{} & \colhead{} & \colhead{[mas]} & \colhead{[pc]} 
    }
    \startdata
    B & 14.68 $\pm$ 0.04 & 12.90 $\pm$ 0.04 & 11.99 $\pm$ 0.04 & 11.12 $\pm$ 0.05 & 0.3393 $\pm$ 0.0266 & $2980^{+446}_{-343}$ \\
    C & 15.20 $\pm$ 0.05 & 14.26 $\pm$ 0.06 & 13.79 $\pm$ 0.05 & 13.32 $\pm$ 0.05 & 1.8078 $\pm$ 0.0213 & $\phn 549^{+3}_{-8}$ \\
    H & 15.81 $\pm$ 0.06 & 14.40 $\pm$ 0.06 & 13.73 $\pm$ 0.06 & 13.07 $\pm$ 0.06 & 0.6922 $\pm$ 0.0928 & $1453^{+111}_{-97}$ \\
    K & 16.36 $\pm$ 0.07 & 15.47 $\pm$ 0.07 & 15.00 $\pm$ 0.07 & 14.54 $\pm$ 0.07 & 0.8994 $\pm$ 0.0188 & $1113^{+40}_{-37}$ \\
    \enddata
    \tablenotetext{a}{From \citet{Bertaud1969}.}
    \tablenotetext{b}{From \citet{Fiorucci1996}.}
\end{deluxetable*}


During the campaign, the WEBT collaboration observed BL Lac 459 times in \emph{B}-band, 492 times in \emph{V}-band, 1417 times in \emph{R}-band, and 507 times in \emph{I}-band. During the same time period, a subset of the WEBT observatories measured the \emph{R}-band polarization a total of 303 times.

\edit1{
\subsubsection{Optical Polarization Observations}
The \emph{R}-band polarization observations were obtained at the five telescopes noted in Table~\ref{tab:OpticalTelescopes}. The Perkins telescope is equipped with the PRISM camera, which includes a polarimeter with a rotating half-wave plate. Each polarization observation consisted of four consecutive measurements at instrumental position angles $0^\circ, 90^\circ, 45^\circ$, and $135^\circ$ of the waveplate to calculate the normalized Stokes parameters \emph{q} and \emph{u}. (For more detail see \citealt{Jorstad2010}.) Polarization
observations at the LX-200 and AZT-8 telescopes were performed in the same manner, each using an
identical photometer-polarimeter, with two Savart plates rotated by $45^\circ$ relative to each other. Swapping the plates allows one to obtain a  normalized Stokes parameter, either \emph{q} or \emph{u} (for more detail see \citealt{Larionov2008}). Several polarization observations
were also performed at the San Pedro Martir Observatory and Skinakas Observatory (Robopol program). Details of these observations can be found in \citet{Lopez2011} and \citet{Ramaprakash2019}, respectively. 
The degree, $P_{\text{R}}$, and position angle, $\chi_{\text{R}}$, of the polarization in all cases are calculated from  normalized Stokes \emph{q} and \emph{u} parameters. Throughout the paper we indicate
the degree of polarization in percent. All polarization data have been corrected for the Rice statistical bias \citep{Vinokur1965}, according to \citet{Wardle1974} [using the Modified Asymptotic Estimator, \citep[MAS;][]{Plaszczynski2014} in the case of the San Pedro Martir data]. The instrumental polarization of each instrument has been estimated to be $\lesssim0.5\%$, based on measurements of unpolarized calibration stars \citep[e.g.,][]{Schmidt1992}. We have calculated that the average uncertainty of a measurement of $P_{\text{R}}$ is $\langle \sigma_{P_{\text{R}}} \rangle =0.23\%$.
}

\edit1{As described above, a molecular cloud lies along the line of sight to BL Lac, accounting for a substantial fraction of the total hydrogen column density \citep[e.g.,][]{Bania1991, Madejski1999}. This molecular cloud could possibly contaminate the measured polarization due to dichroic absorption by aligned dust particles along the line of sight.}

\edit1{The data reduction methods applied to all of the polarization data reported here use field stars near the position of BL Lac to perform both interstellar and instrumental polarization corrections. If the field stars lie beyond the molecular cloud, then their polarization would include the dichroic absorption effects from the cloud, and the effects of the molecular cloud will be subtracted out from the polarization of the source. The distance to the cloud has been estimated to be $\sim 330$ pc based on the Galactic latitude and average distance of molecular clouds in the solar neighborhood above and below the Galactic plane \citep{Lucas1993, Moore1995}.}

\edit1{We have obtained the \emph{Gaia} DR2 parallaxes for the four main comparison stars listed in Table~\ref{tab:BLLacCompStar} \citep{Gaia2016, Gaia2018}. However, simple inversion of the parallax introduces known biases to the calculated distance, especially when the relative uncertainties are large. A proper calculation of distance requires a proper statistical treatment of the data. With the \texttt{pyrallaxes} program}\footnote{\url{https://github.com/agabrown/astrometry-inference-tutorials}} \edit1{\citep{Luri2018}, we have used Bayesian inference to calculate the distances to each of the comparison stars. We have implemented the two recommended priors from \citet{BailerJones2015}: a uniform distance prior (out to 100 kpc) and an exponentially decreasing space density prior (with a characteristic length scale $L = 1.35$ kpc). Both priors generate extremely similar distance measurements to the stars. We list the parallaxes and distances to the stars, with 90\% uncertainty intervals, in Table~\ref{tab:BLLacCompStar}. These distances are calculated with the exponentially decreasing space density prior.}

\edit1{All comparison stars lie beyond the molecular cloud, with the closest one still $\sim 200$ pc beyond the cloud. If we assume that the emission from the stars is intrinsically unpolarized, then the measured polarization to the stars includes the dichroic absorption effects from the cloud. Using 98 of the polarization measurements from the Perkins telescope taken during the monitoring period (during good weather), we estimate the average ISM polarization parameters toward BL Lac to be $P_{\text{R,ISM}} = 0.43\% \pm 0.08\%$ and $\chi_{\text{R,ISM}} = 69\degr \pm 5\degr$. This level of polarization is well below the upper limit of the contribution of the ambient interstellar dust to polarization along the line of sight through the Milky Way, $P_{\text{isp}} \leq 9\% * \text{E}(B-V)$ \citep{Serkowski1975}. }

In addition, contamination from the host galaxy can modify the observed degree of optical linear polarization. We assume that the emission from the host galaxy, dominated by starlight, is unpolarized. If the observed degree of linear polarization is $P_{\text{obs}}$, then the intrinsic degree of polarization, $P_{\text{AGN}}$, is found through a modulating factor

\begin{equation}
    P_{\text{AGN}} = \frac{F_{\text{obs}}}{F_{\text{obs}} - 0.6 F_{\text{host}}} P_{\text{obs}} \ ,
    \label{eqn:2}
\end{equation}
\noindent where $F_{\text{obs}}$ is the observed flux density, $F_{\text{host}}$ is the total flux density of the host galaxy, and the factor of 0.6 is the fraction of host galaxy contamination. We have applied the correction for dilution of the polarization from the host-galaxy starlight to the values of the degree of polarization of BL Lac reported here. The position angle of polarization is unaffected by the light of the host galaxy.

\subsubsection{\TESS\ Data}
\label{subsec:TESSDataReduction}

The \emph{Transiting Exoplanet Survey Satellite} (\TESS) \citep{Ricker2015} continuously monitors sectors of the sky at a wavelength band of 6,000-10,000 \AA, shifting to different sectors after several weeks. This wavelength coverage nearly encompasses the \emph{R}- and \emph{I}-band WEBT coverage.
A summary of the telescope specifications is provided in Table~\ref{tab:TESSSpecs}.
\TESS\ collects full-frame images (FFIs) of the entire field of view (FOV) every 30 minutes, with ``postcard cutouts" of select targets obtained at a higher cadence of 2 minutes.
A list of \TESS\ identification numbers for sources is given by the \TESS\ Input Catalogue \citep[TIC,][]{Stassun2018}.
Data products from the \TESS\ mission are publicly available on the Mikulski Archive for Space Telescopes (MAST).\footnote{\url{https://heasarc.gsfc.nasa.gov/docs/tess/data-access.html}.}

\begin{deluxetable}{ll}[t!]
    \tablecaption{Summary of \TESS\ telescope specifications.\label{tab:TESSSpecs}}
    \tablewidth{0pt}
    \tablehead{
    \colhead{Attribute} & \colhead{Value}
    }
    \startdata
    Single Camera FOV & $24\degr \times 24\degr$ \\
    Combined FOV & $24\degr \times 96\degr (3200\ \text{deg}^2)$ \\
    Single Camera Aperture & 10.5 cm \\
    Focal Ratio ($f/\#$) & $f/1.4$ \\
    Wavelength Range & 6000 - 10000 \AA \\
    Pixel Size on Sky & 21\arcsec\ \\
    FFI Exposure Time & 30 min \\
    Orbital Period & 12-15 days \\
    \enddata
    \tablerefs{\citet{Ricker2015}}
\end{deluxetable}

\begin{figure}[t!]
    \begin{center}
        \includegraphics[width=0.5\textwidth]{{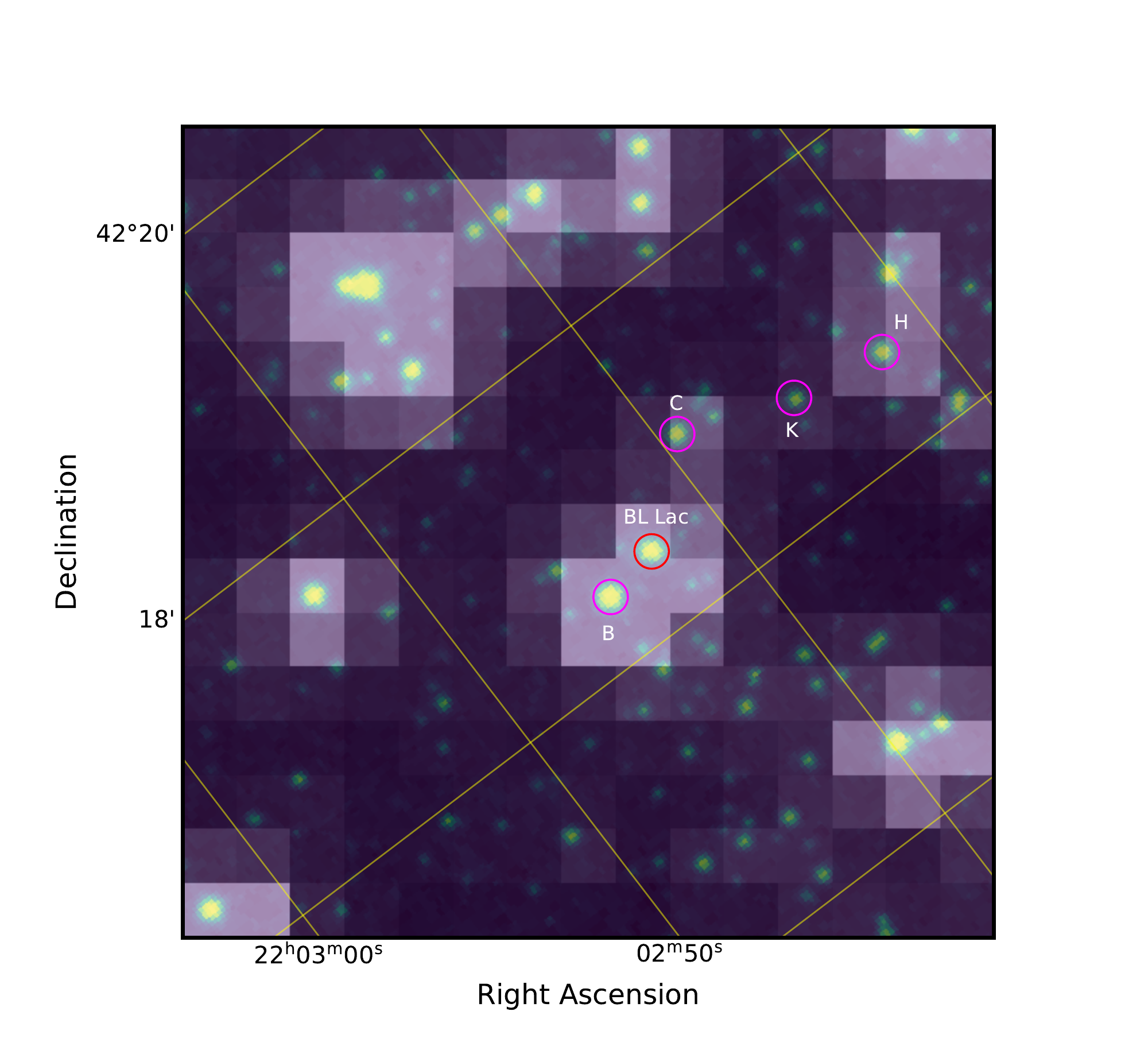}}
        \caption{A $15\times15$ pixel cutout of a \TESS\ FFI centered on BL Lac from an observation on 2019 October 4 07:15:36 UT (large pixels). A Digitized Sky Survey image of the same field is shown in the background.  Magnitudes of labelled primary WEBT comparison stars are given in Table~\ref{tab:BLLacCompStar}. The yellow lines correspond to lines of constant RA and Dec.\label{fig:TESSDSS}}
    \end{center}
\end{figure}

BL Lac (TIC 353622691) was observed by \TESS\ from 2019 September 12 03:29:27 to October 6 19:39:27 UT, with its image located on camera 1 CCD chip 2. Figure~\ref{fig:TESSDSS} shows a cutout of a \TESS\ FFI taken on 2019 October 04 07:15:36 UT, superposed on a Digitized Sky Survey image of the same field. The large pixel sizes and crowded field render photometry of faint targets difficult. However, BL Lac is usually one of the brightest objects in the field, and thus photometry is easier to perform.

\begin{figure}[t!]
    \begin{center}
        \includegraphics[width=0.475\textwidth]{{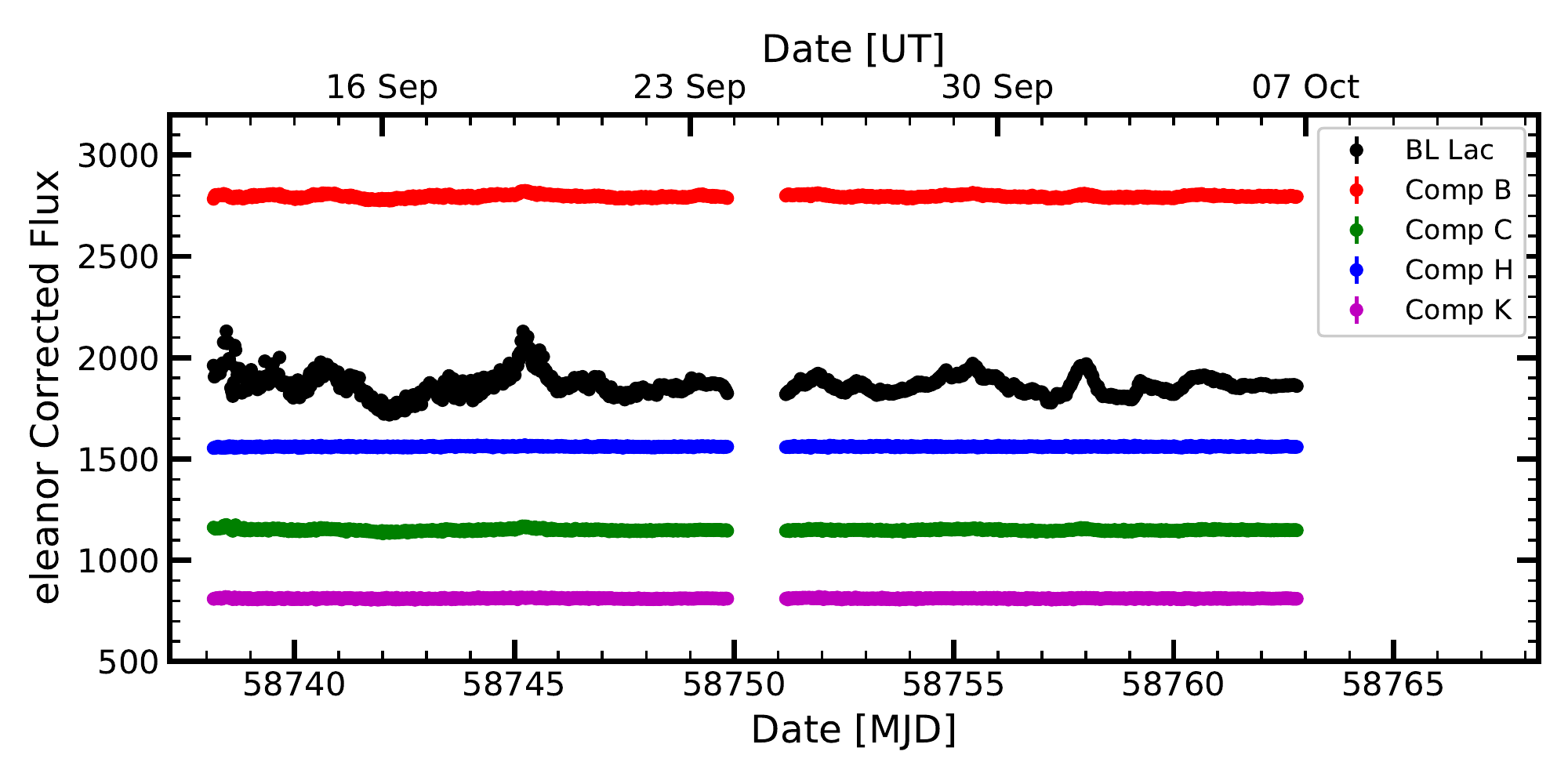}}
        \caption{\TESS\ light curves of BL Lac and four comparison stars.\label{fig:eleanor-lcs}}
    \end{center}
\end{figure}

In order to carry out a preliminary exploration of the \TESS\ data, we have used the \texttt{eleanor} software package \citep{Feinstein2019} to reduce the FFI images for BL Lac and the four main WEBT-recommended comparison stars (see Table~\ref{tab:BLLacCompStar}). The resulting light curves are displayed in Figure~\ref{fig:eleanor-lcs}. We did not make any quality cuts while reducing the data, and for each source we defined a unique 3 or 4 pixel aperture that did not overlap with the aperture of another source of significant flux. The \texttt{eleanor} pipeline is able to remove the majority of systematic effects present in \TESS\ data, as evidenced by the relatively constant flux of the four comparison stars. (Some small variations are seen in comparison stars close to BL Lac, but these variations are due to light from BL Lac leaking into the aperture of the comparison stars.)

The removal of such artificial systematic trends from the data with 30 min cadence is generally successful in stellar light curves made from \TESS\ data. However, long-term variability --- months or longer --- common in AGNs may be mistaken as instrumental effects and removed from the light curve by commonly used \TESS\ data reduction software designed to search for evidence of exoplanets. 
\edit1{Future studies, especially for sources in the CVZ, will need to be more carefully calibrated.} We note that such a procedure is unnecessary for the present study, which focuses on short-term variability with 2-min cadence data, as discussed below. 

BL Lac was selected as a target for monitoring by \TESS\ with a 2 min cadence. These data were processed by the Science Processing Operations Center \citep[SPOC;][]{Jenkins2016}. The pipeline performs both general CCD and pixel-level corrections, computes optimal apertures, completes a photometric analysis of the sources, and performs a ``presearch data conditioning" (PDC) procedure designed to take into account systematic effects. It also removes isolated outliers, corrects the flux of a source for crowding effects, and corrects for the aperture not containing all of the flux from a target source.
The light curve obtained from this method is called the ``PDCSAP" light curve.

There are several aspects of the PDCSAP light curve to check before using it in subsequent analyses. The pipeline calculates the optimal aperture,
which contains only 66\% of the total flux of the blazar (calculated using the pixel response function of the \TESS\ detectors). Also, the flux from BL Lac represents only $\sim30\%$ of the total flux in the aperture from all sources.\footnote{This metric, often used to describe the crowding of a source, may also be susceptible to stray background light entering the telescope. The observing sector containing BL Lac was noted as having high amounts of stray light from the Earth and Moon entering the aperture. A useful depiction of this background light can be seen in the sector video made by Ethan Kruse: \url{https://www.youtube.com/watch?v=MhAtZfMe7oI}.}
We do not consider these issues as reasons to avoid using the PDCSAP light curve. The \texttt{eleanor} analysis in Figure~\ref{fig:eleanor-lcs} shows that the major bright, nearby stars are not variable, and that it is possible to separate the variable emission of BL Lac from the flux of stars within one \TESS\ pixel. Also, we have normalized the light curve \edit1{to its median value} rather than convert the \TESS\ electron counts to an energy flux density for the analysis presented in this paper. 

We find no evidence of large-amplitude exponential decreases in the flux at the beginning of an orbit attributed to uneven heating of the telescope during data transmission modes (dubbed ``thermal ramps") that are often present in \TESS\ light curves. We thus make no cuts of the data near the beginning or end of an orbit.

We have checked the data quality raised by the pipeline for the PDCSAP light curve data, but find an insignificant number of data points with quality issues.
We have, however, eliminated 14 outlier points out of 16,006 observations of BL Lac

The SPOC pipeline is subject to over-fitting similar to the \texttt{eleanor} program (see above). The PDC noise goodness metric (between 0 and 1), present in the header of the light curve data product, is used as an indicator of over-fitting. The PDC noise goodness metric for BL Lac is 0.68, which implies a modest level of removal of intrinsic long-term trends. Since our study focuses on short-term variability of BL Lac, this over-fitting has an insignificant effect on our analysis.



\begin{figure*}[t]
    \begin{center}
        \includegraphics[height=0.925\textheight]{{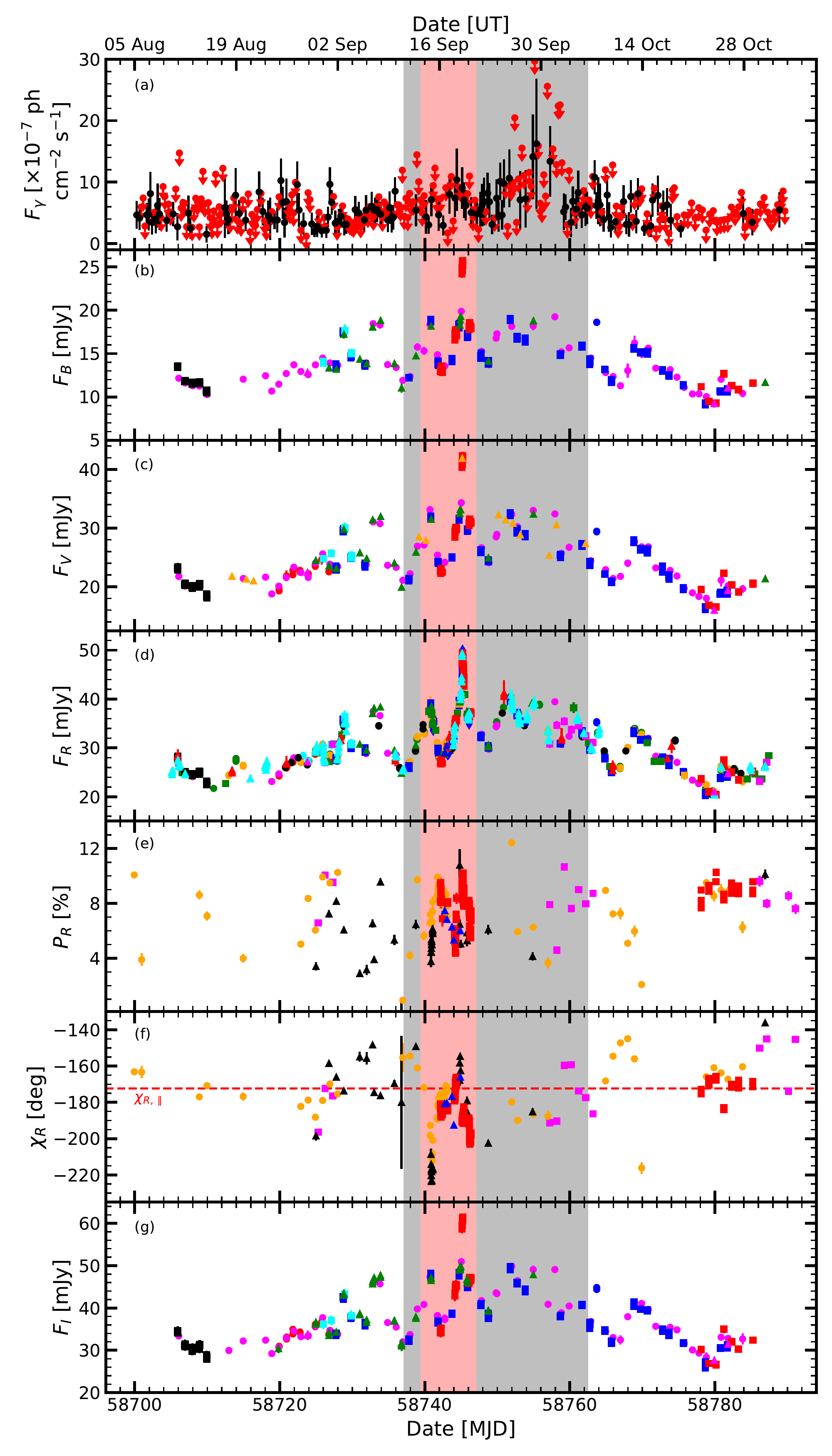}}
        \caption{Light curves and polarization vs.\ time during the WEBT campaign: $(a)$ \fermi-LAT $\gamma$-ray flux, with upper limits denoted by downward-pointing red arrows; $(b-d, g)$ WEBT \emph{BVRI} flux densities. Colors and symbol shapes represent different telescopes, for which a key is provided in Table~\ref{tab:OpticalTelescopes}; $(e)$ degree, $P_{\text{R}}$, and $(f)$ position angle, $\chi_{\text{R}}$, of optical linear polarization in \emph{R}-band; range is selected for comparison with the direction of the parsec-scale jet. 
        In all panels, the gray shaded area indicates the time span of the \TESS\ observations and the red shaded area indicates the period of concurrent NuSTAR and \swift\ observations. Error bars are shown in all panels, but in most cases are smaller than the symbol size.
        \label{fig:EntireCampaignLC}}
    \end{center}
\end{figure*}

\section{Multi-Wavelength Light Curves}
\label{sec:LCs}

The multi-wavelength behavior of BL Lac over the entire WEBT monitoring campaign is shown in Figure~\ref{fig:EntireCampaignLC}. The optical data coverage is dense, especially in $R$-band.

\begin{figure*}[t]
    \begin{center}
        \includegraphics[scale=0.6]{{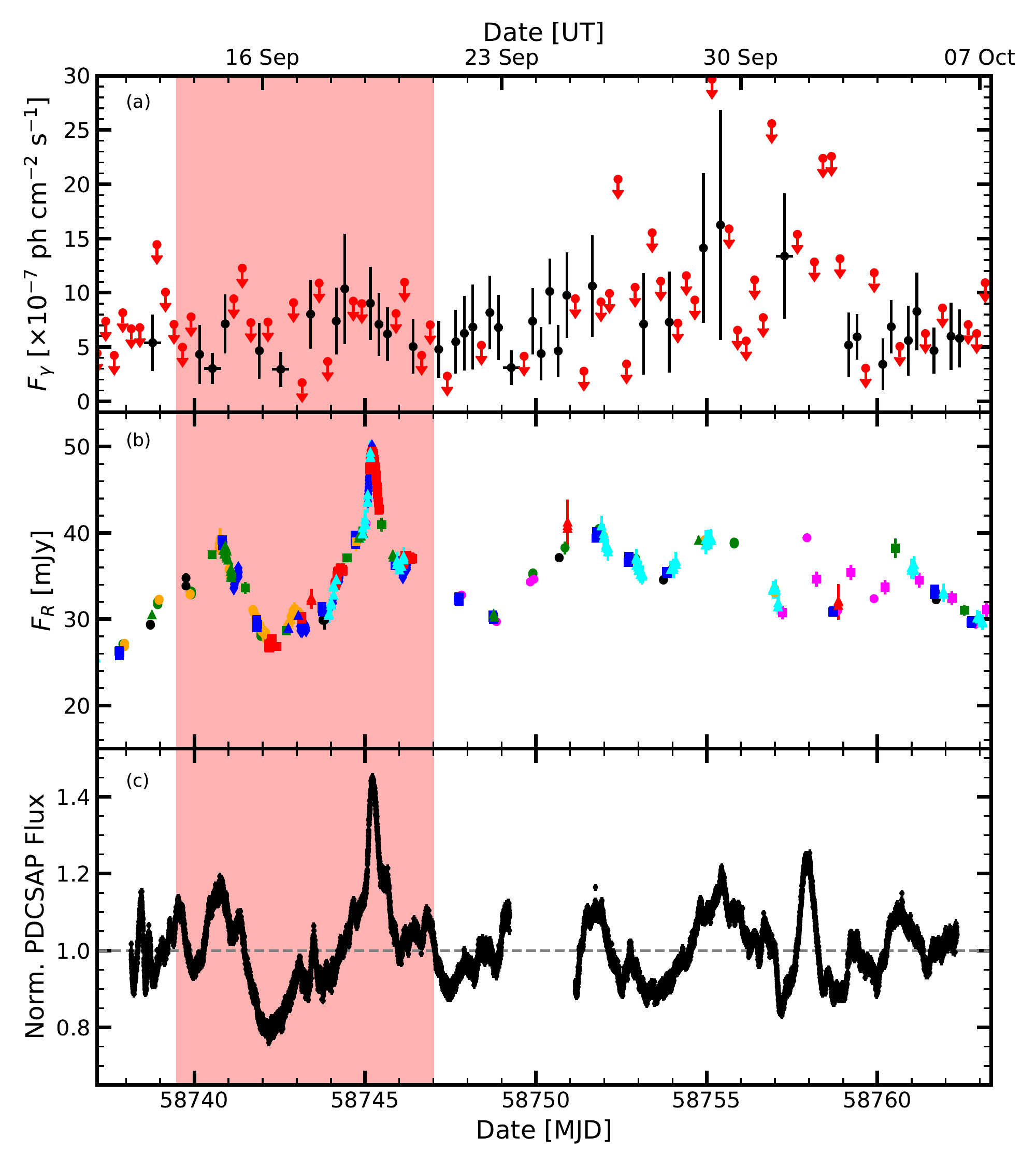}}
        \caption{Flux \edit1{or flux density} vs.\ time of BL Lac during the \TESS\ monitoring period: $(a)$ \fermi-LAT $\gamma$-ray flux, with upper limits denoted by red downward-pointing arrows. $(b)$ WEBT $R$-band flux density; symbol colors and shapes represent different telescopes; a key is provided in Table~\ref{tab:OpticalTelescopes}. $(c)$ \TESS\ 2-min cadence, normalized \TESS\ PDCSAP flux. In all panels, the red shaded region indicates the time of concurrent NuSTAR and \swift\ observations.
        Error bars are shown in all panels, but are often smaller than the symbol size.\label{fig:TESSMonthLC}}
    \end{center}
\end{figure*}

The $\gamma$-ray flux increased from $\sim3\times10^{-7}$ ph cm$^{-2}$ s$^{-1}$ to $1.5\times10^{-6}$ ph cm$^{-2}$ s$^{-1}$ over 10 days, peaking on September 29 (MJD: 58755), and then decayed quickly over the next two days. The optical light curves also rose to a peak in late September, although the details differ from the $\gamma$-ray behavior. In $R$-band the underlying trend (defined by the minima of shorter-timescale variations) corresponded to an increase from $\sim20$ mJy to $\sim35$ mJy before the flux density faded back down to $20$ mJy. Shorter-timescale fluctuations, with durations of hours to days, occurred frequently during the monitoring period at optical wavelengths. Any similar events would be difficult to identify in the $\gamma$-ray light curve owing to the low flux level and large number of non-detections. At least one event is identified at all wavelengths: the rapid brightening and decay on September 19 (MJD: 58745; see below).

The degree of linear polarization, $P_{\text{R}}$, fluctuated between 1\% and 12\% over the monitoring period. The average value, derived from the \edit1{normalized} $q$ and $u$ Stokes parameters, was $\langle P_{\text{R}} \rangle = 6.7\%$ with a standard deviation of 2.1\%. The mean electric-vector position angle was $\langle \chi_{\text{R}} \rangle = -183\degr\pm15\degr$, which is within 1$\sigma$ uncertainty from the average 43 GHz radio jet direction of $-173\degr$ \citep[][marked with a red dashed line in Figure~\ref{fig:EntireCampaignLC}\emph{(f)}]{Jorstad2017}. Significant swings by up to $\sim50\degr$ about this position angle were observed throughout the monitoring period.

Figure~\ref{fig:TESSMonthLC} shows the entire \TESS\ light curve of BL Lac, along with the \fermi-LAT $\gamma$-ray and WEBT $R$-band light curves. The \TESS\ count rates have been normalized to the median value of 1150.39 \edit1{e$^{-}$} s$^{-1}$. \emph{The WEBT and \TESS\ light curves are very similar}, despite the potential over-crowding, aperture, and over-fitting issues present in the PDCSAP light curve. The WEBT light curve, with its sparser sampling, is a reliable tracer of the major events and trends visible in the \TESS\ light curve.

In Figure \ref{fig:TESSMonthLC} the peak of the $\gamma$-ray light curve on September 29 is seen in greater detail. While the WEBT monitoring is sparse around this date, the \TESS\ light curve shows a complicated structure during the $\gamma$-ray brightening. Of particular note is the large increase in optical flux density on September 19, clearly seen in both the WEBT and \TESS\ light curves. This peak is also apparent in the $\gamma$-ray light curve as an increase in flux by a factor of $\sim 2$.

\begin{figure*}
  \begin{center}
    \includegraphics[width=0.9\textwidth]{{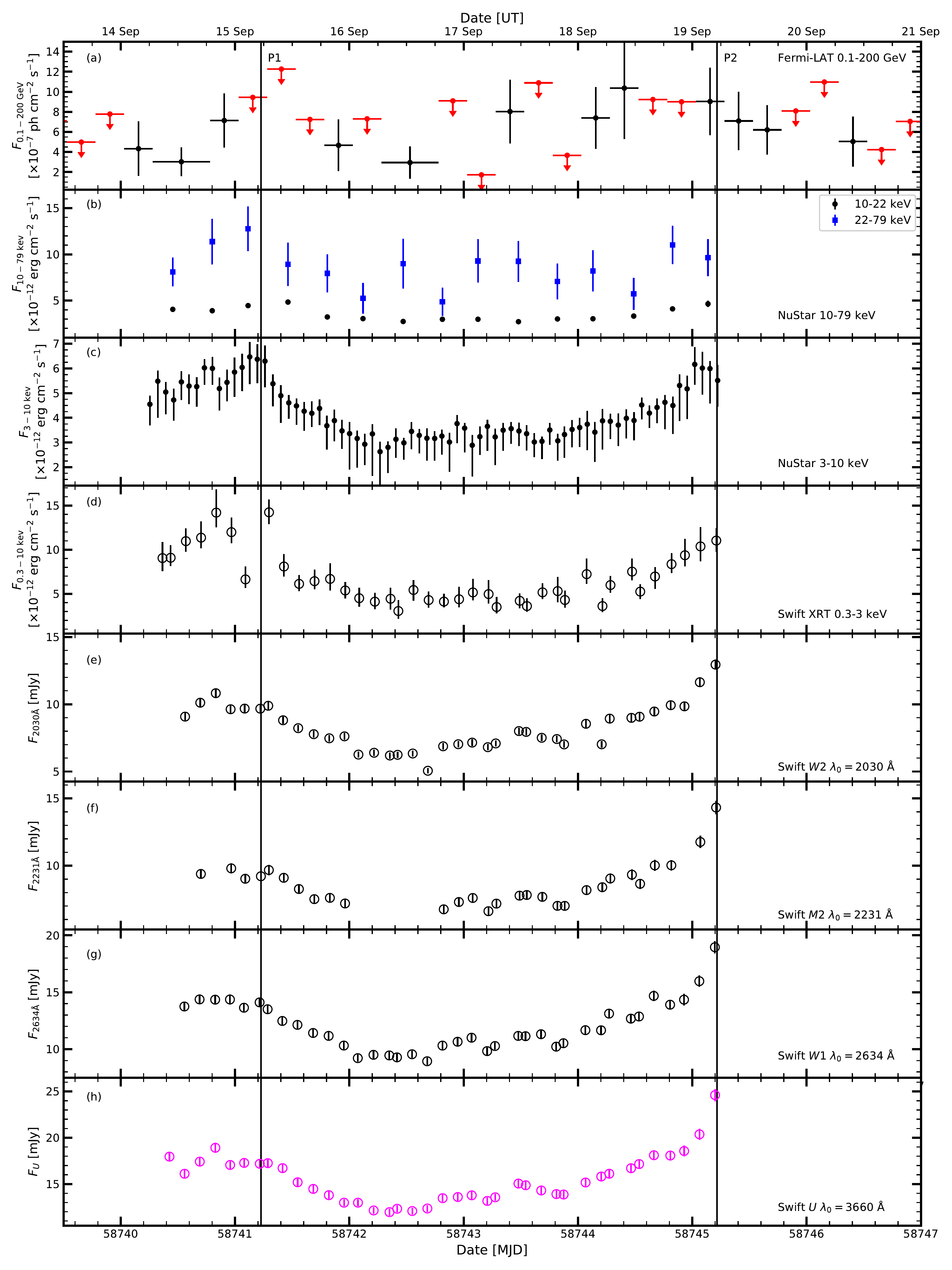}}
    \caption{Flux \edit1{or flux density} vs.\ time of BL Lac during the five days of concurrent monitoring for wavebands at UV and shorter wavelengths. $(a)$ \fermi-LAT $\gamma$-ray flux, with upper limits denoted by red downward-pointing arrows. $(b)$ and $(c)$ NuSTAR X-ray flux. $(d)$ \swift\ X-ray flux. $(e-h)$ \swift\ UV flux densities. 
    In all panels, two solid black lines indicate $P1$, the time of global maximum X-ray flux and $P2$, the time of peak optical flux \edit1{density}. Error bars are shown in all panels, but are smaller than the symbol size in many instances. \label{fig:WeekLC1}}
  \end{center}
\end{figure*}

\begin{figure*}
  \begin{center}
    \includegraphics[width=0.9\textwidth]{{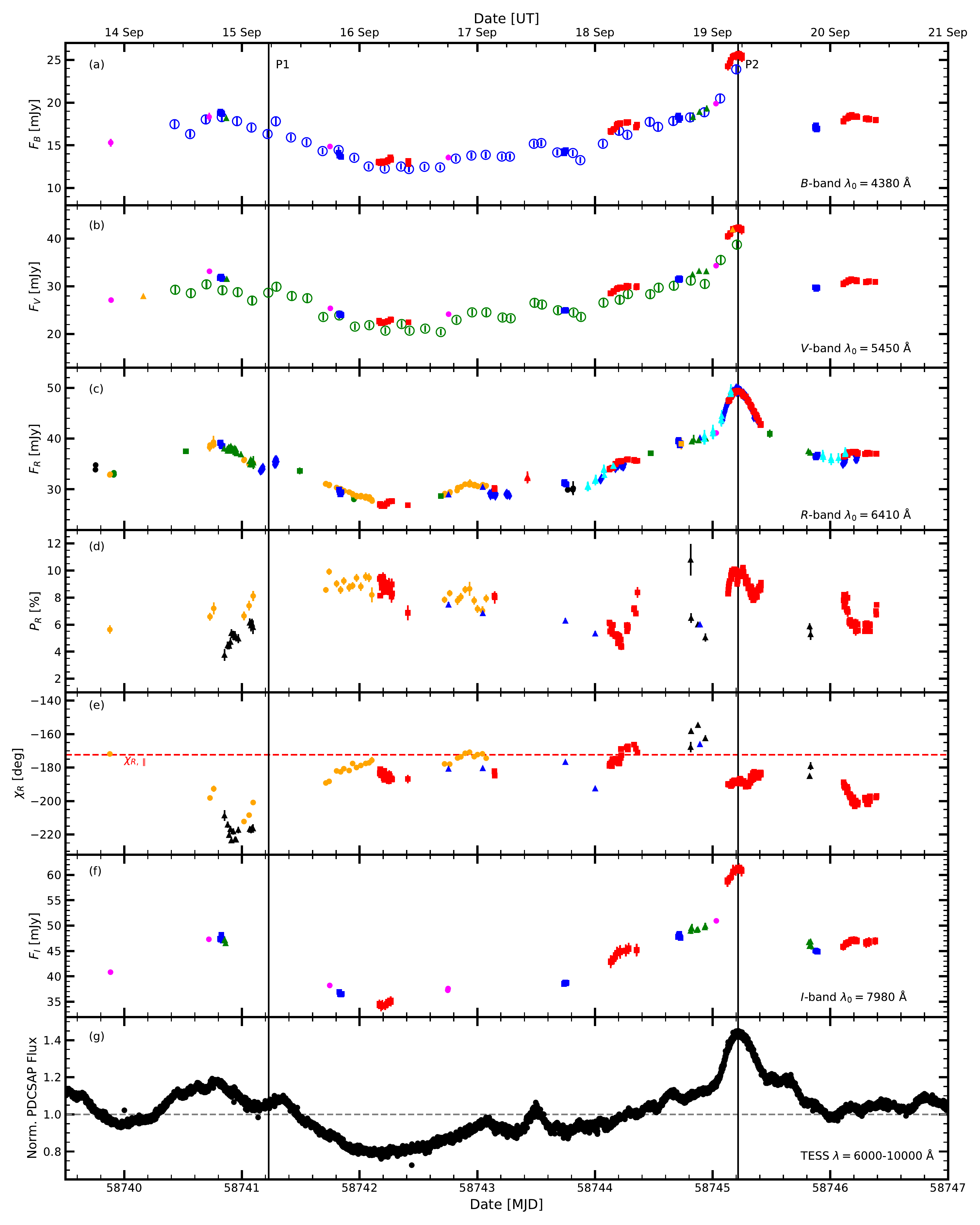}}
    \caption{Flux \edit1{or flux density} vs. time of BL Lac during the five days of concurrent monitoring at optical and near-IR wavebands. $(a-c, f)$ \swift\ (\emph{BV}) and WEBT (\emph{BVRI}) optical flux densities. Open circles denote \swift\ observations, while the other colors and symbols indicate the different ground-based telescopes used; a key is provided in Table~\ref{tab:OpticalTelescopes}. $(d)$ Degree and $(e)$ position angle of optical $R$-band linear polarization. The red dashed line corresponds to the average parsec-scale jet direction.
    $(g)$ \TESS\ 2-min PDCSAP normalized flux.
    As in Figure~\ref{fig:WeekLC1}, the two solid black lines indicate $P1$ and $P2$. Error bars are shown in all panels, but are often smaller than the symbols.\label{fig:WeekLC2}}
  \end{center}
\end{figure*}

Figures~\ref{fig:WeekLC1} and \ref{fig:WeekLC2} display time variations of the flux \edit1{or flux density} and polarization of BL Lac over the five days of concurrent monitoring at all frequencies. 
All light curves show the same general trend of two periods of higher flux near September 15 and 19, labelled \emph{P1} and \emph{P2} respectively, with a period of lower flux in between. The high flux state near \emph{P1} is most easily seen in the X-ray light curves (panels \emph{b-d}).
All of the light curves exhibit similar amplitudes of variability by a factor up to $\sim 2$. The UV and optical light curves only show a moderate increase of \edit1{flux density} during \emph{P1} compared with the higher-amplitude increase of \emph{P2}. The peak of \emph{P2} is very well sampled by both \TESS\ and the WEBT observations, with a smooth rise and fall. The rise of \emph{P2} is also well sampled at higher energies; however, the observations ended before the decline could be detected.

The optical linear polarization varied significantly near the periods \emph{P1} and \emph{P2}, but was generally stable during the intervening low-flux plateau. During the plateau, $P_{\text{R}}$ was high, near 9\%, and $\chi_{\text{R}}$ was essentially parallel to the 43 GHz radio jet \citep{Jorstad2017}. The position angle $\chi_{\text{R}}$ was quite variable during \emph{P1}, undergoing a $\sim 20\degr$ swing, but relatively stable for several hours during \emph{P2}.


\section{Short-Timescale Variability}
\label{sec:ShortVariability}

Inspection of the light curves, especially that from \TESS\, in Figures~\ref{fig:TESSMonthLC} and \ref{fig:WeekLC2}, reveals several periods of rapid changes in flux. In this section we calculate the shortest timescales of variability in the \TESS\ and X-ray data. We discuss the variability of the optical polarization in $\S$\ref{sec:polarization}.

\subsection{Intraday Variability of \TESS\ Data}
\label{subsec:IntradayTESS}

Several statistical methods have been developed and applied to quasar variability in order to quantify the degree of short-timescale, including intraday, flux variations. \citet{deDiego2010} compared several statistical tests using simulated light curves, and determined that a one-way analysis of variance (ANOVA) test is one of the most robust ways to identify statistically significant variability. An ANOVA test is a metric to judge the equivalence of measurements in a sample by breaking the sample into several groups and evaluating the means and variances of those groups.
The null hypothesis for an ANOVA test is equality of all group means. Applied to blazar variability, this null hypothesis can be translated as non-variability of the source over the time-period being tested. Two statistical measures are returned from the test, an $F$ statistic and a $p$ value. The null hypothesis can be rejected if (1) the returned $F$ statistic is greater than a critical value $F_{\text{crit}}$, calculated using the two degrees of freedom ($dk1 = k-1$ and $dk2 = N-k$, where $k$ is the number of groups and $N$ is the total number of measurements in the sample) and a user-selected significance value, and (2) the returned $p$ value is smaller than the significance value.
\citet{deDiego2010} recommends using a number of groups $k \geq 5$ in order for the test to have the most power to detect variability. In the following analyses, we label the critical value with the corresponding degrees of freedom as $F_{dk1,dk2}$.

An ANOVA test has been used to identify and characterize the optical flux density and polarization variability of the BL Mrk 421 \citep{Fraija2017} and the FSRQ 3C454.3 \citep{Weaver2019}, with observed timescales of variability of $\sim 2$ hr in both cases. The sampling rate of observations in these studies was on the order of several minutes between observations, for at most a few hours each night.
In contrast, the \TESS\ light curve of BL Lac, obtained at a 2-min cadence over about 25 days, allows for a much more systematic survey of variability to be performed. This produces robust metrics to be used to test for variability \citep{deDiego2015}. Since the \TESS\ light curves are evenly and densely sampled compared to the timescale of variations being investigated, a simple ANOVA test is sufficient \citep{deDiego2014}.

We have broken the \TESS\ PDCSAP light curve into hour-long sets, starting from the first \TESS\ observation, each containing $\sim30$ data points.\footnote{One set contains 8 data points, two contain 24, and the rest contain 31.} In total, we perform tests on 516 hour-long light-curve segments. All segments were normalized to the PDCSAP median value of 1150 e$^{-}$ s$^{-1}$
in order to avoid issues with the flux scaling present in \TESS\ light curves (see $\S$\ref{subsec:TESSDataReduction}).
Following the recommendation of \citet{deDiego2010}, we passed each light curve through an ANOVA test, breaking these hourly segments into 5 groups ($\sim6$ data points per group; $df1 = 4$ and average $\langle df2 \rangle = 26$). We have chosen a significance value of $p < 0.001$ (3$\sigma$) as the threshold for variabliity over the hour-long periods. Through this method, we have identified 107 hour-long periods during which BL Lac was significantly variable ($\sim20\%$ of the total observation period, after one accounts for the downlink break between orbits).

\begin{figure}[t]
    \begin{center}
        \includegraphics[width=0.45\textwidth]{{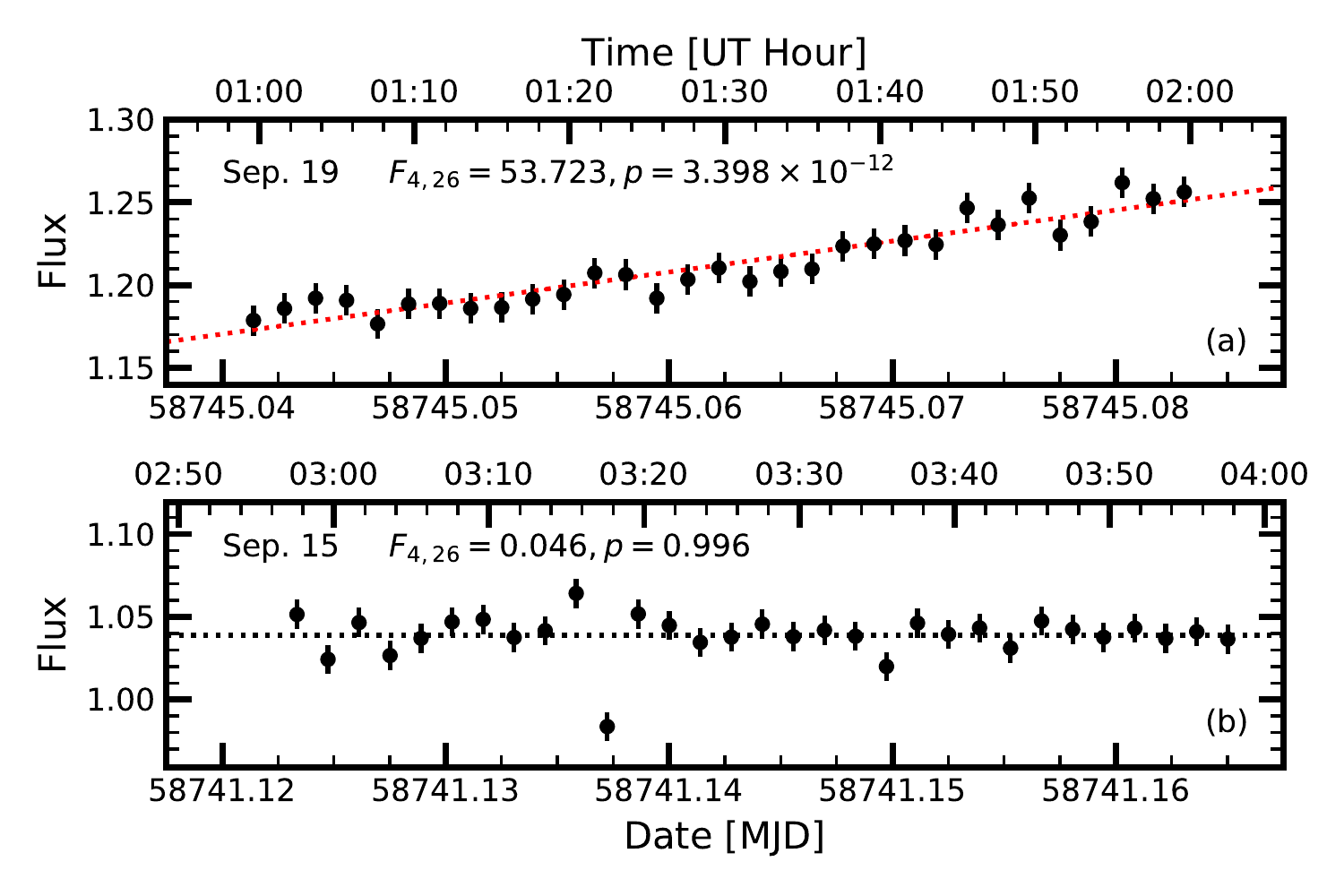}}
        \caption{Examples of hour-long \TESS\ light-curve segments analyzed using the ANOVA test: the most \emph{(a)} and least \emph{(b)} statistically significant \edit1{variability}. The red and black dotted lines show linear fits to the data for cases \emph{(a)} and \emph{(b)}, respectively.
        \label{fig:ANOVANights}}
    \end{center}
\end{figure}

Figure~\ref{fig:ANOVANights} shows two examples of the analyzed hour-long light-curve segments. Panel \emph{(a)} corresponds to the most statistically significant detected variability (starting on September 19 00:59:37 UT; MJD: 58745.0414), with $F_{4,26} = 53.723$ and $p = 3.398 \times 10^{-12}$. During this hour, the flux increased by $\sim10\%$. In contrast, panel \emph{(b)} represents the hour-long period with the smallest $F$-value (starting on September 15 02:49:12 UT; MJD: 58741.1233), with $F_{4,26} = 0.046$ and $p = 0.996$.

We cannot conclusively determine whether the variability is slightly greater when the source is in a higher flux state. The average flux (normalized to the median value) of a variable hour was 1.036, with a standard deviation of 0.104. For the non-variable hours, the average flux was 0.998, with a standard deviation of 0.094. The difference is therefore not statistically significant.

\begin{figure}[t]
    \begin{center}
        \includegraphics[width=0.45\textwidth]{{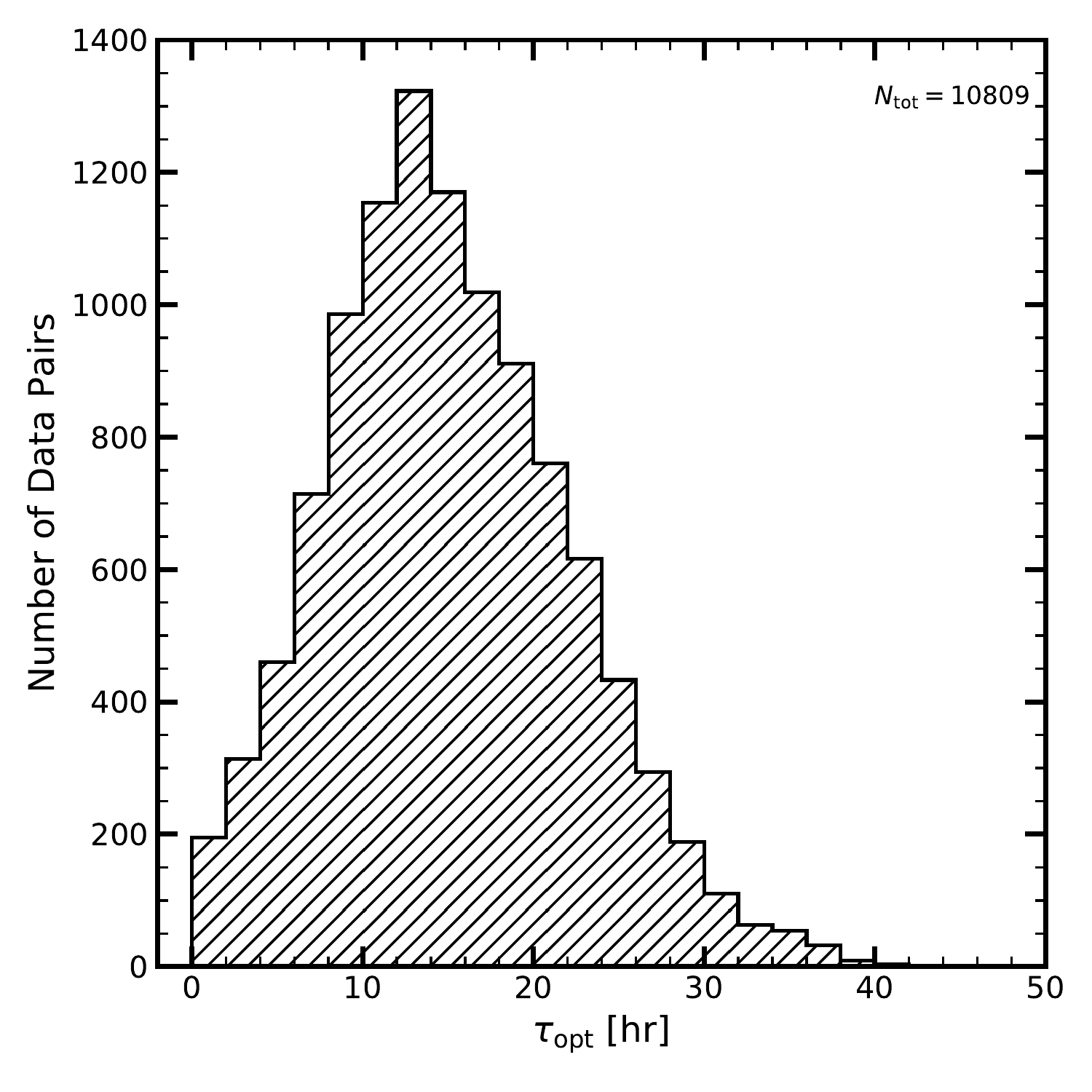}}
        \caption{Histogram of timescales of variability of \TESS\ data. \label{fig:TimescalesTESS}}
    \end{center}
\end{figure}

With the ANOVA test showing that BL Lac is variable on sub-hour timescales during a significant fraction of the monitoring period, we now calculate the timescale of optical flux variability, $\tau_{\text{opt}}$. We consider all pairs of flux measurements within the hour-long sets of data if, for a given pair, $S_2 - S_1 > \frac{3}{2}(\sigma_{S_1} + \sigma_{S_2})$, where $S_i$ and $\sigma_{S_i}$ refer to the flux and associated uncertainty of each measurement. Of the $\sim50,000$ possible pairs of observations, $\sim11,000$ met this uncertainty criterion. For these pairs, we calculate $\tau_{\text{opt}}$ using the formalism suggested by \citet{Burbidge1974}: $\tau = \Delta t / \text{ln}(S_2 / S_1)$ with $S_2 > S_1$, where $\Delta t = |t_2 - t_1|$ is the difference in the time of observation of each measurement. 
The average timescale, is 15 hours, with a standard deviation of 7 hours. This is very similar to the derived minimum timescale of variability in the softer 3-10 keV band (see $\S$\ref{subsec:XRayVariability} below). The minimum calculated timescale of variability of the \TESS\ data is 
31 minutes.
We plot the calculated timescales of variability in Figure~\ref{fig:TimescalesTESS}, with histogram bins of 2 hours. 

\subsection{X-Ray Variability}
\label{subsec:XRayVariability}

We have performed the same test for variability on the \swift\ XRT 0.3-3 and NuSTAR 3.0-10 keV light curves as done on the \TESS\ light curve. Owing to the much lower sampling rate and time-span of the X-ray observations, we first performed an ANOVA test on the entire five-day period. We have separated each sample light curve into 5 equal-sized groups for the ANOVA test. Both the \swift\ XRT and NuSTAR light curves are detected as significantly variable, with  $F$-statistics and $p$-values of $F_{4,35}^{\swift} = 12.2,\ p^{\swift} = 2.7\times10^{-6}$, and $F_{4,70}^{\text{NuSTAR}} = 34.9,\ p^{\text{NuSTAR}} = 5.0\times10^{-16}$. Because the photon indices show evidence for variability in Figures \ref{fig:XRTPhIndvsFlux} and \ref{fig:NuSTARcounts}, we have performed an ANOVA test on the photon index versus time curves for each satellite. 
While the 3-10 keV photon index is determined through the test to be significantly variable, with $F_{4,70}^{\Gamma, \text{NuSTAR}} = 13.0,\ p^{\Gamma, \text{NuSTAR}} = 5.4\times10^{-8}$, the 0.3-3 keV photon index is only moderately variable, with $F_{4,35}^{\Gamma, \swift} = 3.95$,\  $p^{\Gamma, \swift} = 0.009$, slightly above the $3\sigma$ threshold. 

Since both the \swift\ XRT and NuSTAR light curves show variability, we now proceed with the higher time-resolution NuSTAR 3-10 keV light curve to investigate shorter timescales of variability.
To accomplish this, we divided the light curve into five day-long bins, each considered a separate sample containing 15 data points. These five samples were then each passed through the ANOVA test, with 5 groups of 3 data points per sample. This binning resulted in an optimum number of data points per group for the ANOVA test to be effective. Table~\ref{tab:XRayANOVA} gives the time periods of each sample and the calculated $F$ and $p$ statistics. Only two day-long bins are variable at the $p < 0.001$ level. These times correspond to the decay of \emph{P1} and rise of \emph{P2} in the X-ray light curves.

\begin{deluxetable}{ccrl}[t]
    \tablecaption{ANOVA $F$ and $p$ statistics calculated for day-long bins of the NuSTAR 3.0-10.0 keV energy band. The critical $F$-value is $F_{4,10}^{\text{crit}} = 11.283$. \label{tab:XRayANOVA}}
    \tablewidth{0pt}
    \tablehead{
    \colhead{Start Time} & \colhead{End Time} & \colhead{$F$} & \colhead{$p$} \\
    \colhead{[UT]} & \colhead{[UT]} & \colhead{} & \colhead{}
    }
    \startdata
    09-14 06:07:05 & 09-15 04:40:30 &  5.42 & 0.014 \\
    09-15 06:16:46 & 09-16 04:49:50 & 15.9 & $2.5\times10^{-4}$ \\
    09-16 06:26:47 & 09-17 04:59:02 &  1.70 & 0.23 \\
    09-17 06:36:33 & 09-18 05:09:02 &  2.22 & 0.14 \\
    09-18 06:46:57 & 09-19 05:18:44 & 22.5 & $5.5\times10^{-5}$ \\
    \enddata
\end{deluxetable}

We calculate the timescale of variability using the above method for all pairs of data within each day-long bin of observations that was deemed variable by the ANOVA test. In total, 210 pairs of data are available, but only 69 meet the uncertainty requirement. The average timescale of variability for the X-ray light curve is 36 hr, with a standard deviation of 10 hr and a minimum of 14.5 hr.


\section{Multi-Band Behavior} 
\label{sec:MulticolorBehavior}

Analysis of multi-\edit1{wavelength} IR/optical/UV data can identify separate components contributing to emission, each with its own continuum spectrum and variability properties.
In order to isolate the contribution of the component of (likely synchrotron) radiation that is variable on the shortest timescales, we follow a method first suggested by \citet{Choloniewski1981} and later developed by \citet{Hagenthorn1997}. A relative continuum spectrum can be constructed from essentially simultaneous flux density \edit1{measurements} in different bands by considering the slopes of the sets of cross-frequency \edit1{flux density vs.\ flux density (here shortened to ``flux-flux'')} relations.
This method has been successfully applied to a number of blazars \citep{Hagenthorn2008,Larionov2008,Jorstad2010,Larionov2010,Larionov2016,Gaur2019,Larionov2020}.
In the case of BL Lac \citep{Larionov2010, Gaur2019}, the relation between the optical and near-infrared flux densities over long timescales and major changes in flux \edit1{density} cannot be properly fit by a simple linear dependence. These authors obtained a second-order polynomial fit to the flux density of a given band $i$: $F_{\nu,i} =  a_i F_{\text{R}}^2 + b_i F_{\text{R}} + c_i$. They also found that the polynomial fits flatten toward higher frequencies, indicating that BL Lac exhibits a bluer-when-brighter trend, in agreement with other methods of determination of the spectral slope of the variable component \citep{Villata2002, Villata2004, Papadakis2007}. The flux \edit1{density} range available to \citet{Hagenthorn2004} was not wide, hence they did not detect any deviations from a linear dependence in the flux-flux plots.

\begin{figure*}[t]
    \begin{center}
        \includegraphics[width=0.75\textwidth]{{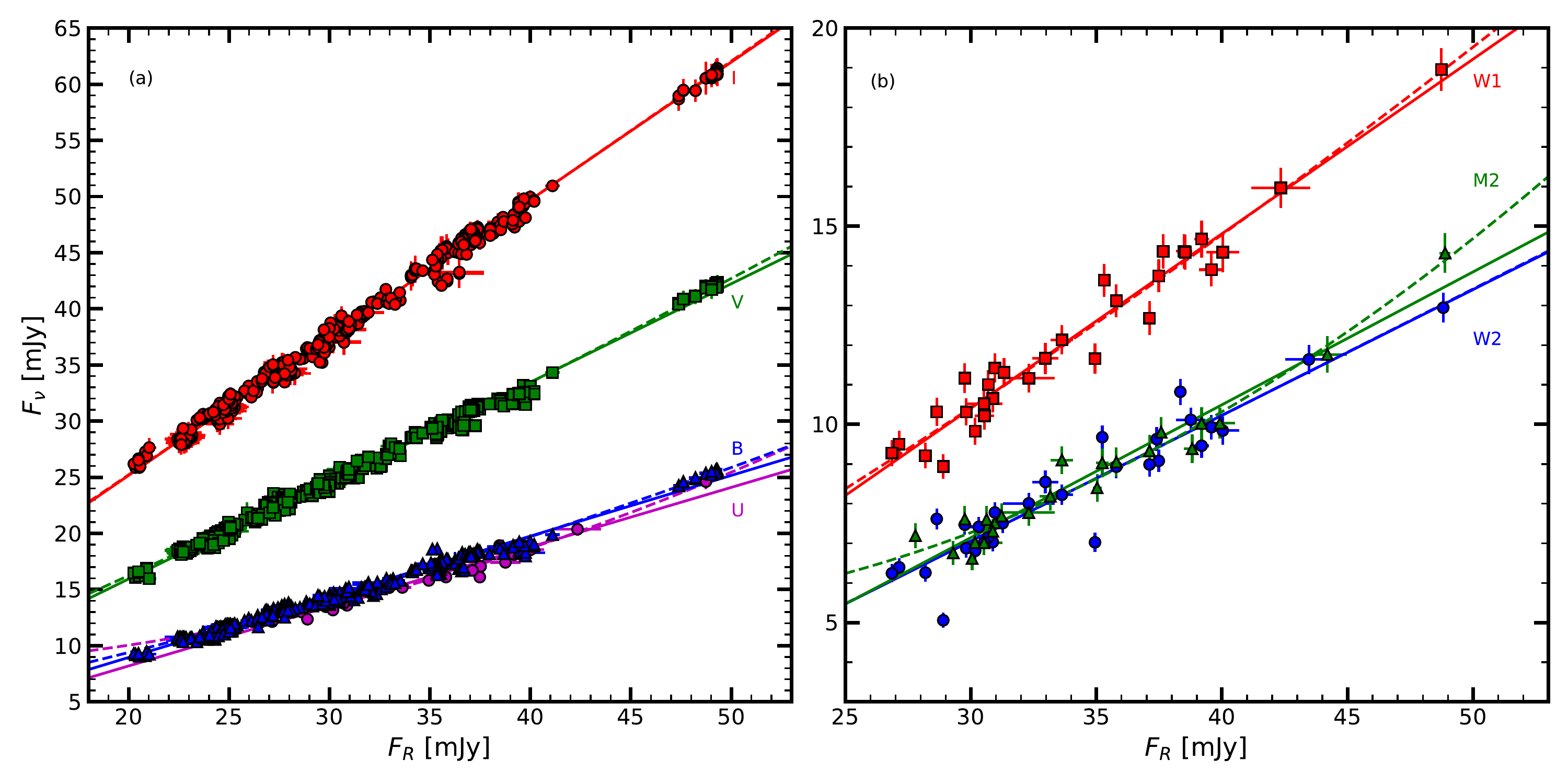}}
        \caption{Dereddened, host galaxy-corrected flux-flux dependencies in the optical-UV region. \emph{(a)} WEBT \emph{BVI} and \swift\ \emph{U} vs. \emph{R}. \emph{(b)} \swift\ \emph{W1, M2,} and \emph{W2} vs. \emph{R}.
        In both panels the solid and dashed lines represent linear and 2nd-degree polynomial fits, respectively.
        \label{fig:HTFluxFlux}}
    \end{center}
\end{figure*}

Figure~\ref{fig:HTFluxFlux} shows the optical and UV flux-flux relations relative to the WEBT \emph{R} band. To obtain this, we associated the UV data with the \emph{R}-band observations that were nearest in time.
For the \emph{BVI} dependencies, only WEBT data from telescopes with quasi-simultaneous multi-band observations of BL Lac were used from the entire time period of observations. The dependencies do not show any changes over time during the 3 months of observations. The optical (\emph{UBVI}) behavior is shown in panel \emph{(a)} of Fig.~\ref{fig:HTFluxFlux}, while the UV behavior is shown in panel \emph{(b)}.

\begin{deluxetable*}{cccDDDD}[t]
    \tablecaption{$\chi^2$ values and coefficients of fits to the flux-flux relations in Figure~\ref{fig:HTFluxFlux}.\label{tab:fluxfluxchi}}
    \tablewidth{0pt}
    \tablehead{
     \colhead{Waveband} & \colhead{N} & \colhead{Degree of Fit} & \multicolumn2c{$\chi^2$} & \multicolumn2c{$a$} & \multicolumn2c{$b$} & \multicolumn2c{$c$}
     }
     \startdata
     W2 &  30 & 1 & 1.3 & &                     & &  0.317 $\pm$ 0.022 & & -2.466 $\pm$ 0.753 & \\
        &     & 2 & 1.3 & & 0.0001 $\pm$ 0.0034 & &  0.311 $\pm$ 0.25 & & -2.357 $\pm$ 4.380 & \\
     \hline
     M2 &  23 & 1 & 0.44 & &                     & &  0.335 $\pm$ 0.017 & & -2.918 $\pm$ 0.603 &  \\
        &     & 2 & 0.34 & & 0.007 $\pm$ 0.002   & & -0.157 $\pm$ 0.177 & &  6.043 $\pm$ 3.245 &  \\
     \hline
     W1 &  30 & 1 & 0.63 & &                     & &  0.441 $\pm$ 0.019 & & -2.808 $\pm$ 0.636 &  \\
        &     & 2 & 0.62 & & 0.002 $\pm$ 0.003   & &  0.308 $\pm$ 0.205 & & -0.478 $\pm$ 3.654 &  \\
     \hline
     U  &  30 & 1 & 0.51 & &                     & &  0.530 $\pm$ 0.020 & & -2.410 $\pm$ 0.689 &  \\
        &     & 2 & 0.40 & & 0.008 $\pm$ 0.003   & & -0.056 $\pm$ 0.194 & &  7.911 $\pm$ 3.447 &  \\
     \hline
     B  & 429 & 1 & 4.4 & &                     & &  0.540 $\pm$ 0.003 & & -1.862 $\pm$ 0.092 &  \\
        &     & 2 & 3.7 & & 0.0034 $\pm$ 0.0003 & &  0.314 $\pm$ 0.021 & &  1.769 $\pm$ 0.340 &  \\
     \hline
     V  & 449 & 1 & 3.0 & &                     & &  0.876 $\pm$ 0.003 & & -1.591 $\pm$ 0.094 &  \\
        &     & 2 & 2.9 & & 0.0022 $\pm$ 0.0003 & &  0.727 $\pm$ 0.023 & &  0.810 $\pm$ 0.380 &  \\
     \hline
     I  & 473 & 1 & 4.3 & &                     & &  1.227 $\pm$ 0.004 & &  0.620 $\pm$ 0.129 & \\
        &     & 2 & 4.3 & & 0.0004 $\pm$ 0.0005 & &  1.201 $\pm$ 0.033 & &  1.042 $\pm$ 0.544 & \\
   \enddata
\end{deluxetable*}

We have fit a straight line ($F_\nu = b F_{\text{R}} + c$) and second-order polynomial ($F_\nu = a F_{\text{R}}^2 + b F_{\text{R}} + c$) to the data. Table~\ref{tab:fluxfluxchi} gives the results of a $\chi^2$ goodness of fit test for both fits for each band. In general, the $\chi^2$ test indicates a slight preference for a second-order polynomial fit for almost every waveband. However, the difference between the $\chi^2$ values for the two fits is small, as are the quadratic coefficients. We attribute this to the relatively modest amplitudes of variability during our observations, as was the case for \citet{Hagenthorn2004}. Therefore, we use linear fits for subsequent analyses.

\begin{figure}[t]
    \begin{center}
        \includegraphics[width=0.45\textwidth]{{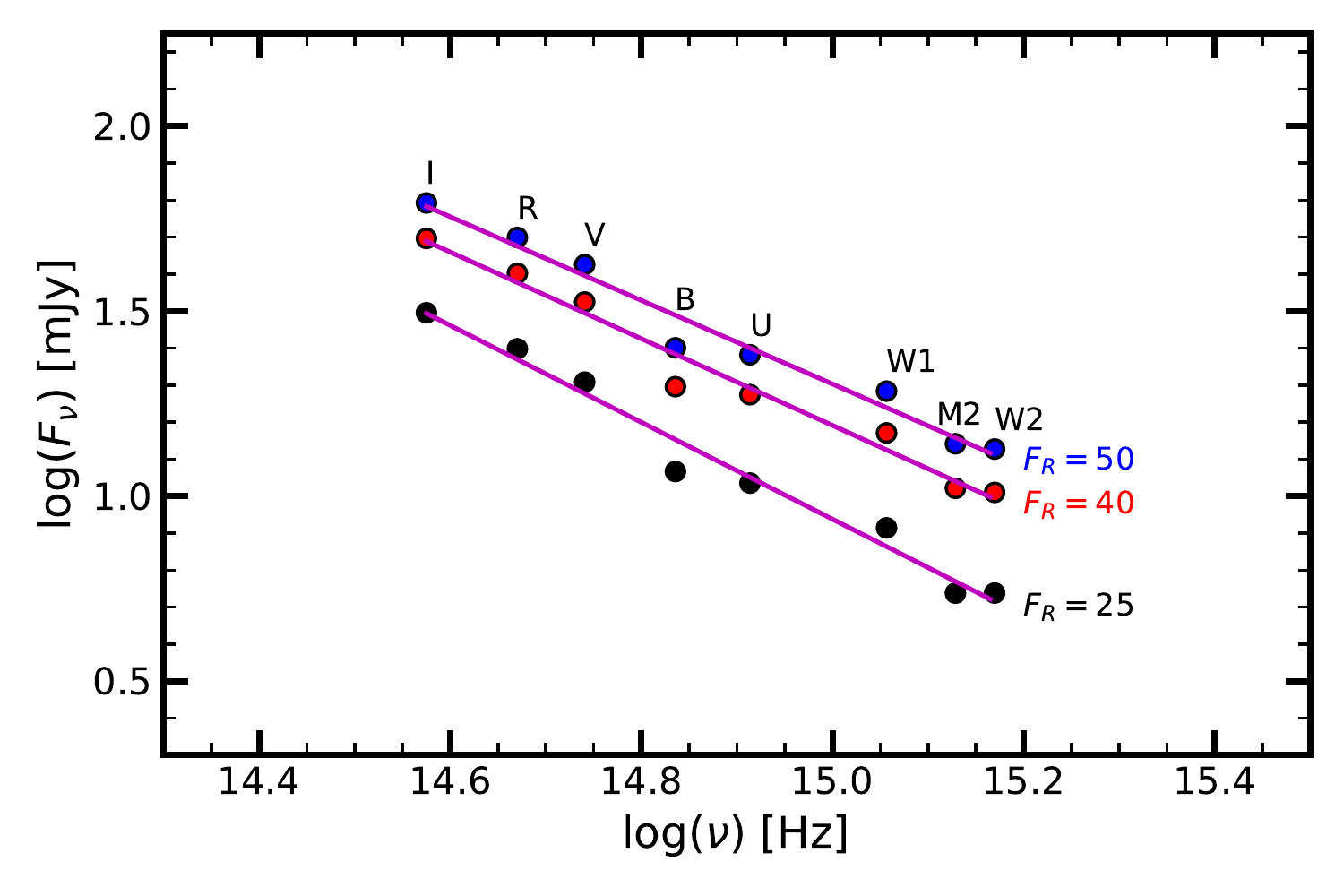}}
        \caption{Synthetic spectra of BL Lac in optical-UV bands obtained from polynomial regressions of flux-flux relations at different brightness levels in \emph{R}-band, 25 mJy (black), 40 mJy (red), and 50 mJy (blue); pink lines show linear fits to the spectra.
        \label{fig:FittedDopplerChange}}
    \end{center}
\end{figure}

\begin{figure}[t]
    \begin{center}
        \includegraphics[width=0.4\textwidth]{{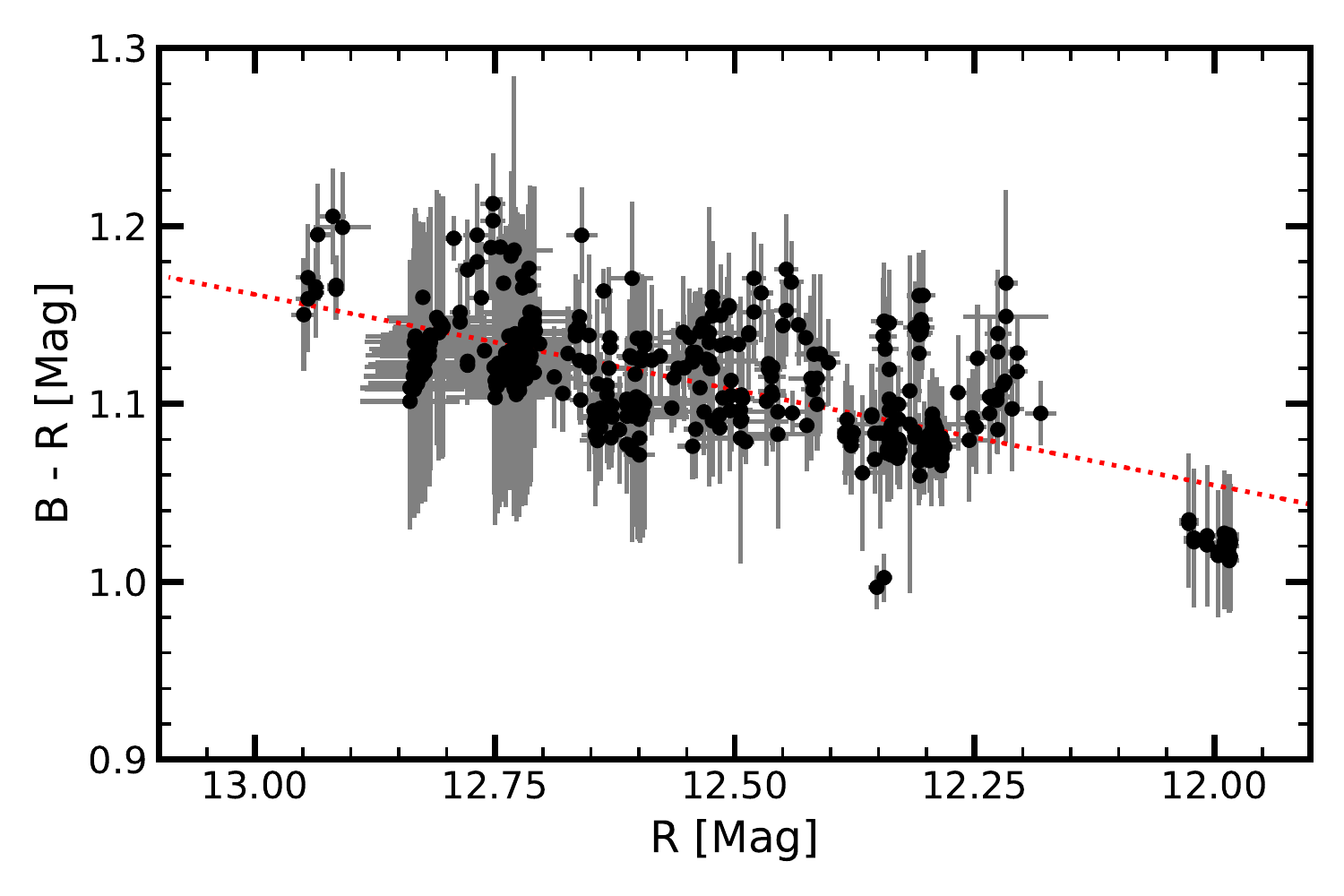}}
        \caption{Color index \emph{B}$-$\emph{R} of BL Lac as a function of \emph{R}-band brightness. The error bars are shown in grey for clarity. The red dotted line is a linear fit to the data.\label{fig:BminR}}
    \end{center}
\end{figure}

In Figure~\ref{fig:FittedDopplerChange} we show the synthetic spectra of BL Lac for \emph{R}-band \edit1{flux densities} $F_{\text{R}} = 25$, 40, and 50 mJy (in black, red, and blue, respectively). Table~\ref{tab:SplineFits} lists the coefficients of power-law fits to the synthetic spectra of the form $\log{(F_\nu)} = \alpha \log{(\nu)} + \beta$,
where $\alpha$ is the optical spectral index. Figure~\ref{fig:FittedDopplerChange} shows that the synthetic spectra corresponding to different brightness levels have slightly different slopes (see also Table~\ref{tab:SplineFits}), with the slope flattening toward higher flux states. This bluer-when-brighter trend is also apparent in Figure~\ref{fig:BminR}, which displays the \emph{B}$-$\emph{R} evolution as a function of \emph{R}-band brightness.
Color indices made with combinations of the other available filters show similar trends.

\begin{deluxetable}{cDD}[t]
    \tablecaption{Coefficients of linear fits to the synthetic spectra in Figure~\ref{fig:FittedDopplerChange} of the form $\log{(F_\nu)} = \alpha \log{(\nu)} + \beta$. \label{tab:SplineFits}}
    \tablehead{0pt}
    \tablehead{
    \colhead{$F_{\text{R}}$ [mJy]} & \multicolumn2c{$\alpha$} & \multicolumn2c{$\beta$}
    }
    \startdata
    25 & -1.31 $\pm$ 0.08 & & 20.6 $\pm$ 1.2 & \\
    40 & -1.17 $\pm$ 0.08 & & 18.6 $\pm$ 1.2 & \\
    50 & -1.13 $\pm$ 0.08 & & 18.3 $\pm$ 1.2 & \\
    \enddata
\end{deluxetable}

\begin{figure}
    \begin{center}
    \includegraphics[width=0.45\textwidth]{{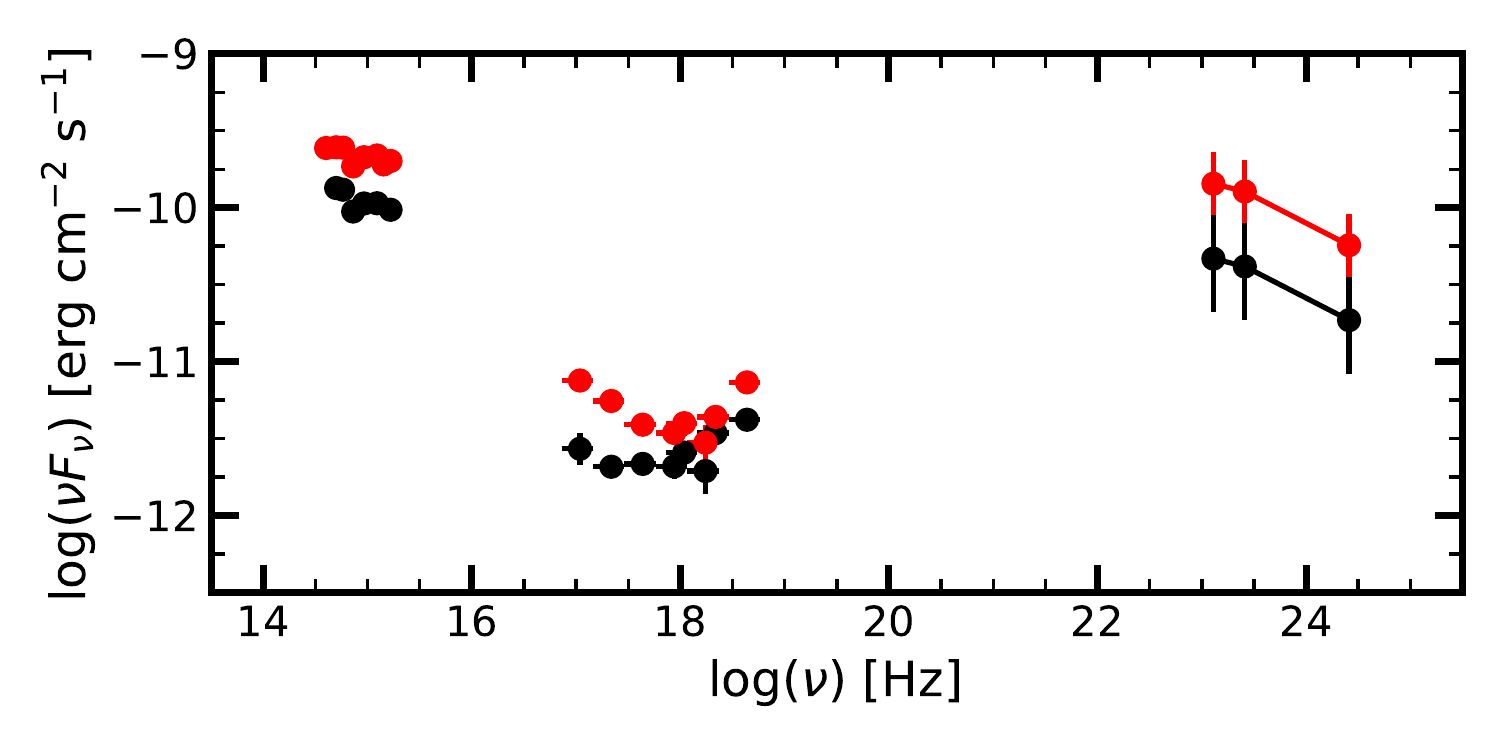}}
    \caption{SED of BL Lac during the low-flux plateau (black, September 16; MJD: 58742.4) and peak \emph{P2} (red, September 19; MJD: 58745.2). The \emph{R}-band flux density at these times was $F_{\text{R}} \approx 25$ and $F_{\text{R}} \approx 50$ mJy, respectively.\label{fig:SED}}
    \end{center}
\end{figure}

Figure~\ref{fig:SED} shows the SED of BL Lac at two different flux states, the low-flux plateau and peak \emph{P2}. To describe the X-ray portion of the SEDs, we have calculated 
fluxes within seven energy intervals: 0.3-0.6, 0.6-1.2, 2.4-4.8, and 4.8-9.6 (\swift), plus 3-6, 6-12 and 12-24 keV (NuSTAR), for simultaneously measured data, as described in $\S$\ref{subsubsec:SimultaneousNuSTARSwift}. BL Lac is brighter across all wavebands during event \emph{P2}. The soft X-ray portion of the SEDs appears to include contributions from both steep-spectrum synchrotron and flatter-spectrum IC emission components. The synchrotron component becomes more prominent at higher flux states, while the IC component dominates at hard X-ray energies. As mentioned in $\S$\ref{subsubsec:SimultaneousNuSTARSwift},
the break energy moves to higher energies as the X-ray flux increases. 

\begin{figure}
    \centering
    \includegraphics[width=0.45\textwidth]{{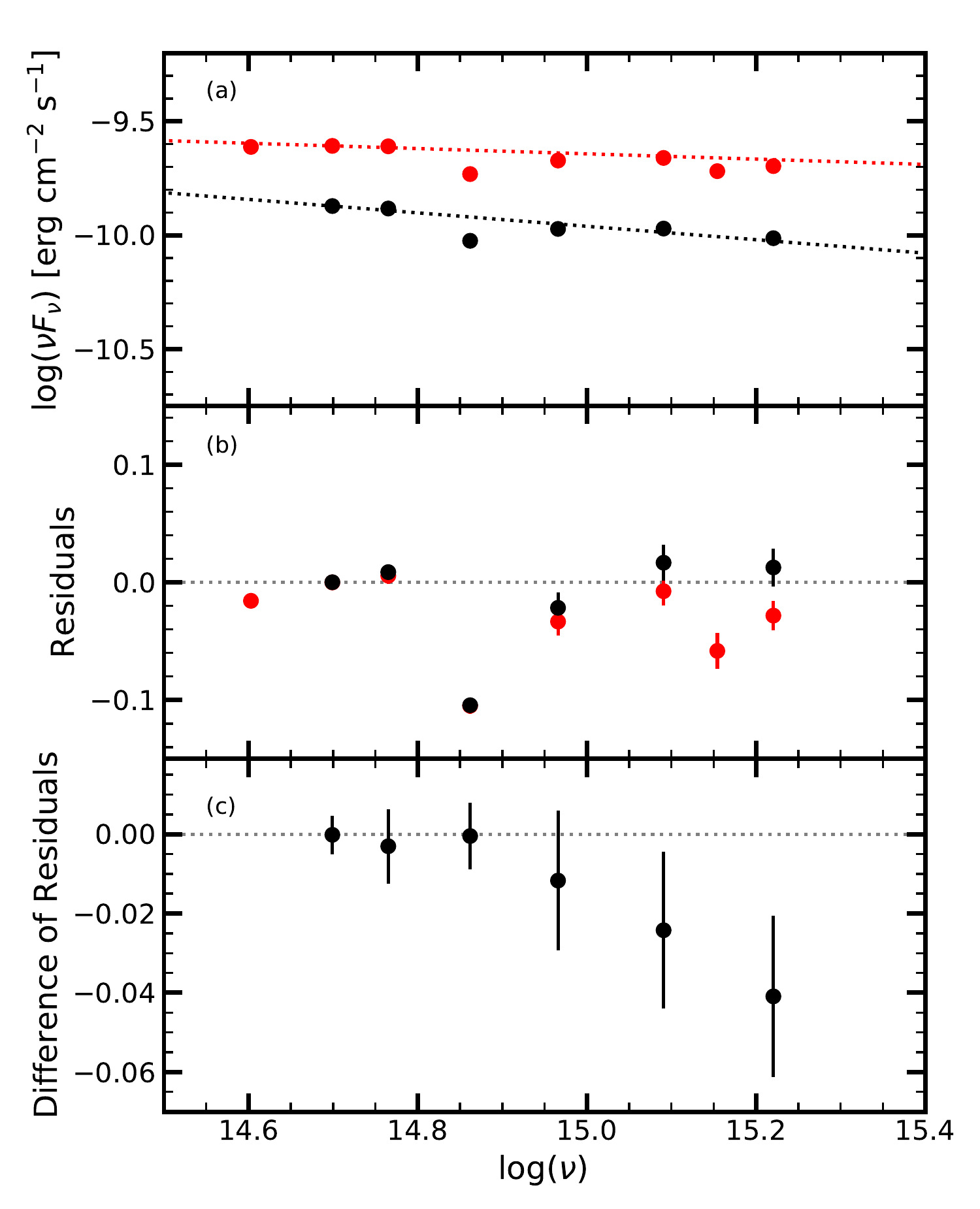}}
    \caption{\emph{(a)} Optical/UV SED of BL Lac for the low-flux plateau (black) and peak \emph{P2} (red), as in Figure~\ref{fig:SED}, with a power-law spectrum. \emph{(b)} Residuals of the power-law model to the data. \emph{(c)} Difference between the high-flux and low-flux residuals of the power-law model. \label{fig:BExcess}}
\end{figure}

In the optical region the spectral indices flatten from B band toward UV bands. \cite{Raiteri2009} have inferred a contribution from thermal emission in the optical/UV spectrum of BL Lac\edit1{, which they attributed to an accretion disk component. They have modeled such a disk as blackbody emission with a temperature $\geq$20,000~K and a luminosity $\geq6\times10^{44}$ erg~s$^{-1}$.} 
To investigate whether such a thermal component contributes to the SED in Figure \ref{fig:SED}, we have restricted the high- and low-flux SEDs to  optical/UV frequencies in Figure \ref{fig:BExcess}\emph{(a)}. We apply a power-law model (drawn with dotted lines) for both states, adopting the spectral indices from Table \ref{tab:SplineFits}. The small residuals of the model (Fig.\ \ref{fig:BExcess}\emph{(b)}), indicate that a single power-law can adequately describe the optical/UV emission.
[In all of our SEDs, the flux \edit1{density} in the \emph{B} filter is low compared to the expected power-law model, regardless of whether the data is taken from the WEBT or \swift\ observations. We suggest that this low flux density might arise from the wide filter bandpass and spectral shape of the source.]
While a single power-law component provides a general description of the observed flux \edit1{density}, we do see an excess in the UV portion of the spectrum of the low-flux compared to the high-flux state. This difference is clearly seen in Figure~\ref{fig:BExcess}\emph{(c)}, with the difference between the high- and low-flux residuals increasing toward shorter wavelengths. This suggests that a UV excess occurs at the low-flux state.  
However, an accretion disk component with the same characteristics as reported by \citet{Raiteri2009} would significantly exceed the optical-UV SED when added to the synchrotron component. Taking into account that accretion disk emission can significantly change on timescales of months to years \citep{Kaspi2000}, this may indicate a weakening of the disk contribution in recent years.

\section{Polarization Behavior}
\label{sec:polarization}

Over the entire WEBT campaign, we measured $P_{\text{R}}$ at 303 times, with the densest sampling during the five days of intensive X-ray monitoring. The data are displayed in Figure (Fig.~\ref{fig:EntireCampaignLC}\emph{(e)} and \emph{(f)}). The degree of polarization $P_{\text{R}}$ fluctuated rapidly between 1\% and 12\% over most of the observing period, although it was stable at $\sim9\%$ from mid-October to the beginning of November. \edit1{The average uncertainty on a measurement of $P_{\text{R}}$ is $\langle \sigma_{P_{\text{R}}} \rangle =0.23\%$.}

\edit1{Results of an ANOVA test can be considered reliable if the variable being tested is approximately normally distributed. However, as mentioned earlier, the degree of polarization follows a Rice distribution. Several Monte Carlo simulations have been performed to show that, as the sample size increases, the ANOVA test is robust against violations from normality \citep{Donaldson1966, Tiku1971}. As the number of measurements of $P_{\text{R}} \sim 300$ and the polarization data have been corrected for the Rice bias, we have thus used an ANOVA test, as described above, on the values of $P_{\text{R}}$ over the entire period. We have} calculated the $F$ statistic for 5 groups to be $F_{4,298} = 13.47$, with a $p$ value of $p = 4.2\times10^{-10}$. This confirms that $P_{\text{R}}$ was variable over the entire time period.

In order to determine the timescale of variability of $P_{\text{R}}$, we have searched for all pairs of measurements between which the values of $P_{\text{R}}$ differed by a factor $\geq2$, in order to calculate the halving or doubling timescale $\tau_2$. We restrict the analysis to the well-sampled observations between September 14 and 21 (MJD: 58740-58747), with $\Delta t < 50$ hr. We then identify the shortest timescale of variability under the constraint that the measurements meet the same uncertainty criterion as we used for the flux density: $|P_i - P_j| > (3/2)(\sigma_{P_i} + \sigma_{P_j})$.
The date restriction reduces the number of measurements to 223, yielding $\sim25,000$ data pairs. Of these, only 341 pairs meet the minimum $\Delta t$ and uncertainty criteria. For each of these data pairs, we assumed a linear change with time to determine $\tau_2$. 

\begin{figure}
    \centering
    \includegraphics[width=0.45\textwidth]{{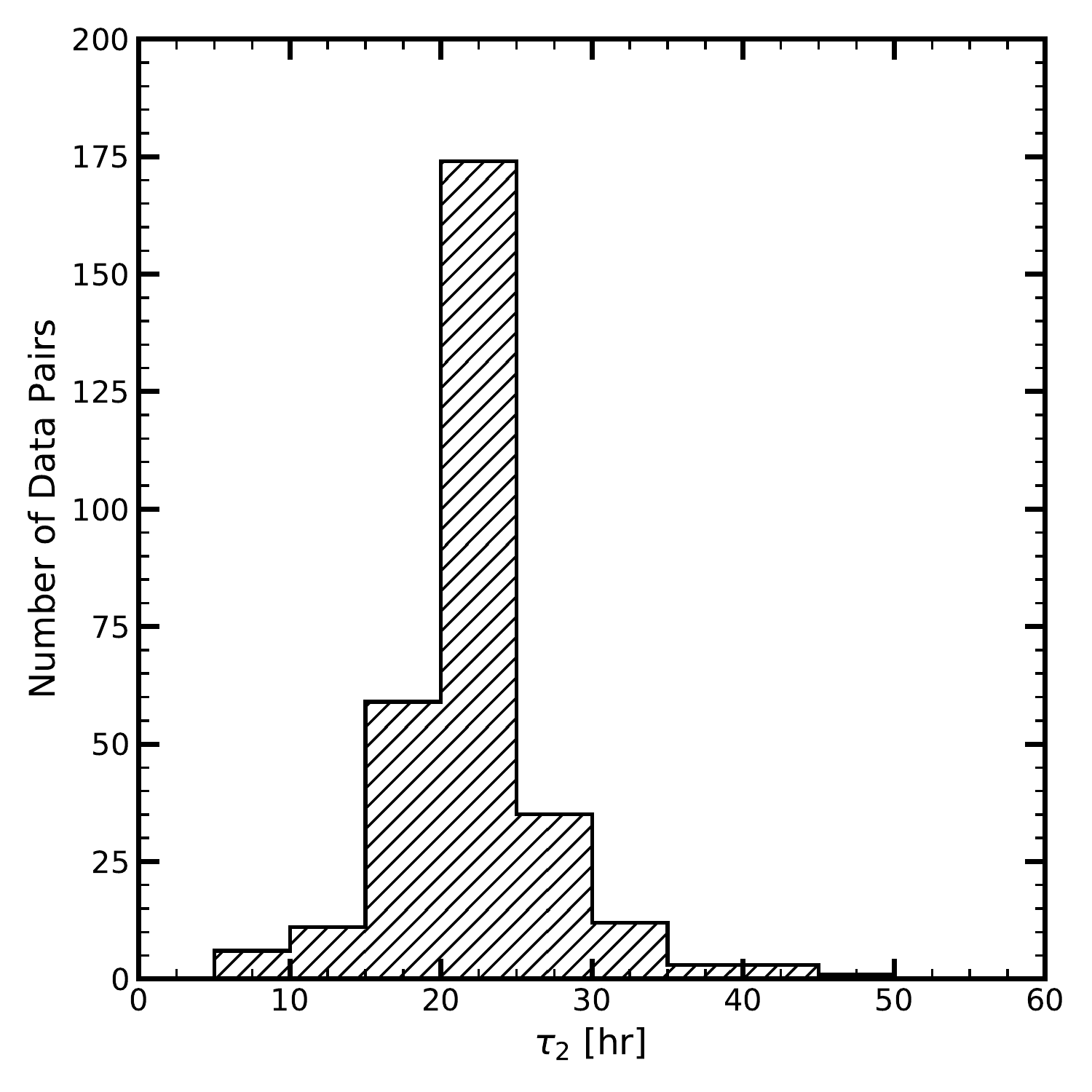}}
    \caption{Histogram of the timescales of changes in $P_{\text{R}}$ by a factor of 2. The minimum observed timescale is $\tau_2 = 5$ hr.\label{fig:PolarizationHist}}
\end{figure}

Figure~\ref{fig:PolarizationHist} shows the distribution of $\tau_2$ obtained from this fitting method. The minimum value is $\tau_2 = 5$ hr. The peak of the distribution lies in the 20-25 hour time bin, longer than that of the optical flux variations but shorter than than that of the X-ray variations. 

The position angle of the polarization $\chi_{\text{R}}$ was aligned with the direction of the 43 GHz parsec-scale radio jet \citep{Jorstad2017} at the beginning of the monitoring period. Swings away from this  polarization angle are seen in September during the high optical flux state.

At the beginning of the week of concurrent observations at all wavelengths, $P_{\text{R}}$ rose from $\sim5\%$ to $\sim9\%$ near in time to the X-ray brightening event \emph{P1}. This was accompanied by a swing in $\chi_{\text{R}}$ from parallel to the radio jet to $\sim45\degr$ away. While $\chi_{\text{R}}$ returned to nearly parallel to the jet direction shortly thereafter, $P_{\text{R}}$ remained high for two days, slowly decreasing from 9\% to 6\%.

Complicated polarization behavior is seen during the periods \emph{P1} and \emph{P2}. A large EVPA swing and increase in $P_{\text{R}}$ accompanied \emph{P1} when only the high-energy light curves showed a pronounced peak. However, a relatively stable EVPA $\sim 15\degr$ from the jet direction and varying $P_{\text{R}}$ occurred during \emph{P2} when pronounced changes in flux occurred at all wavelengths. Such behavior, with a stable EVPA despite changing flux and degree of polarization, has been observed in BL Lac in the past \citep{Gaur2014}, which is able to be incorporated into a shock-in-jet model if the shock Lorentz factor is low.

\begin{figure}[t]
    \begin{center}
        \includegraphics[width=0.45\textwidth]{{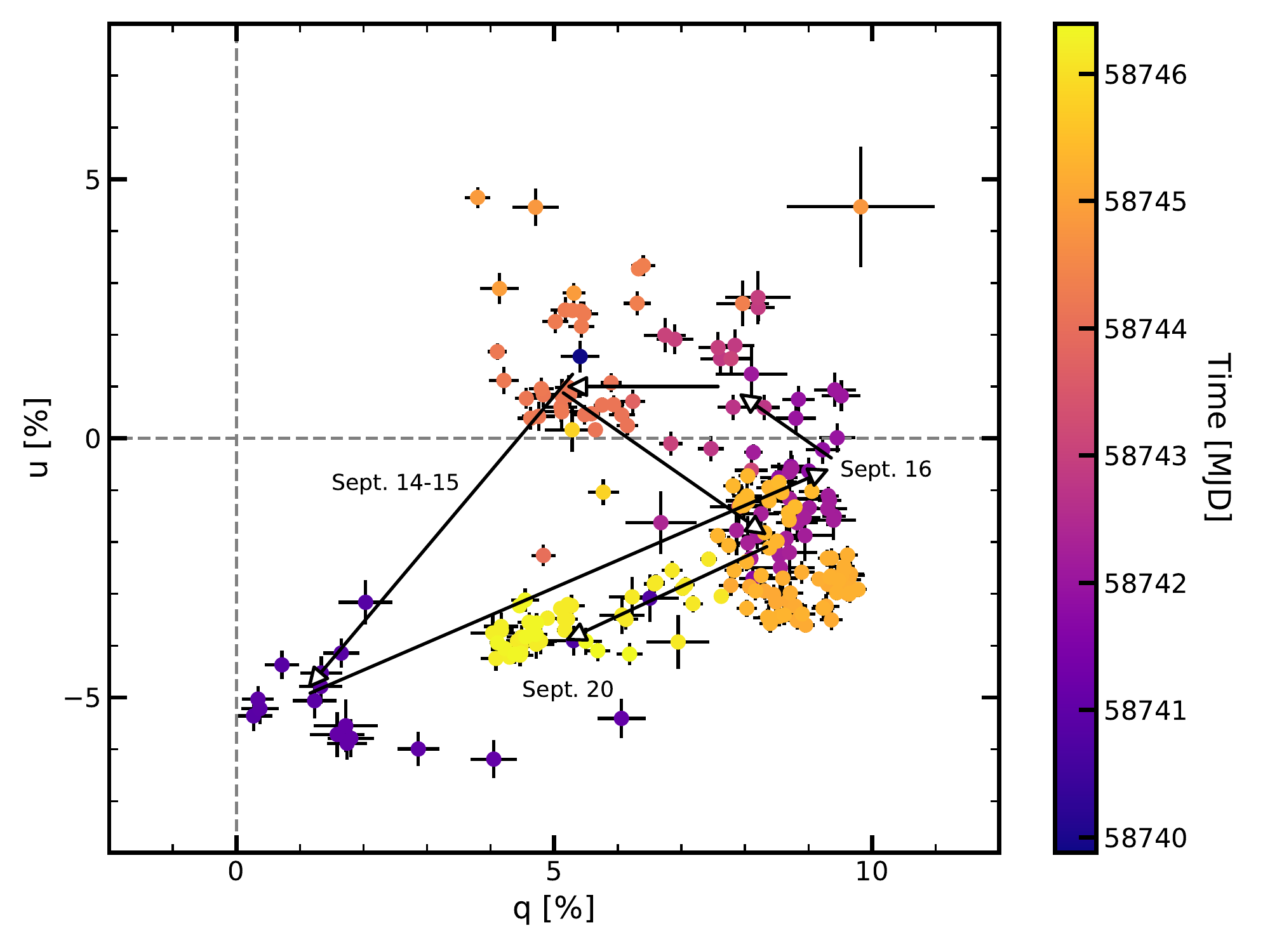}}
        \caption{\edit1{Normalized} Stokes $u$ vs.\ $q$ from September 14-20 (MJD: 58740-58746). The color bar shows the date of each observation. Six arrows trace the general trend over the time period. The gray dashed lines show $q$ or $u$ values of 0\%.\label{fig:stokesweek}}
    \end{center}
\end{figure}

Figure~\ref{fig:stokesweek} shows the polarization changes in the \edit1{normalized} Stokes parameter \emph{q} vs.\ \emph{u} plane for the time period September 14-20 \edit1{based on their measured values}. The polarization of BL Lac during this time was characterized by high positive values of $q$. A smooth connection can be made between the polarization states of BL Lac on different nights, appearing as swings or rotations in the \emph{q}-\emph{u} plane, although this trend does not change $\chi_{\text{R}}$ significantly (see Figure~\ref{fig:WeekLC2}\emph{e}). 

\begin{figure}[t]
    \begin{center}
        \includegraphics[width=0.45\textwidth]{{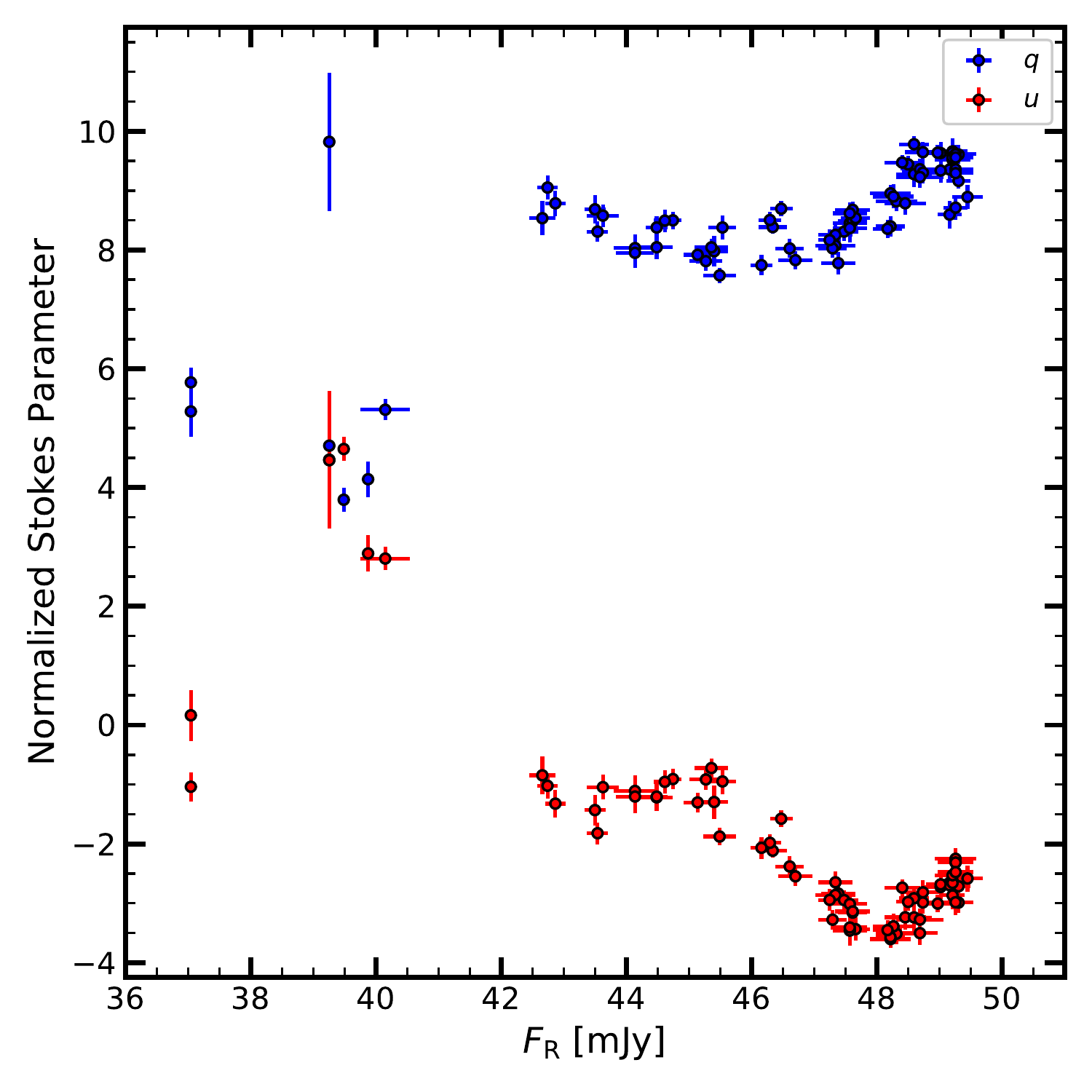}}
        \caption{The relationship between \edit1{normalized} Stokes parameters and \emph{R}-band flux density during the peak flux period \emph{P2}.\label{fig:QUP2}}
    \end{center}
\end{figure}

In order to characterize the origin of the rotation in the \emph{q}-\emph{u} plane, we display in Figure~\ref{fig:QUP2} the change in \emph{q} and \emph{u} as a function of the \emph{R}-band flux density at flux peak \emph{P2} seen across all wavelengths (September 18 18:00:00 - 20 00:00:00 UT). The relationship between \emph{q} and \emph{u} is markedly non-linear. A linear relation would be expected if a single variable source were responsible for the variability of both the flux and polarization \citep{Hagenthorn1999}.
The behavior could instead result from the superposition of emission from a number of components with different flux and polarization parameters, as we discuss in \S\ref{subsec:Interpretation} below. 

\section{Correlation Analysis}
\label{sec:Correlations}

\subsection{Short-Timescale Correlations}
\label{subsec:ShortTimescaleCorrelations}

\begin{deluxetable*}{llCCCcc}
  \tablecaption{Results of Multiwavelength Correlation Analysis.\label{tab:Correlations}}
  \tablewidth{0pt}
  \tablehead{
  \colhead{Band 1} & \colhead{Band 2} & \multicolumn3c{\texttt{PLIKE} Interval [days]} & \colhead{ZDCF Value of Peak} & \colhead{Bootstrap Significance} \\
  \colhead{}       &                  & \colhead{Lower Bound} & \colhead{Peak} & \colhead{Upper Bound}             & \colhead{} & \colhead{of ZDCF Peak}
  }
  \startdata
  Gamma-Ray & \TESS\                       & -0.158 & -0.105 & +0.016 & $0.314^{+0.097}_{-0.100}$ & 3.3 \\
  NuSTAR 10-22 keV & \TESS\                & +0.046 & +0.051 & +0.058 & $0.893^{+0.049}_{-0.063}$ & 4.2 \\
  NuSTAR 3-10 keV & \TESS                & +0.196 & +0.286 & +0.326 & $0.874^{+0.026}_{-0.029}$ & 11.0 \\
  \swift\ XRT 0.3-3 keV & \TESS        & +0.262 & +0.376 & +0.457 & $0.789^{+0.057}_{-0.065}$ & 6.8 \\
  \swift\ UVW2 & \TESS                 & -0.018 & +0.063 & +0.089 & $0.941^{+0.017}_{-0.020}$ & 8.8 \\
  \swift\ UVM2 & \TESS        & -0.021 & -0.001 & +0.035 & $0.952^{+0.016}_{-0.019}$ & 7.3 \\
  \swift\ UVW1 & \TESS     & -0.020 & +0.014 & +0.053 & $0.962^{+0.011}_{-0.013}$ & 8.7 \\
  \swift\ U & \TESS  & -0.036 & -0.013 & +0.017 & $0.965^{+0.010}_{-0.012}$ & 8.8 \\
  WEBT B  & \TESS  & -0.046 & -0.017 & -0.007 & $0.970^{+0.004}_{-0.005}$ & 18.6\\
  WEBT R & \TESS & -0.028 & -0.015 & -0.003 & $0.963^{+0.003}_{-0.003}$ & 32.7\\
  \enddata
  \tablecomments{Positive lags indicate that band 2 leads band 1.}
\end{deluxetable*}

We perform a correlation analysis between each of the light curves presented above and the \TESS\ light curve. We use the \emph{z}-transformed discrete correlation function \citep[ZDCF,][]{Alexander1997}, which provides a properly normalized (between -1 and 1) interval of results, along with uncertainties. The latter are sampling errors based on the noise of the original data, which we calculate using the recommended 100 Monte Carlo draws.

We use the \texttt{Peak Likelihood} algorithm \citep[\texttt{PLIKE},][]{Alexander2013} to estimate the maximum of the cross-correlation function (CCF) and the $1\sigma$ fiducial distribution of the lag interval. The \texttt{PLIKE} algorithm provides an estimate of the CCF peak and the uncertainty of the time lag without any \emph{a priori} assumptions about either the shape of the CCF peak or models of the light curves.

We verify the significance of the correlations by comparing the ZDCF results with the statistics of correlations on 3,000 pairs of bootstrapped artificial light curves (ALCs). We have removed points with excessively large errors per the recommendation of \citet{Alexander1997}, thus eliminating any need to weigh selected points. We generate the 3,000 ALCs by randomly selecting (with replacement) and placing flux points and associated uncertainties on the preserved observational dates. Once the ALCs have been built for each light curve to be compared, we randomly pair the ALCs (without replacement) and compute the ZDCF. Results of this analysis are used to derive a $2\sigma$ probability of obtaining a given coefficient of correlation by chance. In all cases where the ZDCF values of the actual light curve correlation are greater than 0.8, the bootstrap analysis generally gives a lag time consistent with the peak of the ZDCF. When the ZDCF is weaker, the bootstrap analysis still generally agrees, but frequently with less than $2\sigma$ significance. It is important to note that no result smaller than the bin size of the data is meaningful. In the case of the $\gamma$-ray light curve, the shortest meaningful time delay is 0.25 days. To reduce the impact of the upper-limits, we use flux values of half of the 2-$\sigma$ upper-limit values of the $\gamma$-ray data in all the correlations.

\begin{figure*}[t!]
  \begin{center}
    \includegraphics[width=5.25in,angle=90]{{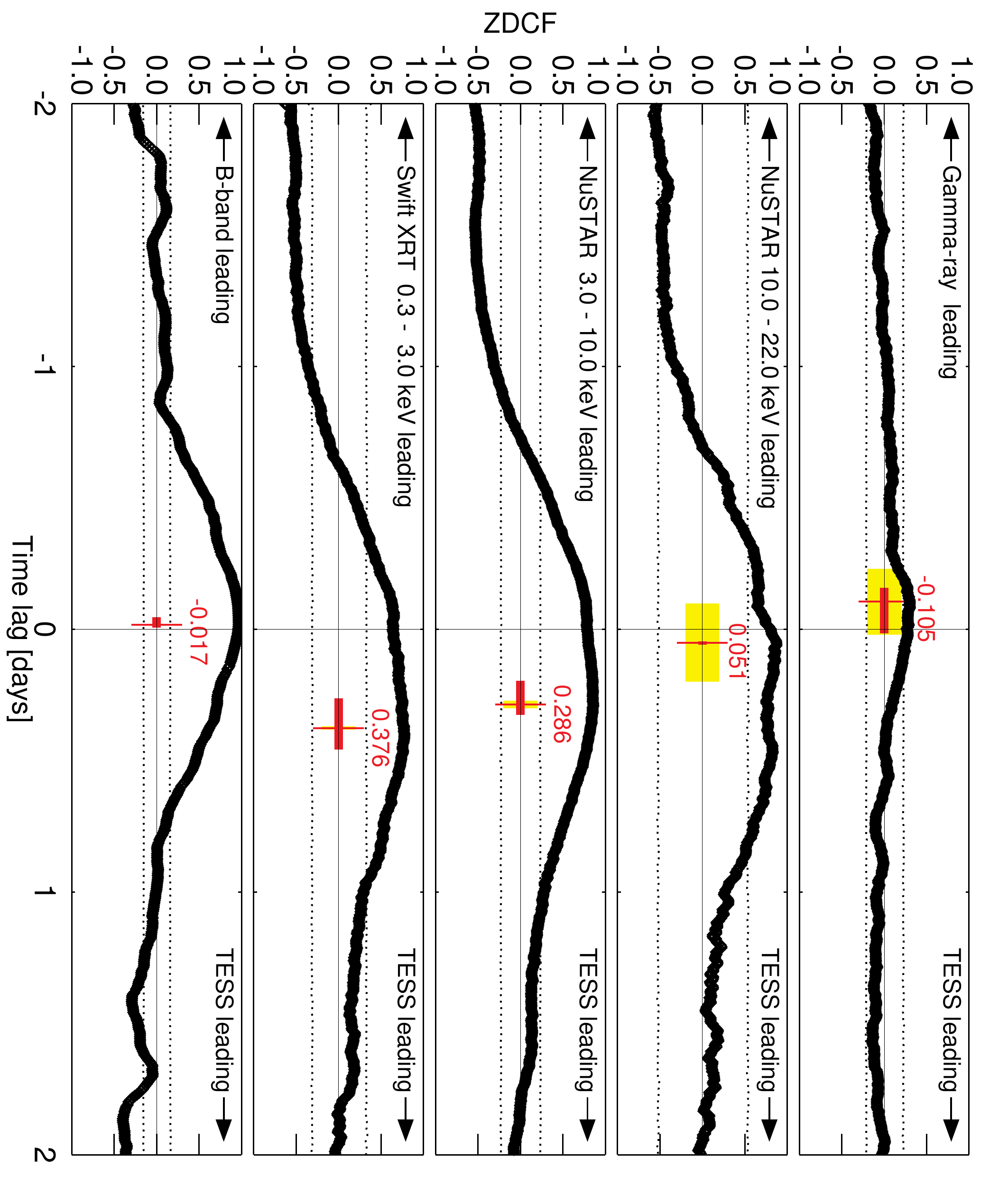}}
    \caption{ZDCF correlations of light curves at
    several wavebands with \TESS\ data. The red vertical line indicates the time lag of the peak of the ZDCF. The red horizontal line is the maximum likelihood \texttt{PLIKE} interval. The yellow shaded region marks the time binning of the data. The horizontal dotted lines indicate the $2\sigma$ probability of a chance correlation. \label{fig:ZDCF}}
  \end{center}
\end{figure*}

Table~\ref{tab:Correlations} lists values of the ZDCF peaks and their \texttt{PLIKE} lag fiducial intervals. A graphical representation of the ZDCF of several light curves with the \TESS\ light curve is shown in Figure~\ref{fig:ZDCF}. 
Due to the time binning of the $\gamma$-ray light curve, the results of the correlation analysis are consistent with zero \edit1{lag between the $\gamma$-ray and optical (\TESS) light curves}. 
However, there is a significant lag between the optical and X-ray variations. In particular, the \TESS\ light curve leads the X-ray variations by $0.38^{+0.08}_{-0.11}, 0.29^{+0.04}_{-0.09},$ and $0.051^{+0.007}_{-0.005}$ days for the \swift\ 0.3-3 keV, NuSTAR 3-10 keV, and NuSTAR 10-22 keV light curves, respectively. This trend of optical leading X-ray variations is seen with the WEBT \emph{B}- and \emph{R}-band and \swift\ UV light curves as well.

\subsection{Long-Timescale Correlations}
\label{subsec:LongTimescaleCorrelations}

It has been noticed in several FSRQs that the long-timescale optical and $\gamma$-ray variations are well correlated, with lags between 0 and $\sim$3 days, with the optical leading the $\gamma$-ray variations in some cases and the opposite in others \citep[e.g., in the well studied case of 3C454.3;][]{Bonning2009, Gaur2012, Kushwaha2017, Gupta2017}. We now use the 3-month long WEBT \emph{R}-band and \fermi\ $\gamma$-ray light curves to investigate whether such a correlation occurred in BL Lac during our monitoring period. We have binned the WEBT \emph{R}-band observations to match the binning of the $\gamma$-ray light curve, taking the average time of the \emph{R}-band observations for the correlation.

\begin{figure}
    \begin{center}
    \includegraphics[angle=90, width=0.45\textwidth]{{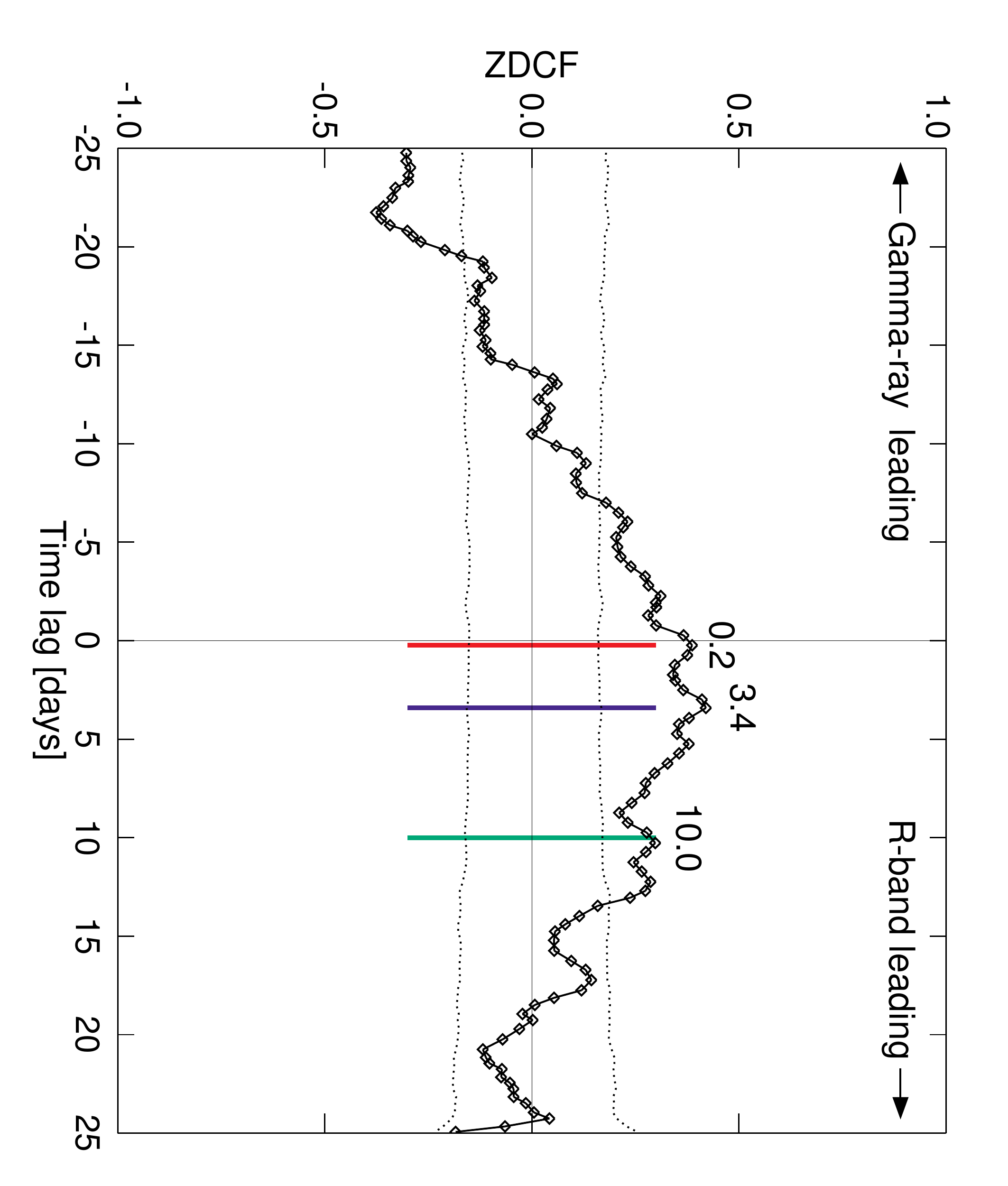}}
    \caption{ZDCF of \fermi\ $\gamma$-ray and WEBT \emph{R}-band light curves, from 2019 August 5 to November 2. The red, blue, and green vertical lines indicate the three major peaks in the correlation (at 0.2, 3.4, and 10.0 days, respectively). The dotted lines indicate a $2\sigma$ probability of a chance correlation.\label{fig:WEBTGammaLong}}
    \end{center}
\end{figure}

The ZDCF of the data is shown in Figure~\ref{fig:WEBTGammaLong}. Three major peaks are seen in the correlation, corresponding to the WEBT \emph{R}-band leading the $\gamma$-ray variations by 0.2, 3.4, and 10.0 days. We have examined each peak more closely, and the \texttt{PLIKE} maximum likelihood interval, value of the ZDCF peak, and bootstrap significance of each peak are given in Table~\ref{tab:WEBTGammaPeaks}. Again, no result smaller than the bin size of the data is meaningful, so the lag of $\sim$0.2 days is essentially the same as zero lag.

\begin{deluxetable*}{ccccc}
  \tablecaption{Results of WEBT \emph{R}-band and $\gamma$-ray Correlation Analysis.\label{tab:WEBTGammaPeaks}}
  \tablewidth{0pt}
  \tablehead{
  \multicolumn3c{\texttt{PLIKE} Interval [days]} & \colhead{ZDCF Value} & \colhead{Bootstrap Significance} \\
  \colhead{Lower Bound} & \colhead{Peak} & \colhead{Upper Bound} & \colhead{of Peak} & \colhead{of ZDCF Peak}
  }
  \startdata
  -0.21 & +0.23 & \phn +2.04 & $0.46^{+0.06}_{-0.06}$ & 8.27 \\
  +0.25 & +3.41 & \phn +5.09 & $0.46^{+0.06}_{-0.64}$ & 7.75 \\
  +9.95 & +9.95 & +12.94     & $0.34^{+0.07}_{-0.07}$ & 5.44 \\
  \enddata
  \tablecomments{Positive lags indicate that the \emph{R}-band leads the $\gamma$-ray band.}
\end{deluxetable*}

\begin{figure*}[t]
    \begin{center}
    \includegraphics[width=\textwidth]{{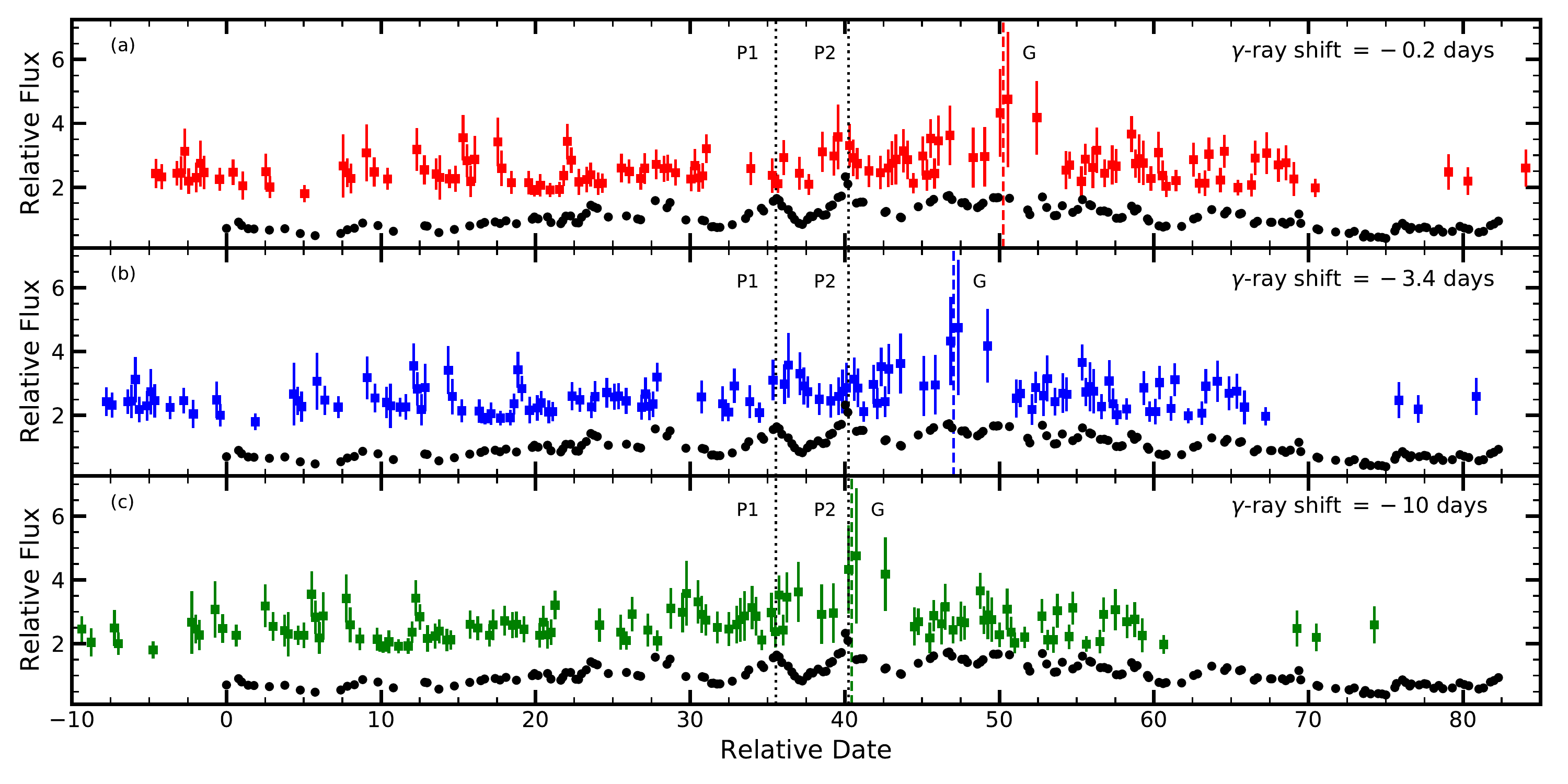}}
    \caption{WEBT \emph{R}-band (black circles) and \fermi\ $\gamma$-ray (colored squares) light curves. See text for scaling. The $\gamma$-ray light curve is shifted by the indicated amount along the $x$-axis in each panel, and upper limits are omitted for clarity. The optical peaks $P1$ and $P2$ have been identified with dotted black lines, and $\gamma$-ray peak \emph{G} with dashed colored lines. \label{fig:WEBTGammaLongCorr}}
    \end{center}
\end{figure*}

To visualize how such local maximize in the ZDCF arise, Figure~\ref{fig:WEBTGammaLongCorr} shows the $\gamma$-ray and WEBT \emph{R}-band light curves, with the $\gamma$-ray light curve shifted by the lags indicated in Table~\ref{tab:WEBTGammaPeaks}. We have normalized the $\gamma$-ray light curve by its median value and the \emph{R}-band light curve by half of its median value to more closely inspect the variations. Both light curves have been vertically shifted to not overlap.

Shifting the $\gamma$-ray light curve in time relative to the optical light curve reveals the cause of the peaks in the correlation. While the majority of the light curves feature low-amplitude flux variations, two large-amplitude variations dominate the correlation: matching the $\gamma$-ray peak near \emph{P2} with the optical peaks near \emph{P1} and\emph{P2} causes the 0.2 and 3.4 day correlations, while matching the large-amplitude $\gamma$-ray peak, \emph{G}, with optical peak \emph{P2} produces the local maximum in the ZDCF near 10 days. In fact, the matching of \emph{G} with \emph{P1} is also seen in the ZDCF as a second, slightly less statistically significant bump at a lag of $\sim13$ days in Figure~\ref{fig:WEBTGammaLong}.

The analysis of the local maxima of the ZDCF given in Table \ref{tab:WEBTGammaPeaks} finds that the statistical significance of the local maximum of the ZDCF at 0.2-day lag is higher than that at the other lags. Therefore, the correlation between the $\gamma$-ray and optical flux variations that is prominent during the 5-day period of intensive monitoring is also present in the 90-day data. The local maxima at the longer lags are less significant, resulting from offsets of peaks in the light curves that may be physically unrelated. 


\section{Discussion\label{sec:discuss}}
\subsection{Summary of Variability of BL Lac}
\label{subsec:SummaryVariability}


The optical, UV, and X-ray light curves of BL Lac derived from our observations are remarkably similar, as is evident from both visual inspection and our correlation analysis. In addition, despite numerous upper limits to the $\gamma$-ray flux, the ZDCF analysis finds a statistically significant, albeit low-amplitude, correlation between the $\gamma$-ray and \TESS\ light curves. The 5 days of concurrent observations with all telescopes feature two maxima in the optical to X-ray fluxes. The amplitudes of these short flares increases with frequency. A low-flux plateau lies between the two high-flux states.

Over the full 90 days of the observations reported here, BL Lac was significantly variable 20\% of the observed time at optical wavelengths, with a characteristic timescale of variability of 12-14 hours and a minimum timescale of $\sim 30$ minutes. Over the 5 days of intensive X-ray monitoring, the minimum timescale of variability was 14.5 hr. The polarization was variable over the entire 90-day monitoring period. We find a timescale of variation by a factor of 2 of the degree of optical {\emph R}-band polarization to be $\tau_2 = 5$ hr.


The optical polarization behavior of BL Lac does not appear to be strongly correlated with the optical flux. Periods of highly variable degree and position angle of polarization occur at times of both strongly variable and stable optical flux, and vice versa. A plot of \edit1{normalized} Stokes \emph{q} versus \emph{u} reveals some order to the variations in polarization, with positive \emph{q} values dominating and changes appearing as rotations or swings in the \emph{q}-\emph{u} plane with only minor changes in the ratio of \emph{u} to \emph{q}, and hence in the EVPA.

\subsection{Interpretation}
\label{subsec:Interpretation}

The non-linear relation of \emph{q} and \emph{u} suggests the superposition of a number of emission components with different flux and polarization parameters. This can be interpreted as evidence for turbulence in the jet \citep[e.g.,][]{Marscher2014},
which can be roughly modeled as $N_{\text{turb}}$ cells, each with a uniform but randomly oriented magnetic field \citep{Burn1966}. 
Under such a model, the degree of linear polarization is $\langle P \rangle \approx P_{\text{max}} N_{\text{turb}}^{-1/2}$ \citep{Burn1966}. Here $P_{\text{max}}$ corresponds to the degree of polarization of incoherent synchrotron emission in a uniform magnetic field, related to the spectral index $\alpha$ as: $P_{\text{max}} = (|\alpha|+1) / (|\alpha|+5/3) * 100$ \edit1{[\%]}.
Adopting the value $|\alpha| = 1.17$ corresponding to the optical spectrum when the flux density was at its median value, $F_{\text{R}} = 40$ mJy (Table~\ref{tab:SplineFits}), we obtain $P_{\text{max}} = 76\%$. The standard deviation of the polarization about this mean is predicted to be $\sigma_P \approx 0.5 P_{\text{max}} N_{\text{turb}}^{-1/2}$ \citep{Jones1988}.
For the observed mean polarization of 6.7\%, the required number of turbulent cells is $N_{\text{turb}} \approx 130$. The measured standard deviation of 2.1\% is somewhat less than the value of 3.4\% expected for the turbulence model. This can be reconciled by the existence of partial ordering of the magnetic field. For example, the partial ordering along the jet direction inferred from the mean EVPA can be the consequence of compression of the turbulent plasma by a shock \citep[e.g.,][]{Hughes1985,Cawthorne2006,Marscher2014} or the superposition of turbulence and a helical magnetic field \citep[e.g.,][]{Raiteri2010,Gabuzda2018}.

The analysis presented in \S\ref{sec:Correlations} reveals correlations between the \TESS\ and all other light curves. The X-ray variations lag behind the \TESS\ light curve by $\sim 0.38^{+0.08}_{-0.11}$ days at 0.3-3 keV, $0.29^{+0.04}_{-0.09}$ days at 3-10 keV, and $0.051^{+0.007}_{-0.005}$ days at 10-22 keV, respectively. The cross-frequency correlations and time lags imply that the emission regions in the various optical, X-ray, and $\gamma$-ray bands are partially co-spatial.
Such correlations and lags are expected if the variable X-ray emission in BL Lac is mainly caused by inverse Compton scattering of synchrotron seed photons in the frequency range of $\sim10^{12}$ to  $\sim10^{15}$ Hz by relativistic electrons that are energized at a front in the jet, e.g., a shock \citep{Marscher1985}. In this case, higher-energy electrons maintain their energy over a shorter distance beyond the shock than do their lower-energy counterparts. This leads to progressive longer lags of the synchrotron or inverse Compton variations toward lower frequencies that can be derived as follows.

\begin{figure}
  \begin{center}
    \includegraphics[width=0.45\textwidth]{{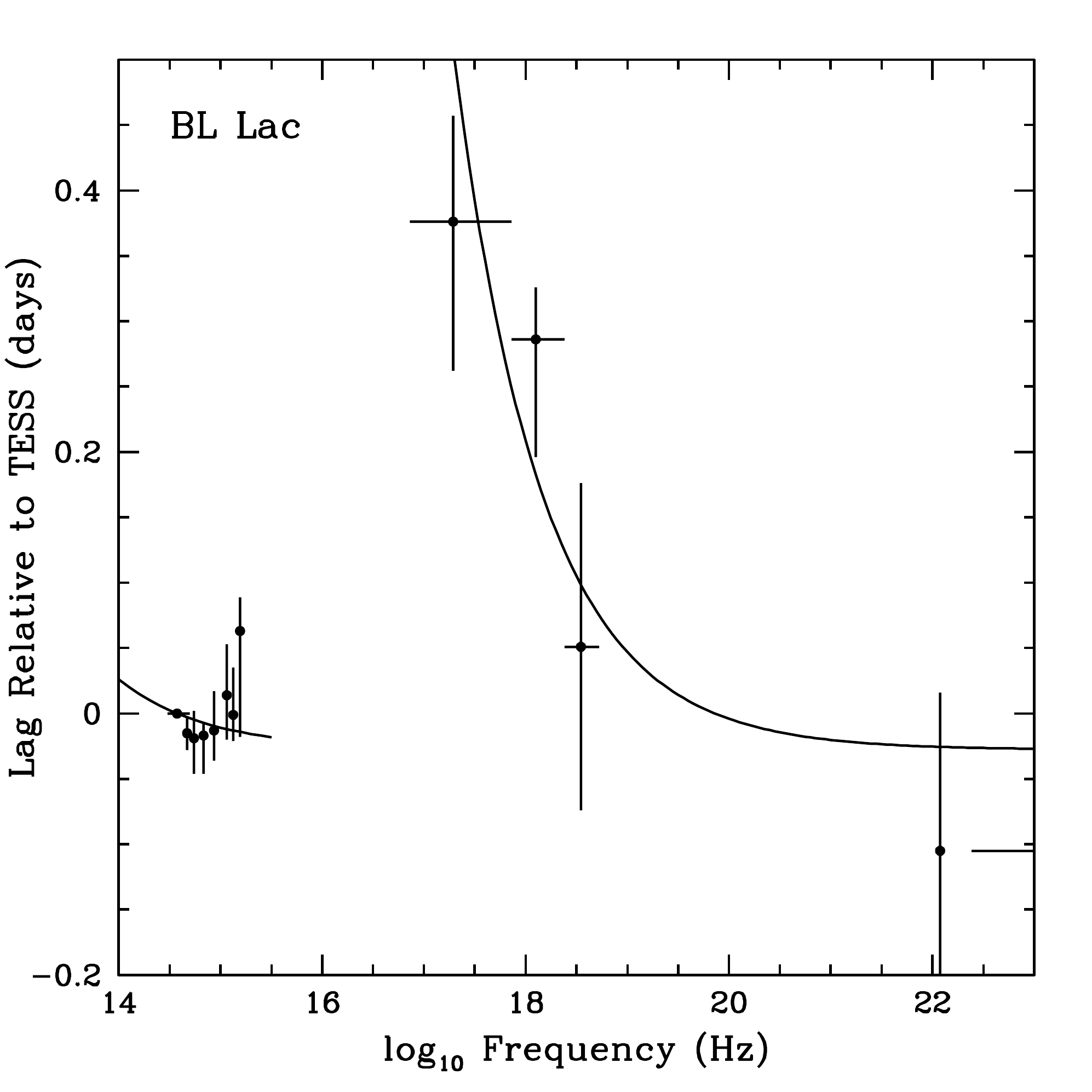}}
    \caption{Time lag of variations of the multi-frequency light curves with respect to the \TESS\ light curve. The solid curve represents the best-fit \citet{Marscher1985} model (see text for details).\label{fig:MarscherGearModel}}
  \end{center}
\end{figure}

The time lag between acceleration and energy loss is given in the rest frame of the emitting plasma by

\begin{equation}
  t'_{\text{loss}} \sim 7.7 \times 10^{8} \left[ \left( B^2 + 8\pi u_{\text{ph}} \right) \gamma \right]^{-1}\,{\text{s}},
  \label{eqn:RadiativeLosses}
\end{equation}

\noindent where $B$ is the magnetic field strength in Gauss, $u_{\text{ph}}$ is the energy density of seed photons for inverse Compton scattering in erg cm$^{-3}$, and $\gamma$ is the electron energy in rest-mass units. In the observer's frame, $t_{\text{loss}} = t'_{\text{loss}}(1+z)/\delta$, where $\delta$ is the Doppler factor. The SED displayed in Figure~\ref{fig:SED} indicates that the synchrotron IR and inverse Compton $\gamma$-ray luminosities are comparable. This implies that $B^2 \sim 8\pi u_{\text{ph}}$, which we assume to be the case. The value of $\gamma$ relates to the frequency of observation $\nu$ (in Hz) as

\begin{equation}
  \gamma \sim 6.0\times10^{-4} \left[ \nu(1+z) / (B\delta) \right]^{-1/2}
  \label{eqn:electronenergy}
\end{equation}

\noindent for synchrotron radiation, where $\delta\approx8$ \citep{Jorstad2017} is the Doppler factor and $z = 0.069$ is the host galaxy redshift. The mean value of $\gamma$ for inverse Compton scattering also depends on frequency as $\nu^{1/2}$. We then obtain

\begin{equation}
t_{\text{loss}} \sim 0.8 \left[{{B(1+z)}\over{\delta}}\right]^{1/2}\left({{\nu} \over{\nu_{\text{TESS}}}}\right)^{-1/2}
(B^2 + 8\pi u_{\text{ph}})^{-1}\, 
  \label{eqn:radlosstime}
\end{equation}
where $\nu_{\text{TESS}}= 4\times10^{14}$ Hz is the median frequency of the \TESS\ band and $t_{\text{loss}}$ is in days. 
Figure~\ref{fig:MarscherGearModel} presents the cross-frequency lag data relative to the \TESS\ light curve, along with the best fit to the equivalent frequency dependence from equation (\ref{eqn:radlosstime}), i.e., $t_{\text{loss}}(\nu)- t_{\text{loss}}(\nu_{\text{TESS}})$. In the fit, there is zero delay between the \TESS\ light curve and that at 300 keV (a free parameter, since we did not observe at an X-ray energy where there was zero lag). The magnetic field value of the fit is
$\sim3$ G. 
Despite the uncertainties in the lags between \TESS\ and the optical-UV light curves, the magnetic field strength is well specified by only the X-ray lags, which reflect the radiative energy losses. The $\nu^{-1/2}$ relation provides a good fit to the lag data, with a reduced $\chi^2 = 0.9$.

We can also equate the radiative energy loss timescale to the minimum timescale of variability in the \TESS\ band, 0.5 hr, and solve equation (\ref{eqn:radlosstime}) for $B$. This independent calculation also results in a value of $\sim3$ G.

Although the X-ray lags relative to the \TESS\ light curve agree with the above model, the SEDs displayed in Figure \ref{fig:SED} pose a problem. Although the hard X-ray spectrum has a spectral index $|\alpha|<1$ expected for inverse Compton scattering, the soft X-ray slope is steep during the high-flux states, $\alpha\approx -1.4$ (see Figure
\ref{fig:XRTPhIndvsFlux}). This suggests that the soft (0.3-3 keV) X-rays arise from a combination of synchrotron radiation by the highest-energy electrons whose energy distribution is steepened by radiative energy losses, and inverse Compton scattering by electrons with lower energies for which the radiative losses are modest. The above explanation of the X-ray time lag requires inverse Compton scattering to be the dominant emission process of the \emph{variable} component of the flux. The X-ray spectral slope during the first maximum of the light curves, \emph{P1}, is quite steep, with $\alpha\approx -1.5$, which strongly suggests that the soft X-ray flare represents the high-frequency end of the synchrotron spectrum. This conclusion, which agrees with previous studies \citep[e.g.,][]{Raiteri2010}, is supported by the pronounced variability in the 0.3-3 keV energy band near the peak (see Fig.\ \ref{fig:WeekLC1}), since the highest-energy electrons that emit synchrotron radiation at X-ray frequencies have the shortest timescales of energy loss. If this is true, then the soft X-ray variations should \emph{lead} the \TESS\ light curve during the flare. Indeed, Figures \ref{fig:WeekLC1} and \ref{fig:WeekLC2} show that a local maximum in the \TESS\ light curve lags behind the soft X-ray maximum. The correlation that gives a delay between the \TESS\ and 0.3-3 keV light curve then must arise from the long lower-flux period, when the soft X-ray spectrum was flatter and therefore the contribution of the inverse Compton component was more important. This is corroborated by the broken power-law fits to the combined \swift\ and NuSTAR spectra: the break energy shifts from $\sim6$ to $\sim2$ keV between the high and low X-ray flux states.

The magnetic field value that we infer from the time lags, $\sim 3$ G, is $\sim10$ times higher than that derived via the ``core shift'' method \citep{Lobanov1998} by \citet{OSullivan2009} \edit1{for the 43~GHz VLBI core located} at a distance of $\sim0.5$ pc from the vertex of the jet. However, the core shift method applies to the ambient jet rather than a shock that energizes electrons as in the above scenario. Compression of the magnetic field in such a shock is expected to increase the field strength by a factor roughly equal to the Lorentz factor, which is estimated to be $\sim 6$ in BL Lac \citep{Jorstad2017}. This implies that the variable emission reported here occurs $\sim0.3$ pc from the jet vertex, upstream of the 43 GHz ``core.''




\section{Conclusions}
\label{sec:Conclusions}

We have carried out a high time-resolution, multi-wavelength observing campaign of BL Lacertae, including monitoring at 2-min cadence with \TESS, in order to investigate the short-timescale variability of the blazar. Our dataset includes: (1) three months of observations with the \fermi-LAT and ground-based WEBT-affiliated telescopes, (2) 25 days of monitoring with \TESS, and (3) five days of densely-sampled NuSTAR and \swift\ measurements. 


All of the optical, UV, and X-ray light curves exhibit a similar trend during the five days of concurrent monitoring. Two high-flux states are separated by a low-flux plateau. The fractional amplitude of the variations is proportional to frequency up to at least the NuSTAR hard X-ray band.
The minimum timescale at optical wavelengths is very short, $\sim30$ min, while the average is 15 hr, very similar to the minimum observed X-ray timescale of 14.5 hr.


Our analysis of the observations confirms statistically significant correlations among the light curves at all frequencies. Frequency-dependent time lags relative to the \TESS\ variations can be explained by a model involving energization of the radiating electrons at a front, such as a shock, beyond which radiative energy losses restrict the emission to smaller volumes at higher frequencies \citep{Marscher1985}. Both the minimum timescale of variability in the \TESS\ band and the values of the time lags agree with such a model if the magnetic field is $\sim3$ G.

Consistent patterns of light curves, SEDs, and polarization versus time have proven elusive to find in blazar data. This is a consequence of both complexity in the physical processes in blazar jets and gaps in time and frequency coverage of monitoring programs. As our study demonstrates, the latter deficiency can be overcome by organizing intensive monitoring programs with current space- and ground-based facilities. Of particular importance to such efforts are instruments capable of essentially continuous monitoring, such as \TESS. Future similar campaigns with even longer duration are likely to provide further valuable insights into the time-variable phenomena that occur in relativistic jets.

\section{Acknowledgements}
{
We gratefully acknowledge the comments and suggestions provided by the anonymous referee that have improved this work.
The data collected by the WEBT Collaboration are stored in the WEBT archive at the Osservatorio Astrofisico di Torino - INAF (\url{https://www.oato.inaf.it/blazars/webt/}); for questions regarding their availability, please contact the WEBT President Massimo Villata (\url{massimo.villata@inaf.it}).
The research at Boston University was supported by NASA grants 80NSSC19K1731 (TESS Guest Investigator Program), 80NSSC20K0080 (NuSTAR Guest Investigator Program), 80NSSC17K0649 (Fermi Guest Investigator Program), and Massachusetts Space Grant 316080, as well as by Boston University Hariri Institute Research Incubation Award 2019-03-007. 
This research was partially supported by the Bulgarian National Science Fund of the Ministry of Education and Science under grants DN 18-10/2017, DN 18-13/2017, KP-06-H28/3 (2018) and KP-06-PN38/4 (2019). The Skinakas Observatory is a collaborative project of the University of Crete, the Foundation for Research and Technology -- Hellas, and the Max-Planck-Institut f\"ur Extraterrestrische Physik.
The work at the University of Sofia was supported by Bulgarian National Science Fund under grant DN18-10/2017 and National RI Roadmap Projects DO1-277/16.12.2019 and DO1-268/16.12.2019 of the Ministry of Education and Science of the Republic of Bulgaria.
ES and SI were supported by funds of the project RD-08-122/2020 of the University of Shumen, Bulgaria.
KM acknowledges JSPS KAKENHI grant no. JP19K03930.
GD gratefully acknowledges the observing grant support from the Institute of Astronomy and NAO Rozhen, BAS, via the bilateral joint research project ``Gaia Celestial Reference Frame (CRF) and Fast Variable Astronomical Objects." This work is a part of the following projects: no. 176011 ``Dynamics and Kinematics of Celestial Bodies and Systems," no. 176004 ``Stellar Physics," and no. 176021 "Visible and Invisible Matter in Nearby Galaxies: Theory and Observations," supported by the Ministry of Education, Science, and Technological Development of the Republic of Serbia.
S.O.K. acknowledges financial support by Shota Rustaveli National Science Foundation of Georgia under contract PHDF-18-354.
This work is partly based upon observations carried out at the Observatorio Astron\'omico Nacional on the Sierra San Pedro Martir (OAN-SPM), Baja California, Mexico.
E.B. acknowledges financial support from UNAM-DGAPA-PAPIIT through grant IN113320.
Support for KLS was provided by NASA through Einstein Postdoctoral Fellowship Award Number PF7-180168.
This article is partly based on observations made with the LCOGT Telescopes, one of whose nodes is located at the Observatorios de Canarias del IAC on the island of La Tenerife in the Observatorio del Teide.
The work at Colgate University was supported by the Colgate University Research Council.
We gratefully acknowledge the contribution of data from D. Blinov and the Robopol program.
This study includes data collected by the \TESS\ mission. Funding for the \TESS\ mission is provided by the NASA Explorer Program. The \TESS\ data analysis benefited from conversations and developmental work that took place at the \emph{Expanding the Science of \TESS} meeting at the University of Sydney. 
This study has used Digitized Sky Survey images based on photographic data obtained with the Oschin Schmidt Telescope on Mount Palomar to aid in the reduction and visualization of the \TESS\ data. The Palomar Observatory Sky Survey was funded by the National Geographic Society. The Oschin Schmidt Telescope is operated by the California Institute for Technology and Palomar Observatory. The plates were processed into the present compressed digital format with their permission. The Digitized Sky Survey was produced at the Space Telescope Science Institute (STScI) under U.S. Government Grant NAG W-2166. 
This work has made use of data from the European Space Agency (ESA) mission
{\it Gaia} (\url{https://www.cosmos.esa.int/gaia}), processed by the {\it Gaia}
Data Processing and Analysis Consortium (DPAC,
\url{https://www.cosmos.esa.int/web/gaia/dpac/consortium}). Funding for the DPAC
has been provided by national institutions, in particular the institutions
participating in the {\it Gaia} Multilateral Agreement.
This research made use of Astropy,\footnote{\url{http://www.astropy.org}} a community-developed core Python package for Astronomy \citep{Astropy2013, Astropy2018}, and astroquery, an astronomical web-querying package in Python \citep{Astroquery2019}.
}




\vspace{5mm}
\facilities{Fermi, NuSTAR, Swift, TESS, WEBT, Gaia}

\software{Astropy \citep{Astropy2013, Astropy2018}, astroquery \citep{Astroquery2019}, eleanor \citep{Feinstein2019}, HEAsoft \citep{HEASARC2014}, Fermi Science Tools \citep{FermiScienceTools2019}, pyrallaxes \citep{Luri2018}}

\bibliography{BLLacBibliography}{}
\bibliographystyle{aasjournal}

\end{document}